\documentclass[prb,twocolumn,superscriptaddress]{revtex4}
\usepackage[dvips]{graphicx}
\usepackage{latexsym}
\begin{document}

\newcommand{\fig}[2]{\includegraphics[width=#1]{#2}}

\title{The Roton Fermi Liquid}

\author{Leon Balents}
\affiliation{Department of  Physics, University of California,
Santa Barbara, CA 93106--4030}
\author{Matthew P. A. Fisher}
\affiliation{Institute for Theoretical Physics,
University of California, Santa Barbara, CA 93106--4030}

\date{\today}

\begin{abstract}
  We introduce and analyze a novel metallic phase of two-dimensional
  (2d) electrons, the Roton Fermi Liquid (RFL), which, in contrast to
  the Landau Fermi liquid, supports both gapless fermionic and bosonic
  quasiparticle excitations.  The RFL is accessed using a re-formulation
  of 2d electrons consisting of fermionic quasiparticles and $hc/2e$
  vortices interacting with a mutual long-ranged statistical
  interaction.  In the presence of a strong vortex-antivortex (i.e.
  roton) hopping term, the RFL phase emerges as an exotic yet eminently
  tractable new quantum ground state.  The RFL phase exhibits a ``Bose
  surface'' of gapless roton excitations describing transverse current
  fluctuations, has off-diagonal quasi-long-ranged order (ODQLRO) at
  zero temperature ($T=0$), but is not superconducting, having zero
  superfluid density and no Meissner effect.  The electrical resistance
  {\it vanishes} as $T \rightarrow 0$ with a power of temperature (and
  frequency), $R(T) \sim T^\gamma$ (with $\gamma >1$), independent of
  the impurity concentration.  The RFL phase also has a full Fermi
  surface of quasiparticle excitations just as in a Landau Fermi liquid.
  Electrons can, however, scatter anomalously from rotonic ``current
  fluctuations'' and ``superconducting fluctuations''.  Current
  fluctuations induced by the gapless rotons scatter anomalously only at
  ``hot spots'' on the Fermi surface (with tangents parallel to the
  crystalline axes), while superconducting fluctuations give rise to an
  anomalous lifetime over the entire Fermi surface {\sl except} near the
  (incipient) nodal points (``cold spots'').
  Fermionic quasiparticles dominate the Hall electrical transport.
  We also find three dominant instabilities of the RFL phase: an
  instability to a conventional Fermi liquid phase driven by vortex
  condensation, a BCS-type instability towards fermion pairing and a
  (non-pairing) superconducting instability.  Precisely {\it at} the
  instability into the Fermi liquid state, the exponent $\gamma$
  saturates the bound, $\gamma =1$, so that $R(T) \sim T$.  Upon
  entering the superconducting state the rotons are gapped out, and the
  anomalous quasiparticle scattering is strongly suppressed.  We discuss
  how the RFL phase might underlie the strange metallic state of the
  cuprates near optimal doping, and outline a phenomenological picture
  to accomodate the underdoped pseudo-gap regime and the overdoped
  Landau Fermi liquid phase.
\end{abstract}

\maketitle

\vspace{0.15cm}

\section{Introduction}
\label{sec:introduction}

Despite the appeal of spin-charge separation as an underpinning to
superconductivity in the cuprates, there are seemingly fatal obstacles
with this approach.  Ever since Anderson's initial
suggestion\cite{PWA1} of a spinon Fermi surface in the normal state at
optimal doping, there have been nagging questions about the {\it
  charge} sector of the theory.  The concept of the ``holon'', a
charge $e$ spinless boson, was introduced in the context of the doped
spin liquid state\cite{KRS}, and was presumed to be responsible for
the electrical conduction.  In the simplest theoretical scenario,
spin-charge separation occurs on electronic energy scales, thereby
liberating the electron's charge from its Fermi statistics.  This
idea has been actively investigated over the past 15 years -- see
Ref.~\onlinecite{SubirReview} for a review.  A
pervasive challenge to this perspective, however, is the difficulty of avoiding
holon condensation and superconductivity at {\it inappropriately} high
temperatures.  In addition, recent theoretical work, which has
elucidated the phenomenology of putative spin-charge separated states,
has led to further conflicts with observations.  One class of theories
have shown how spin-charge separation can emerge from a
superconducting phase by pairing and condensing vortices.\cite{NL}\
Following this work, a $Z_2$ gauge theory formulation greatly
clarified the nature of fractionalization of electronic (and other)
quantum numbers.\cite{Z2}\ It has become clear that a necessary
requirement for true spin charge separation in two dimensions is the
existence of a ``vison'' excitation with a
gap.\cite{earlyvison1,earlyvison2,vison1,vison2}\ The vison is perhaps
most simplest thought of within the vortex pairing picture, as the
remnant of an unbound single vortex.  If these ideas were to apply to
the cuprates, one would expect this gap to be of order the pseudogap
scale $k_B T^*$.  Unfortunately for spin-charge separation advocates,
experiments designed to detect the vison and measure its
gap\cite{visexp1,visexp2}\ have determined an unnaturally low upper
bound of approximately $150K$ for the vison gap in underdoped YBCO. Is
this the death knell for spin-charge separation?

In a very recent paper focusing on the effects of ring-exchange in
simple models of bosons hopping on a 2d square lattice\cite{EBL}, we
have identified a novel zero temperature {\it normal fluid} phase -
(re-)named the ``Exciton Bose Liquid'' (EBL).  In the EBL phase
boson-antiboson pairs (ie an exciton) are mobile, being carried by a
set of gapless collective excitations, while single bosons cannot
propagate.  The resulting quantum state is ``almost an insulator'',
with the d.c. conductivity vanishing as a power of temperature,
$\sigma(T) \sim T^\alpha$ with $\alpha \ge 1$.  This is in contrast to
the ``strange metallic phase'' in the optimally doped cuprates, which
is ``almost superconducting''\cite{PWAbook} with an extrapolated zero
resistance at $T=0$, as if it were a superconductor with $T_c=0$.
This phenomenology suggests the need for a non-superconducting quantum
phase in which the vortices are strongly immobilized at low
temperatures.

Motivated by this, we revisit the $Z_2$ vortex-spinon field theory of
interacting electrons\cite{NL,Z2}, in which the $hc/2e$ vortices and
the spinons have a long-ranged statistical interaction mediated by two
$Z_2$ gauge fields.  Rather than gapping out single vortices while
condensing pairs (which leads to a spin-charge separated
insulator)\cite{Z2}, we would like to find a quantum phase in which
the vortices are gapless but nevertheless immobile.  To this end, we
add an additional ``vortex-ring'' term to the earlier vortex field
theory.  This term is effectively a kinetic energy for
vortex-antivortex pairs, that is for rotons.  To access the limit of
strong roton hopping requires a further reformulation of the $Z_2$
vortex-spinon field theory, replacing the $Z_2$ gauge fields by $U(1)$
gauge fields.  Similar $U(1)$ vortex-fermion field theories have been
explored in Refs.~\onlinecite{tesanovic}.  The resulting $U(1)$ vortex-spinon
formulation is 
tractable in the limit of a very strong roton hopping, and describes
the ``Roton Fermi Liquid'' (RFL) phase, a novel metallic ground
state qualitatively different than Landau's Fermi liquid phase.

Both the $Z_2$ vortex-spinon field theory developed in
Refs.~[\onlinecite{NL,Z2}] and our $U(1)$ vortex-fermion field theory,
described in Section II below, are constructed in terms of operators
which create the excitations of a conventional BCS superconductor: the
Bogoliubov quasiparticles, the $hc/2e$ vortices and the collective
plasmon mode.  This choice of ``basis'', however, does not presume
that the system is necessarily superconducting at low temperatures,
and indeed we intend to employ such a formulation to describe
non-superconducting states.  When superconductivity {\it is} present
at low temperatures, the formulation will also be employed to describe
the ``normal'' state above $T_c$.  In these approaches, the Bogoliubov
quasiparticle is electrically neutralized\cite{NL}, and the resulting
``spinon'' excitation transported around an $hc/2e$ vortex within an
ordinary 2d BCS superconductor, acquire a Berry's phase of $\pi$.
Within the $Z_2$ and $U(1)$ vortex-spinon formulations, one introduces
spinon creation operators, $f^\dagger_{{\bf r}\sigma}$, where we let
${\bf r}$ denote the sites of a 2d square lattice and with $\sigma$
the spin.  Vortex creation operators are also introduced, conveniently
represented in a ``rotor'' representation as $e^{i\theta}$, which live
on the plaquettes of the 2d lattice.  The vortices are minimally
coupled to a gauge field, living on the links of the dual lattice.
The ``flux'' in the gauge field describes charge density fluctuations
on the original lattice sites, and for example encapsulates the
plasmon mode inside the superconducting phase\cite{Duality}.  Finally,
the long-ranged statistical interaction between the spinons and
vortices is incorporated by introducing two Chern-Simons $Z_2$ or
$U(1)$ gauge fields.  The important new element in the present paper
is the inclusion of a roton hopping term.  As we shall see, the RFL
phase is readily accessed within the $U(1)$ formulation when the
magnitude of this roton hopping term is taken signifigantly larger
than the single vortex hopping strength.

Here, we briefly summarize the main results established in the following
Sections.  In Sec. II the $Z_2$ vortex-spinon field theory formulation
of Ref.~[\onlinecite{Z2}] is recast in terms of a lattice Hamiltonian.
Via a sequence of exact unitary transformations on the Hamiltonian, we
demonstrate that it is possible to exchange the $Z_2$ Chern-Simons gauge
fields for their $U(1)$ counterparts.  Within a Lagrangian
representation of the resulting $U(1)$ Hamiltonian which we employ
throughout the paper, it is possible to choose a gauge for one of the
$U(1)$ Chern-Simons fields so that the ``spinon'' is recharged, and has
finite overlap with a bare electron.  (In App.~\ref{ap:elecform} we show
that this gauge choice can effectively be made at the Hamiltonian level,
and construct a Hamiltonian theory in terms of vortices and fermionic
operators which carry the electron charge and have a finite overlap with
a bare electron.)

Initially, in Sec. III, we ignore the fermions entirely, and focus on
the bosonic charge (or vortex) sector of the theory.  A ``spin-wave''
expansion valid in the presence of a large roton hopping term, leads
to a simple theory which is quadratic - {\it except} for a single
vortex hopping term.  Dropping this vortex hopping term then leads to
a soluble harmonic theory of the ``Roton-Liquid'' (RL) phase.  In
addition to the gapless 2d plasmon, the RL phase is shown to support a
``Bose surface'' of gapless roton excitations.  We compute the Cooper
pair propagator in the RL phase, and show that it exhibits
off-diagonal quasi-long-ranged order (ODQLRO) at zero temperature, but
{\it not} a Meissner effect.  The RL phase exhibits a high degree of
``emergent'' symmetry - the number of vortices on every row and column
of the 2d dual lattice is asymptotically conserved at low energies.
This symmetry implies that the harmonic ``fixed point'' theory of the
RL phase has an infinite conductivity at any temperature.

In Sec. IV we study the legitimacy of the approximations used to
arrive at the harmonic RL theory, focussing first on the neglected
vortex hopping term.  We show that for a range of parameters the
vortex hopping term is ``irrelevant'', scaling to zero at low energies
whenever its associated scaling dimension satisfies $\Delta_v \ge 2$.
Nevertheless, at finite temperatures vortex hopping leads to
dissipation, giving a resistance which vanishing as a power law in
temperature, $R(T) \sim T^\gamma$ with $\gamma = 2 \Delta_v -3 \ge 1$.
A ``plaquette duality'' transformation\cite{EBL} allows us to next
address the legitimacy of the initial ``spin-wave'' expansion, used to
obtain the harmonic RL theory. Of paramount importance is the presence
of a term in the dual theory which hops a ``charged'' quasiparticle
excitation, a term {\it not} present in the harmonic fixed-point
theory. We find that the ``charge'' hopping process is irrelevant over
a range of parameters - approximately the complement of the range
where vortex hopping was irrelevant - implying stability of the RL
phase.  When relevant, on the other hand, the ``charge'' quasiparticle
condenses, leading to a superconducting ground state.

The fermions are re-introduced back into the theory in Section V, where
we argue for the stability of the Bose surface of rotons and the 2d
plasmon in the presence of a {\sl gapless} Fermi sea of fermionic
quasiparticles.  We denote the corresponding phase by the Roton Fermi
Liquid (RFL).  The gaplessness of the fermions is somewhat surprising,
and deserves some comment.  Indeed, it is in sharp contrast to the
gapped nature of the quasiparticles in both the superconducting phase
and $Z_2$ fractionalized insulator, in which the spinons experience a
BCS-like ``pair field''.  The cause of this difference is the existence
of gapless single vortex excitations (and fluctuations) in the RFL,
which according to our analysis leads to the ``irrelevance'' of the
fermion pairing term.  Crudely, because the bosonic pair field exhibits
only OD{\sl Quasi-}LRO rather than ODLRO, there is no average pair field
felt by the quasiparticles, and hence no gap.  In this sense, the RFL is
in fact closer to a Fermi-liquid state than it is a superconductor.

Section VI is devoted to an analysis of the properties of the RFL phase.
We study both the longitudinal and Hall conductivities, and find that
the dissipative electrical resistance vanishes with a power of
temperature, $R(T) \sim T^\gamma$ with $\gamma \ge1$, similar to the
behavior without fermions present.  But the fermions are found to
dominate the Hall response, leading within a na\"ive Drude treatment to
an inverse Hall angle varying as $\Theta_H^{-1} \sim 1/(\tau_f^2
T^\gamma)$, with $\tau_f$ the fermionic quasiparticle transport
lifetime.  Due to the presence of the ``Bose surface'' of gapless
rotons, electrons at finite energy $\omega$ experience anomalous
scattering, {\it not} present in a Landau Fermi liquid.  Specifically,
quasiparticles scatter due to rotonic ``current fluctuations'' and
``superconducting fluctuations'', which contribute additively to the
electron decay rate.  The former gives rise to especially strong
electron scattering at ``hot spots'' -- points on the Fermi surface with
tangents parallel to the axes of the square lattice.  At such hot spots
the associated electron decay rate varies with an anomalous power of
energy, $\omega^{(\gamma + 2)/2}$, for $1 < \gamma < 2$.  The decay rate
from superconducting fluctuations is present everywhere along the Fermi
surface {\sl except} near ``cold spots'' at the incipient $d$-wave nodal
points.  This contribution grows strongly with decreasing
energy/temperature, although it has a smaller overall amplitude than the
current fluctuation contribution.   Upon entering a superconducting state, the
rotons -- gapless within the RFL phase -- become gapped out, and all
anomalous scattering is strongly suppressed.

Finally, in Section VII we briefly discuss possible connections of the
present work to the cuprates.  We suggest that the RFL phase might
underlie the unusual behavior observed near optimal doping in the
cuprates, in particular the ``strange metal'' normal state above $T_c$.
A scenario is outlined which also incorporates the pseudo-gap regime and
the conventional Fermi liquid behavior in the strongly overdoped limit.

\section{The Model}
\label{sec:model}

We are interested in electrons hopping on a 2d square lattice, with
electron creation operators $c^\dagger_{{\bf r}\sigma}$.  Here, the
sites are denoted (in bold roman characters) as, ${\bf r}=x_1 {\bf \hat{x}}
+ x_2 {\bf \hat{y}}$, where $x_1,x_2$ are integers, $\hat{\bf
  x}_1=\hat{\bf x}$, $\hat{\bf x}_2=\hat{\bf y}$ are unit vectors
along the $x$ and $y$ axes, and $\sigma = \uparrow,\downarrow$
denotes the two spin polarizations of the electron (such a spin
index will be distinguished from Pauli matrices $\sigma^\mu$ by
the lack of any superscript).

\subsection{$Z_2$ Chargon-spinon formulation}
\label{sec:z_2-chargon-spinon}

We begin by formulating the electron problem in spin-charge
separated variables using the $Z_2$ gauge theory Hamiltonian.  We
emphasize that this formulation does not imply that spin-charge
separated excitations are deconfined, and indeed this formulation
correctly describes the low-energy physics of conventional
confined phases as well.

The $Z_2$ gauge theory is most readily formulated in terms of a
charge $e$ singlet bosonic chargon $b_{\bf r},b_{\bf r}^\dagger$,
a neutral spin-$1/2$ fermionic spinon $f_{{\bf r}\sigma}, f_{{\bf
r}\sigma}^\dagger$, and an Ising gauge field (Pauli matrix)
$\sigma_j^\mu({\bf r})$ residing on the link between sites ${\bf
r}$ and ${\bf r}+\hat{\bf x}_j$.  It is convenient to use a rotor
representation for the chargons, $b_{\bf r} = e^{-i\phi_{\bf r}}$,
$b^\dagger_{\bf r} b^{\vphantom\dagger}_{\bf r} = n_{\bf r}$, with
$[\phi_{\bf r},n_{\bf
  r'}]=i\delta_{\bf r,r'}$.

The $Z_2$ Hamiltonian is conveniently expressed in terms of the
Hamiltonian density, $H_{Z_2}=\sum_{\bf r} {\cal H}_{Z_2}$, which
in turn is decomposed into a bosonic charge sector, a fermionic
sector and a gauge field contribution, ${\cal H}_{Z_2}={\cal H}_c
+ {\cal H}^{Z_2}_f + {\cal H}_g$:
\begin{eqnarray}
  \label{eq:Z2chargeham}
  {\cal H}_c & = & - t_c \sum_j \sigma_j^z({\bf r}) \cos(\partial_j\phi_{\bf
    r}-A_j({\bf r})) + u_c (n_{\bf r}-\rho_0)^2 , \\
  {\cal H}_g & = & - t_v \sum_j \sigma^x_j({\bf r}) \!-\!
  K\!\!\!\!\prod_{\Box({\bf r}+{\bf w})} \!\!\!\!\sigma^z \!\!\!
  - \kappa_r\! \prod_j \sigma_j^x({\bf r}) \sigma_j^x({\bf
    r}\!+\!\hat{\bf x}_j), \\
  {\cal H}^{Z_2}_f & = &  -  \sum_j \sigma_j^z({\bf r})
  [ t_s f^\dagger_{{\bf r} +  {\bf \hat{x}}_j \sigma}
  f^{\vphantom\dagger}_{{\bf r} \sigma}+\Delta_j f_{{\bf r}+\hat{\bf
      x}_j \sigma} \epsilon_{\sigma\sigma'}f_{{\bf r}\sigma'}+ {\rm
    h.c.}] \nonumber \\
  & & - t_e \sum_j f^\dagger_{{\bf r} +  {\bf \hat{x}}_j
    \sigma} e^{i (\partial_j \phi_{{\bf r}} -A_j({\bf r}))}
  f^{\vphantom\dagger}_{{\bf r} \sigma}+ h.c. . \label{HfZ2}
\end{eqnarray}
Here, $\rho_0 \equiv 1 - x$ is the electron (charge) density with $x$
measuring deviations from half-filling. Throughout the paper,
$\partial_j$ with $j=1,2$ denotes a {\it discrete} (forward) spatial
lattice derivatives in the $x_1$ and $x_2$ directions, for example,
$\partial_1 \phi_{\bf r} = \partial_x \phi_{\bf r} = \phi_{{\bf r} +
  {\bf \hat{x}}} - \phi_{\bf r}$.  We have included an external
(physical) vector potential $A_j({\bf r})$ in order to calculate
electromagnetic response and to include applied fields. The
Hamiltonian ${\cal H}_c$ describes the dynamics of the chargons
hopping with strength, $t_c$, which are minimally coupled to the $Z_2$
gauge field. The dynamics of the gauge fields is primarily determined
from ${\cal H}_g$, the first two terms of which constitute the
standard pure $Z_2$ gauge theory Hamiltonian.  The ``magnetic''
contribution involves the plaquette product,
\begin{equation}
\prod_{\Box({\bf r}+{\bf w})} \sigma^z \equiv \sigma^z_1({\bf r})
\sigma^z_1({\bf
    r}+ \hat{\bf y}) \sigma^z_2({\bf r}) \sigma^z_2({\bf r}+\hat{\bf
    x}),
\end{equation}
which is the $Z_2$ analog of the lattice curl. Here we have
defined, ${\bf w} = (1/2)(\hat{x}+\hat{y})$, and ${\bf r} +{\bf
w}$ denotes the center of the plaquette.  We have also included an
additional contribution bi-linear in $\sigma^x$, which in the dual
vortex representation below will become a ``roton'' hopping term.
In the fermion Hamiltonian, ${\cal H}^{Z_2}_f$, we have defined
the antisymmetric matrix
$\epsilon_{\sigma\sigma'}=i\sigma^y_{\sigma\sigma'}$, and take
$\Delta_j=(-1)^j\Delta$, which describes a nearest-neighbor pair
field with d-wave symmetry.  Apart from the first two terms
familiar to aficionados of the $Z_2$ gauge theory\cite{Z2}, we
have included a less exotic bare electron hopping amplitude $t_e$.
We will primarily be interested in the limit that the spinon
hopping strength is significantly larger than the electron hopping
strength, $t_s \gg t_e$.

In most of the analysis of this paper, we will consider the limit of
small spinon pairing $\Delta_j \rightarrow 0$.  This can be justified
either by the assumption $\Delta_j \ll t_s$, or by the irrelevance in
the renormalization group (RG) sense, which will occur occur in some
regimes.  If strictly $\Delta_j=0$, both fermion (spinon) number and
boson charge are conserved, and in principle may be separately fixed.
However, for $\Delta_j \rightarrow 0$, even infinitesimal, this is not
the case.  Instead the spinons will {\sl equilibrate} in some time that
diverges as $\Delta_j\rightarrow 0$ but is otherwise finite, and the
system will choose a unique fermion density to minimize its (free)
energy.  We will return to this point in the $U(1)$ formulation in
Sec.~\ref{sec:u1-vortex-spinon-1}.

The full Hamiltonian above has a set of local $Z_2$ gauge
symmetries, commuting with each of the local operators,
\begin{equation}
  \label{eq:Z2constraint}
  {\cal C}_{\bf r}^1 = (-1)^{n_{\bf r}+ n^f_{{\bf r}}} \prod_{+({\bf r})} \sigma^x ,
\end{equation}
where $n^f_{{\bf r}} = f_{{\bf r}\sigma}^\dagger f_{{\bf
r}\sigma}^{\vphantom\dagger}$ is the fermion density and the $Z_2$
lattice divergence is defined as:
\begin{equation} \prod_{+({\bf r})}
\sigma^x \equiv \prod_j \sigma^x_j({\bf
    r}) \sigma^x_j({\bf r}-\hat{\bf x}_j) .
\end{equation}
Physical states are required to be gauge invariant, which is
specified by the set of local {\it constraints}: ${\cal C}^1_{\bf
r} \equiv 1$.  This is the $Z_2$ analog of Gauss law for
conventional electromagnetism.

The connection between the $Z_2$ gauge theory and a theory of
interacting electrons, is most apparent in the limit that $t_v$ is
taken to be much larger than all other couplings. In this limit
the electron creation operator is equivalent to the product of the
chargon and spinon creation operators, $c^\dagger_{{\bf r \sigma}}
= b^\dagger_{\bf r} f^\dagger_{{\bf r} \sigma}$.  Indeed, when
$t_v \rightarrow \infty$ the $Z_2$ ``electric'' field becomes
frozen, $\sigma^x_j \approx 1$, and the gauge constraints then
imply that on each site of the lattice the sum of the chargon and
spinon numbers, $n_{\bf r} + n^f_{\bf r}$, is even. Moreover, for
large $t_v$, the chargon and spinon hopping terms are strongly
suppressed, and can be considered perturbatively. Upon integrating
out the gauge fields, one will thereby generate an additional electron
kinetic energy term with amplitude of order $t_c t_s/t_v$.  A
brief discussion is given in Appendix~\ref{sec:fermiliquid}.

In what follows, we will study the $Z_2$ gauge theory more
generally, away from the large $t_v$ limit. Of interest is the
electron Greens function,
\begin{equation}
{\cal G}_e({\bf r}_1, \tau_1; {\bf r}_2, \tau_2) = - \langle
T_\tau \hat{c}_{{\bf r}_1\sigma}(\tau_1) \hat{c}^\dagger_{{\bf
r}_2 \sigma}(\tau_2) \rangle  .
\end{equation}
We will express the electron operators as,
\begin{equation}
c_{{\bf r}\sigma} \equiv b_{\bf r} f_{{\bf r}\sigma} ,
\end{equation}
which is exact as $t_v \rightarrow \infty$, but more generally
should be sufficient to extract the universal low energy and long
length scale behavior of the electron Green's function. We will
also be interested in correlation functions involving the Cooper
pair creation and destruction operators, $B^\dagger_{\bf r},B_{\bf
r}$, with $B_{\bf r} = (b_{\bf r})^2 = e^{-2i\phi_{\bf r}}$.

\subsection{$Z_2$ Vortex-spinon formulation}
\label{sec:z_2-vortex-spinon}

In what follows, it will prove particularly convenient to work
with vortex degrees of freedom, rather than the chargon fields. To
arrive at such a description, we use the U(1) duality
transformation\cite{Duality}, in which the dual variables sit
naturally on the 2d dual lattice. We denote the sites of the 2d
dual lattice by sans serif characters as ${\sf r}=x_1 {\bf
\hat{x}} + x_2 {\bf \hat{y}} + {\bf
  w}$, with ${\bf w} = (1/2)({\bf \hat{x}} + {\bf \hat{y}})$ and
integer $x_1,x_2$.  The duality transformation itself defines two
conjugate gauge fields $[a_i({\sf r}), e_j({\sf r}')] = i\delta_{ij}\delta_{\sf
  r,r'}$, where
\begin{eqnarray}
  \label{eq:duality}
  n_{\bf r} & = & \frac{1}{\pi} \epsilon_{ij} \partial_i a_j({\bf
    r-w}), \\
  \partial_i \phi_{\bf r} & = & \pi \epsilon_{ij} e_j({\bf
    r+w}-\hat{\bf x}_j).
\end{eqnarray}
Here, as defined, due to the discreteness of the $n_{\bf r}$
variables, $a_j({\sf r})$ takes on values that are integer multiples
of $\pi$, while $e_j({\sf r})$ is a periodic variable with period
$2$.  This transformation is faithful provided the constraint
\begin{equation}
  \label{eq:U1constraint}
  (\vec{\nabla}\cdot \vec{e})({\sf r}) \equiv  \sum_j \partial_j
  e_j({\sf r}-{\bf x}_j) = 0 \;\; ({\rm mod}\, 2),
\end{equation}
or equivalently
\begin{equation}
  \label{eq:U1constrainta}
  {\cal C}^2_{\sf r} = e^{i\pi (\vec{\nabla}\cdot \vec{e})({\sf r})} = 1
\end{equation}
is imposed at every site ${\sf r}$ of the dual lattice.
Rewriting the charge Hamiltonian, one has
\begin{eqnarray}
  \label{eq:DualChargeHam}
  {\cal H}_c & = & - t_c\sum_i \sigma_i^z({\bf r}) \cos(\pi
  \epsilon_{ij} e_j({\bf r+w}-\hat{\bf x}_j)-A_i({\bf r}) )\nonumber \\
  &&  + \frac{u_c}{\pi^2}(\epsilon_{ij}\partial_i
  a_j- \pi\rho_0)^2.
\end{eqnarray}

Conventional $hc/2e$ superconducting vortices are composites of a
``vison'' (topological excitation in $\sigma_j^z$) and a
half-vortex in $\phi$.  To describe them, we perform a unitary
transformation to a new Hamiltonian $\tilde{H}_{Z_2}$ with new
constraints $\tilde{\cal C}^a$ ,
\begin{equation}
  \label{eq:Hdual1}
  \tilde{H}_{Z_2} = U^\dagger H_{Z_2} U^{\vphantom\dagger} , \qquad \tilde{\cal
    C}^a = U^\dagger {\cal
    C}^a U^{\vphantom\dagger},
\end{equation}
with the unitary operator,
\begin{eqnarray}
  \label{eq:unitary}
  U & = & \exp[\frac{i}{2} \sum_{{\bf r},i,j} \epsilon_{ij}a_i({\bf r+w}-\hat{\bf
    x}_i) (\sigma_j^z({\bf r})-1)] \nonumber\\
  & = & \prod_{\bf r} (\sigma^z_1({\bf r}))^{\frac{a_2({\bf r}+\overline{\bf w})}{\pi}}
  (\sigma^z_2({\bf r}))^{\frac{a_1({\bf r}-\overline{\bf w})}{\pi}},
\end{eqnarray}
with $\overline{\bf w} = (1/2)(\hat{\bf x} - \hat{\bf y}) = {\bf
  w}-\hat{\bf y}$.  The transformed constraints are,
\begin{eqnarray}
  \label{eq:Z2constraint2a}
  \tilde{\cal C}_{\bf r}^1 & = & (-1)^{n^f_{{\bf r}}} \prod_{+({\bf
      r})} \sigma^x = 1,  \\
  \tilde{\cal C}^2_{\sf r} & = & (-1)^{(\vec{\nabla}\cdot
    \vec{e})({\sf r})} \prod_{\Box({\sf r})} \sigma^z = 1.
    \label{eq:Z2constraint2b}
\end{eqnarray}
Under this unitary transformation,
\begin{eqnarray}
  \label{eq:Z2ham2}
  \tilde{\cal H}_c & = & - t_c \sum_j \cos( \pi e_j({\sf
    r})+\epsilon_{jk}A_k({\sf r}+{\bf\overline{w}})) \nonumber \\ && +
  \frac{u_c}{\pi^2}(\epsilon_{ij} \partial_i a_j -
  \pi\rho_0)^2 , \\
  \tilde{\cal H}_g & = & \!\!\!\!- t_v \!\sum_j
  \!\overline{\sigma}_j^x({\sf r}) 
  \cos(a_j({\sf r})) \!-\! K
  (-1)^{(\vec{\nabla}\cdot \vec{e})({\sf r})} + {\cal H}^{Z_2}_r.
\end{eqnarray}
Here we have defined, $\overline{\sigma}^x_1({\sf r}) =
\sigma^x_2({\sf
  r}+\overline{\bf w}), \overline{\sigma}^x_2({\sf r}) =
\sigma^x_1({\sf r}-\overline{\bf w})$, and have used
Eq.~(\ref{eq:Z2constraint2b}).  The transformed ``roton'' hopping
term becomes,
\begin{equation}
  {\cal H}^{Z_2}_r = -\kappa_r \overline{\sigma}_1^x({\sf
    r}) \overline{\sigma}_1^x({\sf
    r}+\hat{\bf x}_2)\cos(\partial_y a_x({\sf r})) + (x\leftrightarrow
  y) .
\end{equation}
The fermion Hamiltonian is almost unchanged in the dual
vortex-spinon representation,
\begin{eqnarray}
  \label{eq:fdv}
  \tilde{\cal H}^{Z_2}_f & = &  -  \sum_j \sigma_j^z({\bf r})
  [ t_s f^\dagger_{{\bf r} +  {\bf \hat{x}}_j \sigma}
  f^{\vphantom\dagger}_{{\bf r} \sigma}+\Delta_j f_{{\bf r}+\hat{\bf
      x}_j \sigma} \epsilon_{\sigma\sigma'}f_{{\bf r}\sigma'}+ {\rm
    h.c.}] \nonumber \\
  & & - t_e \sum_j \sigma_j^z({\bf r}) e^{i(\pi \overline{e}_j({\bf
      r})-A_j({\bf r}))}
    f^\dagger_{{\bf r} +  {\bf \hat{x}}_j
      \sigma} f^{\vphantom\dagger}_{{\bf r} \sigma}+ h.c. ,
      \label{HtildefermZ2}
\end{eqnarray}
the changes appearing only in the electron hopping (the last term).
Here we have defined,
\begin{equation}
  \overline{e}_i({\bf r}) = \epsilon_{ij} e_j({\bf r} + {\bf w} - \hat{x}_j)  .
\label{eq:defineebar}
\end{equation}
Notice that $\pi \overline{e}_j$ couples ``like a gauge field'' to
the spinons in the final electron term.

To arrive at the final $Z_2$ vortex-spinon theory, we split the
electric and magnetic fields into longitudinal and transverse parts,
$a_j = a_j^l + a_j^t$, $e_j=e_j^l+ e_j^t$, with $\vec{\nabla}\cdot
\vec{a}^t=\vec{\nabla}\cdot\vec{e}^t=0$, $\epsilon_{ij}\partial_i
a_j^l=\epsilon_{ij}\partial_i e_j^l= 0$.  For future purposes, we note
that the longitudinal part of $e_j$ is related to the transverse part
of $\overline{e}_j$ and vice versa, i.e. $\overline{e}_i^{l/t}({\bf
  r}) = \epsilon_{ij} e_j^{t/l}({\bf r} + {\bf w} - \hat{x}_j)$.  It
is convenient then to solve the constraint for the longitudinal
fields,
\begin{eqnarray}
  \label{eq:vortexvarstheta}
  a_j^l({\sf r}) &  = & -\partial_j \theta_{\sf r}, \\
  (\vec{\nabla}\cdot\vec{e}^l)({\sf r}) & = & N_{\sf r}.\label{eq:vortexvarsN}
\end{eqnarray}
The fields $e^{i\theta_{\sf r}}$ and $N_{\sf r}$ have the
interpretation of vortex creation and number operators, as can be seen
by tracking the circulation defined from the original chargon phase
variable $\phi$.  One finds canonical commutation relations,
\begin{equation}
  [\hat{\theta}_{\sf r}, \hat{N}_{\sf r^\prime}] = i \delta_{{\sf r}{\sf
      r^\prime}}  .\label{eq:Nthetacom}
\end{equation}

At this stage, $N_{\sf r}$ is a periodic variable with period $2$,
and $-\partial_j \theta_{\sf r} + a^t_j({\sf r})$ is constrained
to be an integer multiple of $\pi$.  It is convenient to soften
the latter constraint, and in order to respect the uncertainty
relation implied by Eq.~(\ref{eq:Nthetacom}), at the same time
relax the periodicity of $N_{\sf r}$.  Formally, this is
accomplished by replacing,
\begin{equation}
  \label{eq:replacecosine}
  -t_c \sum_j \cos (\pi e_j+\epsilon_{jk}A_k) \rightarrow {\rm Const.} +
  \frac{u_v}{2} 
  \sum_j (e_j+\epsilon_{jk} A_k/\pi)^2,
\end{equation}
with $u_v \approx t_c \pi^2$, and adding a term to the
Hamiltonian, $\tilde{H}_{Z_2} \rightarrow \tilde{H}_{Z_2} +
\sum_{\sf r} {\cal H}^{Z_2}_{2v}$, with
\begin{equation}
  \label{eq:pairvortex}
  {\cal H}^{Z_2}_{2v}  =  - t_{2v} \sum_j \cos (2\partial_j \theta-
  2a_j^t).
  \label{pairvhop}
\end{equation}
The constraint is recovered for large $t_{2v}$, but we will
consider a renormalized theory in which $t_{2v}$ may be considered
a small perturbation.

It is convenient to regroup the longitudinal contribution to
$\tilde{H}_c$ along with the terms in $\tilde{H}_g$  and ${\cal
H}^{Z_2}_{2v}$ into vortex ``potential'' and ``kinetic'' terms,
$H_N$ and $H^{Z_2}_{kin}$. We thereby arrive at the final form for
the $Z_2$ vortex-spinon Hamiltonian,
\begin{equation}
\tilde{H}_{Z_2} = H_{pl} + H_N + H^{Z_2}_{kin} +
\tilde{H}_f^{Z_2}, \label{HamZ2vs}
\end{equation}
with the fermion Hamiltonian $\tilde{\cal H}_s^{Z_2}$ given in
Eq.~(\ref{HtildefermZ2}) and with,
\begin{eqnarray}
  \label{eq:z2vortexspinon}
    {\cal H}_{pl} & = & {u_v \over 2} [e^t_j({\sf
      r})\!+\!\epsilon_{jk}A^l_k({\sf r}+\overline{\bf w})]^2 \nonumber
    \\ && +
    \frac{v_0^2}{2u_v} 
  [\epsilon_{ij} \partial_i a^t_j({\sf r})\! - \pi \rho_0]^2  , \\
  H_{N} & = &   {u_v \over 2} \sum_{{\sf r},{\sf r}^\prime}
  (\hat{N}_{\sf r}-\frac{B_{\sf r}}{\pi})
  (\hat{N}_{{\sf r}^\prime}-\frac{B_{\sf r}}{\pi}) V({\sf r} - {\sf
    r}^\prime) , \\ 
  {\cal H}^{Z_2}_{kin} & = & {\cal H}^{Z_2}_v  + {\cal H}^{Z_2}_{2v}
  + {\cal H}^{Z_2}_r.
\end{eqnarray}
Here $B_{\sf r} = \epsilon_{ij}\partial_i A_j({\sf r}-{\bf w})$ is the
physical flux through the plaquette of the original lattice located at
the site ${\sf r}$ of the dual lattice.
In the ``plasmon'' Hamiltonian, ${\cal H}_{pl}$, an implicit sum
over $j=1,2$ is understood and we have defined a (bare) plasmon
velocity, $v_0$, as $v_0^2 = 2 u_ct_c = 2u_c t_v /\pi^2$. Inside
the superconducting phase, this Hamiltonian describes the
Goldstone mode - or sound mode - and can be readily diagonalized
to give the dispersion, $\omega^2_{pl}({\bf k}) = v_0^2 \sum_j [2
\sin(k_j/2)]^2$ with $|k_j| < \pi$ in the first Brillouin zone.
Since the electron charge density is given by, ${1 \over \pi}
\epsilon_{ij}
 \partial_i a_j$, one can readily include long-ranged Coulomb interactions
which will modify the plasmon at small ${\bf k}$.

In $H_N$ above we have set $K=0$, dropping henceforth
the term $\delta H_N(K) = - K\sum_{\sf r}
(-1)^{N_{\sf r}}$, since it will not play an important role in the
phases of interest.
The vortex-vortex interaction energy $V({\sf r})$ is the Fourier
transformation of the discrete inverse Laplacian operator,
$V^{-1}({\bf k}) \equiv {\cal K}^2({\bf k})$, with ${\cal
K}^2({\bf k}) = \sum_j 2(1-\cos k_j)$, and has the expected
logarithmic behavior at large distances, $V({\sf r}) \sim {1 \over
2 \pi} \ln(|{\sf r}|)$.  The vortex kinetic energy, ${\cal
H}^{Z_2}_{kin}$, is a sum of three contributions - a single vortex
hopping term,
\begin{equation}
{\cal H}^{Z_2}_v =  - t_{v} \sum_j \overline{\sigma}_j^x({\sf
    r})\cos (\partial_j \theta_{\sf r}- a_j^t) ,
\end{equation}
a pair-vortex hopping term ${\cal H}^{Z_2}_{2v}$ given in
Eq.~(\ref{pairvhop}), and a ``roton'' hopping term,
\begin{equation}
{\cal H}^{Z_2}_r = -\kappa_r \overline{\sigma}_2^x({\sf
    r}) \overline{\sigma}_2^x({\sf
    r}+\hat{\bf x}_1)\cos(\Delta_{xy} \theta_{\sf r} -
    \partial_x a^t_y({\sf r})) + (x\leftrightarrow y) ,
  \end{equation}
where we have defined,
\begin{equation}
\Delta_{xy} \theta_{{\sf r}} = \Delta_{yx} \theta_{{\sf r}} \equiv
\sum_{e_1,e_2 = 0,1} (-1)^{e_1 + e_2} \theta_{{\sf r}+ e_1 {\bf
\hat{x}}+ e_2 {\bf \hat{y}} }  . \label{deltaxy}
\end{equation}
Notice that ${\cal H}^{Z_2}_r$ hops two vortices, originally at
sites of the dual lattice on opposite corners of an elementary
square, to the other two sites. Equivalently, this term can be
interpreted as hopping a vortex-antivortex pair on neighboring
sites (ie. a vortex ``dipole'' or more simply a {\it roton}) one
lattice spacing in a direction perpendicular to the dipole.  Such
a roton is a 2d analog of a 3d vortex ring, and in a Galilean
invariant superfluid (such as $4-He$) vortex rings propagate in
precisely this manner. Henceforth we shall refer to this process
as a ``roton hopping'' process.

The above $Z_2$ vortex-spinon Hamiltonian must be supplemented by
the two gauge constraints, which from
Eqs.~(\ref{eq:Z2constraint2a},\ref{eq:Z2constraint2b}) and
(\ref{eq:vortexvarsN}) can be cast into an appealingly simple and
symmetrical form:
\begin{equation}
\tilde{\cal C}_{\bf r}^1  =  (-1)^{n^f_{\bf r}} \prod_{\Box({\bf
r})} \overline{\sigma}^x = 1,
  \label{Z2con1}
\end{equation}
\begin{equation}
\tilde{\cal C}^2_{\sf r}  =  (-1)^{N_{\sf r}} \prod_{\Box({\sf
        r})} \sigma^z = 1.
\label{Z2con2}
\end{equation}

These constraints correspond to an attachment of a $Z_2$ flux in
$\sigma^z$ and $\overline{\sigma}^x$ to the vortex number {\it
parity} and the spinon number parity, respectively.  Since the
spinons are minimally coupled to $\sigma^z$ and the vortices to
$\overline{\sigma}^x$, this implies a sign change upon hopping a
spinon around a vortex -- or vice versa. Indeed, if the partition
function for the vortex-spinon Hamiltonian (without ${\cal
H}^{Z_2}_r$ and with $t_e=0$) is expressed as an imaginary time
path integral with the $Z_2$ constraints in
Eqs.~(\ref{Z2con1},~\ref{Z2con2}) imposed, the resulting Euclidean
action becomes identical to Eqs (109)-(113) in
Ref.~[\onlinecite{Z2}].

It is instructive to obtain explicit expressions for the electron
and Cooper pair creation operators in terms of the dualized vortex
degrees of freedom.  From Eq.~(\ref{eq:duality}) we can directly
obtain an expression for the Cooper pair destruction operator,
$B_{\bf r} = e^{-2i \phi_{\bf r}}$, as
\begin{equation}
B_{\bf r} = \prod_{\bf r}^{\infty} e^{2\pi i  \overline{e}_j({\bf r}')
  d{\bf r}'_j} ,
\label{pairop}
\end{equation}
where the $\prod$ symbol here denotes a product along a semi-infinite
``directed'' string running on the links of the original lattice,
originating at ${\bf r}$ and terminating at spatial infinity, with
$d{\bf r}'$ the unit vector from the point ${\bf r}'$ along the string
to the next point.  In terms of $e_j$ (rather than $\overline{e}_j$),
the product contains a factor of $\exp(\pm 2\pi i e_j({\sf r}))$ for
each link of the dual lattice that crosses the string, taking the
positive/negative sign for directed links crossing the string from
right/left to left/right proceeding from ${\bf r}$ to $\infty$.  We
will use the above notation when possible to present precise analytic
expressions for such strings.
The path independence of the string is assured by the second gauge
constraint, ${\cal C}^2_{\sf r} = 1$.  Since the unitary
transformation, $U$ in Eq.~(\ref{eq:unitary}), commutes with
$e^{2\pi i e_j}$, this is the correct expression for the Cooper
pair operator within the $Z_2$ vortex-spinon theory.

An expression for the electron operator, $c_{{\bf r}\sigma} =
b_{\bf r} f_{{\bf r}\sigma}$, in the dual vortex-spinon theory can
be extracted by moreover re-expressing the ``chargon'' operator
$b_{\bf r} = e^{-i\phi_{\bf r}}$ as a string,
\begin{equation}
b_{\bf r} = \prod_{\bf r}^\infty  e^{i\pi \overline{e}_j d{\bf r}'_j} = {\cal
S}_{vort}({\bf r}) {\cal S}_\phi({\bf r})  . \label{bstring}
\end{equation}
For later convenience, we have here decomposed this expression
into a piece depending on the vortex configurations through the
longitudinal ``electric'' field, $e_j^{\ell}$, and a contribution
depending on the smooth part of the phase, $\phi$, through the
transverse field, $e_j^t$:
\begin{equation}
{\cal S}_{vort}({\bf r}) = \prod_{\bf r}^\infty
 e^{i \pi \overline{e}_j^{t}d{\bf r}'_j} ; \hskip1cm  {\cal
   S}_\phi({\bf r}) = \prod_{\bf 
r}^\infty e^{i\pi \overline{e}_j^l d{\bf r}'_j} . \label{SvortSphi}
\end{equation}
But unlike the Cooper pair operator, the ``chargon'' operator
transforms non-trivially under the unitary transformation in
Eq.~(\ref{eq:unitary}):
\begin{equation}
\tilde{b}_{\bf r} = U^\dagger b_{\bf r} U = \prod_{\bf r}^\infty
[\sigma^z_i e^{i\pi \overline{e}_j d{\bf r}'_j} ] , \label{btilde}
\end{equation}
now including a factor of $\sigma^z_j({\bf r})$ for each link of
the string.  Again, the path independence is guaranteed by the
gauge constraint, $\tilde{\cal C}^2_{\sf r} = 1$.  
The final expression for the electron operator within the dualized
$Z_2$ vortex-spinon field theory follows simply as,
\begin{equation}
\tilde{c}_{{\bf r}\sigma} = U^\dagger c_{{\bf r}\sigma} U =
\tilde{b}_{\bf r} f_{{\bf r}\sigma} .
\label{electronZ2}
\end{equation}

As discussed at length in Ref.~[\onlinecite{Z2}], the $Z_2$
vortex-spinon formulation is particularly well suited for
accessing spin-charge separated insulating states. Specifically,
when pairs of vortices hop around they ``see'' an average gauge
flux of, $2 \epsilon_{ij}
\partial_i a_j = 2\pi \rho_0$.  Thus at half-filling with
$\rho_0=1$, vortex-pairs are effectively moving in zero flux and
at large pair-hopping strength, $t_{2v} \rightarrow \infty$, will
readily condense - driving the system into an insulating state
with a charge gap.  In the simplifying limit with vanishing {\it
single} vortex hopping strength, $t_v=0$, all of the vortices are
paired, $(-1)^N =1$, and the $Z_2$ gauge constraint in
Eq.~(\ref{Z2con2}) reduces to $\prod_{\Box} \sigma^z
=1$. The ``vison'' excitations (plaquettes with $\prod_ {\Box}
\sigma^z = -1$) are gapped out of the ground state, and the
spinons, being minimally coupled to $\sigma^z$, can propagate as
deconfined excitations.  The charge sector supports gapped but
deconfined chargon excitations, which can be viewed as topological
defects in the pair-vortex condensate.

But as we shall see below, to access the new Roton Fermi Liquid phase
requires taking the strength of the roton hopping strength large, and
the $Z_2$ formulation proves inadequate.  To remedy this, we introduce
in subsection C below, a new $U(1)$ formulation of the vortex-spinon
field theory.  As we shall demonstrate, the $U(1)$ and $Z_2$
vortex-spinon formulations are formally equivalent, and by a sequence
of unitary transformations it is possible to pass from one
representation to the other.  Care should be taken when considering
operators which transform non-trivially under the unitary operations
related different representations, however.  A third dual vortex
formulation involving electron (rather than spinon) operators is
briefly discussed below in Appendix \ref{ap:elecform}. The Hamiltonian
in this ``vortex-electron'' formulation is equivalent under a sequence
of unitary transformations to both the $Z_2$ and $U(1)$ vortex-spinon
Hamiltonians.

To establish these equivalences, it is convenient to ``choose a
gauge'' in the $Z_2$ theory.  As detailed in Appendix
\ref{ap:enslaveZ2}, it is possible to unitarily transform to a
basis in which the $Z_2$ gauge fields are completely ``slaved'' to
the vortex and spinon operators, and can be eliminated completely
from the theory.  Specifically, in the chosen gauge the
$x-$components of both $\sigma^z$ and $\overline{\sigma}^x$ are
set to unity on every link of the lattice:
$\overline{\sigma}_1^{x}({\sf r})= \sigma_1^{z}({\bf r}) = 1$. As
we shall see in subsection C below, the $U(1)$ gauge fields in the
$U(1)$ vortex-spinon formulation can be similarly enslaved.
Remarkably one arrives at the {\it identical} ``enslaved''
Hamiltonian in both cases, thereby establishing the formal
equivalence between the two formulations.

\subsection{U(1) Vortex-spinon formulation}
\label{sec:u1-vortex-spinon}

In the $U(1)$ formulation of the vortex-spinon field theory, the
Pauli matrices $\sigma^z, \overline{\sigma}^x$ which live as $Z_2$
gauge fields on the links of the original and dual lattice,
respectively, are
effectively replaced by exponentials of two
$U(1)$ gauge fields:  $\overline{\sigma}^x_j({\sf r}) \rightarrow
exp[i \alpha_j({\sf r})]$ and  $\sigma^z_j({\bf r})
 \rightarrow
exp[i\beta_j({\bf r})]$. These two $U(1)$ gauge fields are
canonically conjugate variables taken to satisfy,
\begin{equation}
[\alpha_i({\sf r} - {\bf \hat{x}}_i),\beta_j({\bf r}^\prime)] = i
\pi \epsilon_{ij} \delta^2({\bf r} - {\bf r}^\prime)  ,
\end{equation}
with ${\sf r} = {\bf r}   + {\bf w}$. For two ``crossing'' links
these commutation relations imply that the exponentials,
$e^{i\alpha}, e^{i\beta}$, anticommute with one another, $[e^{i
\alpha}, e^{i \beta}]_- =0$.

\subsubsection{$U(1)$ Vortex-spinon Hamiltonian}
\label{sec:u1-vortex-spinon-1}

The full Hamiltonian for the $U(1)$ vortex-spinon field theory
takes the same form as the $Z_2$ vortex-spinon Hamiltonian in
Eq.~(\ref{HamZ2vs}),
\begin{equation}
H = H_{pl} + H_N + H_{kin} + H_f,
\label{HamU1vs}
\end{equation}
with $H_{pl}$ and $H_N$ given as before in
Eq.~(\ref{eq:z2vortexspinon}). Only the vortex kinetic terms and
the fermion Hamiltonian are modified. Once again the vortex
kinetic energy terms are decomposed into single vortex,
pair-vortex and roton hopping processes:
\begin{equation}
  {\cal H}_{kin}  =  {\cal H}_v  + {\cal H}_{2v}
  + {\cal H}_r .
\end{equation}
Since the vortices in the $U(1)$ formulation are minimally coupled
to the $U(1)$ gauge field, $\alpha_j({\sf r})$, each of these
three terms will be modified from their $Z_2$ forms. Specifically,
in terms of the associated Hamiltonian densities we have,
\begin{equation}
{\cal H}_v = -t_v \sum_{j=1,2} \cos(\partial_j \theta - a^t_j + \alpha_j)  ,
\label{vorthop}
\end{equation}
\begin{equation}
{\cal H}_{2v} = -t_{2v} \sum_{j=1,2}
\cos(2\partial_j \theta - 2a^t_j + 2\alpha_j) ,
\label{vortpairhop}
\end{equation}
\begin{equation}
{\cal H}_{r} = -{\kappa_r \over 2}  \cos[\Delta_{xy} \theta -
\partial_x(a^t_y - \alpha_y)]   + (x \leftrightarrow y)   ,
\label{rothop}
\end{equation}
with $\Delta_{xy} \theta_{{\sf r}}$ defined in
Eq.~({\ref{deltaxy}).

The Hamiltonian density for the fermions in the $U(1)$ formulation
is given by
\begin{eqnarray}
{\cal H}_f  & = & -  \sum_j e^{i\beta_j({\bf r})} [ (t_s + t_e e^{i(\pi
  \overline{e}_j({\bf r})-A_j({\bf r}))})
f^\dagger_{{\bf r} +  {\bf \hat{x}}_j \sigma}
    f^{\vphantom\dagger}_{{\bf r} \sigma}  \nonumber \\
    & & +  \Delta_j
    [{\cal S}_{\bf r}]^2 f_{{\bf r}+\hat{\bf
        x}_j \sigma} \epsilon_{\sigma\sigma'}f_{{\bf r}\sigma'}+ {\rm
      h.c.}]
        \label{eq:Hf}
\end{eqnarray}
In the $U(1)$ formulation, the (average) density of spinons,
$\langle f^\dagger_{\sigma} f^{\vphantom\dagger}_{\sigma} \rangle
= \langle {1 \over \pi} \epsilon_{ij} \partial_i \alpha_j \rangle
$ (as follows from Eq.~(\ref{U1constraint})) is taken to be {\it
equal} to the (average) charge density, $\langle {1 \over \pi}
\epsilon_{ij}
\partial_i a_j \rangle$.
A new element, not present in the $Z_2$ fermion Hamiltonian,
${\tilde{\cal H}}^{Z_2}_s$ in Eq.~(\ref{HtildefermZ2}), is the
``string operator'', ${\cal S}_{\bf r}$. The string operator is
given as a product of $e^{i\beta}$ running along directed links of
the original lattice from the site ${\bf r}$ to spatial infinity:
\begin{equation}
{\cal S}_{\bf r} = \prod_{\bf r}^\infty e^{i\beta_j d{\bf r}'_j} .
\end{equation}
As we shall discuss below, once we restrict the Hilbert space to
gauge invariant states other choices for the ``path'' of the
string are formally equivalent.

In addition to global spin and charge conservation, the full
$U(1)$ vortex-spinon Hamiltonian, ${\cal H}$ above, has a number
of {\it local} gauge symmetries. To fully define the model we need
to specify the set of gauge invariant states that are allowed.
Associated with each of the ``Chern-Simons'' fields, $\alpha$ and
$\beta$,  is a $U(1)$ gauge symmetry. Specifically, the full
Hamiltonian is invariant under {\it independent} gauge
transformations:
  \begin{eqnarray}
    e^{-i \theta_{\sf r}} & \rightarrow &   e^{-i \theta_{\sf r}} e^{i \chi_r}, \nonumber \\
    \alpha_j({\sf r}) & \rightarrow & \alpha_j({\sf r}) + \partial_j \chi_{\sf r} ,
    \label{Hgauge1}
  \end{eqnarray}
and
  \begin{eqnarray}
    f_{{\bf r}\sigma} & \rightarrow & f_{{\bf r}\sigma}e^{i\Lambda_{\bf
        r}}, \nonumber \\
    \beta_j({\bf r}) & \rightarrow & \beta_j({\bf r}) +
\partial_j \Lambda_{\bf r},
    \label{Hgauge2}
  \end{eqnarray}
for arbitrary real functions, $\Lambda_{\bf r}$ and $\chi_{\sf r}$,
living on the original and dual lattices, respectively.
The corresponding operators which
transform the fields in this way are,
\begin{equation}
{\cal G}_v(\chi_{\sf r}) = e^{-i \sum_{\sf r} \chi_{\sf r} [
N_{\sf r} - {1 \over \pi} \epsilon_{ij} \partial_i \beta_j({\sf r}
- {\bf w}) ]} , \label{U1opvortex}
\end{equation}
and
\begin{equation}
{\cal G}_f(\Lambda_{\sf r}) = e^{i \sum_{\bf r} \Lambda_{\bf r}
[ f^\dagger_{{\bf r} \sigma}
f_{{\bf r} \sigma} - {1 \over \pi}
\epsilon_{ij} \partial_i \alpha_j({\bf r} - {\bf w}) ]} .
\label{U1opspinon}
\end{equation}
Both of these operators commute with the full Hamiltonian
$H$. The $U(1)$ sector is specified by simply setting ${\cal
G}_v = {\cal G}_f =1$ for arbitrary $\chi_{\sf
  r}$ and $\Lambda_{\bf r}$.  From
Eqs.~(\ref{U1opvortex},\ref{U1opspinon}) this is equivalent to
attaching $\pi$ flux in the statistical gauge fields $\alpha$ and
$\beta$ to the spinons and vortices:
\begin{equation}
\epsilon_{ij} \partial_i \alpha_j({\bf r} - {\bf w}) = \pi f^\dagger_{{\bf r} \sigma}
f_{{\bf r} \sigma}  ;   \hskip0.5cm  \epsilon_{ij} \partial_i \beta_j({\sf r} - {\bf w}) = \pi N_{\sf r}   .
\label{U1constraint}
\end{equation}
Notice that this is simply the $U(1)$ analog of the $Z_2$ flux
attachment in Eqs.~(\ref{Z2con1},\ref{Z2con2}),
and implies the same sign change
when a spinon is transported around a vortex or vice versa. The
only difference is that in the $U(1)$ formulation the phase of
$\pi$ is accumulated gradually when the spinon is taken around the
vortex, whereas the sign change in the $Z_2$ theory can occur when
the spinon hops across a single link. The formal equivalence of
the $Z_2$ and $U(1)$ formulations will be established below.

As we shall see, the ``smearing'' of the accumulated $\pi$ phase
change, makes the theory in the $U(1)$ formulation eminently more
tractable. The one notable complication is the square of the ``string operator'',
which in the $U(1)$ sector is a non-trivial function of
$e^{2i\beta}$ along the string, rather than equaling unity as in
the $Z_2$ sector. However, it is worth emphasizing that within the
$U(1)$ sector of the theory, the value of the operator ${\cal O}_{\bf r}
\equiv {\cal S}_{\bf r}^2$ is
{\it independent} of the chosen path.  Specifically, consider two
(unitary) string operators, denoted ${\cal O}_1, {\cal O}_2$, with
different paths running from the same site ${\bf r}$ off to
spatial infinity.  The ``difference'' between the two string
operators, ${\cal O}_1^{-1}{\cal O}_2$, is a product of
$e^{2i\beta}$ around {\it closed loops}.  But due to the $U(1)$
gauge constraint in Eq.~(\ref{U1constraint}), $\epsilon_{ij}
\partial_i \beta_j = \pi N$, this product is an exponential
of the total vorticity $N_{tot}$ inside the closed loops, ${\cal
O}_1^{-1}{\cal O}_2 = exp(2\pi i N_{tot})$. Since the vorticity is
integer, one deduces that the string operator is indeed path
independent, ${\cal O}_1 = {\cal O}_2$.

It will prove useful to obtain expressions for the electron and
Cooper pair creation operators within the $U(1)$ vortex-spinon
formulation.  The Cooper pair operator has the same form as in the
$Z_2$ vortex-spinon formulation, given explicitly in
Eq.~(\ref{pairop}), but the electron operator is modified in a
non-trivial way. As we shall check explicitly below, the electron
operator within the $U(1)$ formulation involves a string depending
on both the dual ``electric field'', $e_j$, as well as the
statistical gauge field, $\beta_j$:
\begin{equation}
c_{{\bf r}\sigma} = f_{{\bf r}\sigma} \prod_{\bf r}^\infty
e^{i[\pi \overline{e}_i+ \beta_i]d{\bf r}'_i}  , \label{electronU1}
\end{equation}
with $\overline{e}_i({\bf r})$ defined in Eq.~(\ref{eq:defineebar}).
The path independence follows from the condition in
Eq.~(\ref{eq:vortexvarsN}), together with the second $U(1)$ gauge
constraint in Eq.~(\ref{U1constraint}) above;
\begin{equation}
\pi \partial_j e_j({\sf r} - {\bf x}_j) = \epsilon_{ij} \partial_i
\beta_j({\sf r} - {\bf w}).
\end{equation}
This implies an equality between the longitudinal ``electric
field'' and the transverse statistical field:
\begin{equation}
\beta^t_1({\bf r}) = \pi e^l_2({\bf r} + \overline{\bf w}) ;
\hskip1cm \beta^t_2({\bf r}) = - e^l_1({\bf r} -
\overline{\bf w}) ,
\end{equation}
or
\begin{equation}
  \label{eq:bteqebt}
  \beta_i^t = -\pi \overline{e}_i^t,
\end{equation}
and enables the electron destruction operator to be re-expressed
as,
\begin{equation}
c_{{\bf r}\sigma} = f_{{\bf r}\sigma} {\cal S}_\phi({\bf r})
\prod_{\bf r}^\infty e^{i\beta_j^l d{\bf r}'_j} \label{elecspinon}
\end{equation}
with ${\cal S}_\phi({\bf r})$ defined in Eq.~(\ref{SvortSphi}).
Similarly, the string operator that enters into the $U(1)$ fermion
Hamiltonian, $H_f$ in Eq.~(\ref{eq:Hf}), can be written as,
\begin{equation}
{\cal S}_{\bf r} = {\cal S}_{vort}({\bf r})
\prod_{\bf r}^\infty e^{i\beta^l_j d{\bf r}'_j}  .
\end{equation}
Upon combining the above two equations, and recalling the identity for
the ``chargon'' operator in the original $Z_2$ theory, $b_{\bf r} =
{\cal S}_{vort}({\bf r}) {\cal S}_\phi({\bf r})$ in
Eq.~({\ref{bstring}), implies that, ${\cal S}^\dagger_{\bf r} f_{{\bf
      r}\sigma} = b^\dagger_{\bf r} c_{{\bf r}\sigma}$.  Consequently,
  ``spinon pairing'' terms in the $Z_2$ gauge theory are seen to be
  equivalent to the usual Bogoliubov-deGennes form:
\begin{equation}
[{\cal S}_{\bf r} ]^2 \epsilon_{\sigma \sigma^\prime}
f_{{\bf r}\sigma} f_{{\bf r}\sigma^\prime} = B^\dagger_{\bf r}
\epsilon_{\sigma \sigma^\prime} c_{{\bf r}\sigma} c_{{\bf
r}\sigma^\prime} ,
\label{eq:bdgform}
\end{equation}
with $B^\dagger_{\bf r}$ the Cooper pair creation operator.  For
$d$-wave pairing, the pair field lives on links, and a similar identity
obtains: 
\begin{eqnarray}
  \label{eq:dwaveid}
\!\!\!  e^{i\beta_j({\bf r})} 
    [{\cal S}_{\bf r}]^2 f_{{\bf r}+\hat{\bf
        x}_j \sigma} \epsilon_{\sigma\sigma'}f_{{\bf r}\sigma'} & = &
    \!\!\!B_{{\bf r},{\bf r}+\hat{\bf x}_j}^\dagger 
    c_{{\bf r}+\hat{\bf 
        x}_j \sigma} \epsilon_{\sigma\sigma'}c_{{\bf
        r}\sigma'}. \end{eqnarray} 
Here we have introduced a bond-Cooper pair operator,
\begin{eqnarray}
  \label{eq:bondcp}
  B_{{\bf r},{\bf r}+\hat{\bf x}_j}^\dagger & = & \tilde{b}^\dagger_{\bf
    r} \tilde{b}^\dagger_{{\bf r}+\hat{\bf x}_j}, \\
  \tilde{b}^\dagger_{\bf r} & = & \prod_{\bf r}^\infty e^{-i\pi
    \overline{e}_i d{\bf r}'_i}.
\end{eqnarray}
Note that in the un-transformed chargon variables of the  $Z_2$
gauge theory formulation, $B_{{\bf r},{\bf r}+\hat{\bf
    x}_j}^\dagger=\sigma^z_j({\bf r})b^\dagger_{\bf r} b^\dagger_{{\bf
    r}+\hat{\bf x}_j}$.

Recalling the discussion in Sec.~\ref{sec:z_2-chargon-spinon}, it is
still the case that for $\Delta_j=0$, when this pairing term is absent,
the fermion number is conserved, and naively may be chosen arbitrarily.
However, in the limit $\Delta_j\rightarrow 0$, which we consider here,
this conservation is weakly violated, and only the total charge is
conserved.  The physics at work is clear from Eq.~(\ref{eq:bdgform}):
for non-vanishing $\Delta_j$, quasiparticle pairs and boson pairs are
interchanged, and the two charged fluids come to equilibrium.  Thus in
what follows, we should choose to divide the charge density amongst the
fermions and bosons in such a way as to minimize the total (free)
energy.  This division will therefore shift as parameters of the
Hamiltonian are changed.  How it does so is crucial to the ultimate
low energy physical properties of the system, as is clear from
Eqs.~(\ref{vorthop})-~(\ref{rothop}), which show that the vortices
experience an effective ``flux'' proportional to the difference of the
total charge density ($\epsilon_{ij}\partial_i a_j/\pi$) and the
fermionic density $n_f=\epsilon_{ij}\partial_i\alpha_j/\pi$.  As the
fermion density is varied, the effective flux seen by the vortices
changes.  Significantly, in the limit $t_v\rightarrow \infty$, vortex
hopping dominates the energetics, and is minimized when the fermion
density equals the total charge density, $n_f =
\epsilon_{ij}\partial_i a_j/\pi$.  This naturally
recovers the Fermi liquid phase (App.~\ref{sec:fermiliquid}) by binding
charge $e$ firmly to each fermion, fully accommodating all the electrical
charge.  

In the rest of the paper we work exclusively within the $U(1)$
vortex-spinon formulation, which is particularly suitable for
extracting the properties of the Roton Fermi liquid. Before
embarking on that, we first establish the formal equivalence
between the two formulations by enslaving the $U(1)$ gauge fields.
As detailed in Appendix ~\ref{ap:enslaveU1}, it is possible to
unitarily transform to a gauge with $\vec{\nabla}\cdot\vec{\alpha}
= \vec{\nabla}\cdot\vec{\beta}^{slave} = 0$.  In this gauge, both
$\alpha_j$ and $\beta_j$ are ``enslaved", being fully expressible
in terms of the spinon and vortex densities, $n^f_{\bf r}$ and
$N_{\sf r}$, respectively.  Moreover, the enslaved $U(1)$
Hamiltonian is found to be {\it identical} to the enslaved $Z_2$
Hamiltonian obtained in Appendix ~\ref{ap:enslaveZ2}, and the
enslaved expressions for the electron operators also
coincide.

Having thereby established the equivalence between the $Z_2$ and
$U(1)$ vortex-spinon field theories, in the remainder of the paper
we choose to work exclusively within the $U(1)$ formulation,
employing the Hamiltonian $H$ defined in Eq.~(\ref{HamU1vs}),
together with the gauge constraints in Eq.~(\ref{U1constraint}).
In practice, it is far simpler to work within a Lagrangian
formulation, where the gauge constraints can be imposed explicitly
within a path integral, as detailed in the next subsection.

\subsubsection{Lagrangian for $U(1)$ Vortex-spinon theory }
\label{sec:lagr-form-u1}

In order to impose the $U(1)$ gauge constraints in
Eq.~(\ref{U1constraint}) on the Hilbert space of the full
vortex-spinon Hamiltonian, $H$, we will pass to a Euclidean
path integral representation of the partition function. The
associated Euclidean Lagrangian is readily obtained as a sum of
three contributions,
\begin{equation}
S = \int d \tau [ H  + L_{B} + L_{con}],
\label{Sfulla}
\end{equation}
with $L_{B}$ involving the generalized coordinates and conjugate momenta,
\begin{eqnarray}
L_{B} & = &
\sum_{{\bf r} = {\sf r} + {\bf w}}
[i N_{\sf r} \partial_0 \theta_{\sf r}
-  {i \over \pi} \beta_i({\bf r}) \epsilon_{ij} \partial_0 \alpha_j ({\sf r} + \hat{x}_i )]  \nonumber \\
& + & \sum_{{\bf r} = {\sf r} + {\bf w}}
[i e^t_j({\sf r}) \partial_0 a^t_j({\sf r})
+ f_{{\bf r} \sigma}^\dagger \partial_0
f^{\vphantom\dagger}_{{\bf r} \sigma} ] ,
\end{eqnarray}
with $\partial_0 \equiv \partial/\partial \tau$ denoting an
imaginary time derivative and $L_{con}$ is a Lagrange multiplier
term imposing the two independent $U(1)$ gauge constraints,
\begin{eqnarray}
L_{con} & = & {i \over \pi} \sum_{{\bf r} = {\sf r} - {\bf w}}
\alpha_0({\sf r}) [\epsilon_{ij} \partial_i \beta_j({\bf r}) - \pi N_{\sf r} ] \nonumber \\
& +& {i \over \pi} \sum_{{\bf r} = {\sf r} + {\bf w}}
 \beta_0({\bf r})
[ \epsilon_{ij} \partial_i \alpha_j({\sf r}) - \pi f_{{\bf r}
\sigma}^\dagger f^{\vphantom\dagger}_{{\bf r} \sigma} ] .
\end{eqnarray}
Here we have introduced two Chern-Simons scalar potentials as
Lagrange multipliers, denoted $\alpha_0({\sf r})$ and
$\beta_0({\bf r})$, which live on the dual and original lattice
sites respectively.

Upon introducing another scalar potential, $a_0({\sf r})$, living on the sites
of the dual lattice, and collecting together the longitudinal and transverse parts
of $a_j$ and $e_j$, the full Euclidean action
can be compactly cast into a simple form.
In order to make the vortex physics more explicit we choose to re-introduce the
vortex phase-field $\theta$ within the Lagrangian formulation.
Specifically, we shift
$a_\mu \rightarrow a_\mu + \partial_\mu \theta$
with $\mu = 0,x,y$,
and then integrate over {\it both} $a_\mu$ and $\theta$.
In this way we arrive at the final form for the full Euclidean Lagrangian:
\begin{equation}
S = S_c + S_f + S_{cs} + S_A  .
\label{Sfullb}
\end{equation}
The charge sector action $S_c =\int_\tau \sum_{\sf r} {\cal
L}_c$ can be expressed in terms of the Lagrangian density,
\begin{equation}
{\cal L}_c= {\cal L}_{a} + {1 \over 2u_0} (\partial_0 \theta - a_0 + \alpha_0)^2 +  {\cal L}_{kin},
\label{Lc}
\end{equation}
with the $u_0 \rightarrow 0$ limit enslaving $a_0 = \partial_0 \theta  +
\alpha_0$ and
\begin{equation}
{\cal L}_{a} = {u_v \over 2} (\overline{e}_j-\frac{A_j}{\pi})^2 - i e_j
(\partial_0 a_j - 
\partial_j a_0) + {v_0^2 \over 2 u_v} (\epsilon_{ij} \partial_i a_j - \pi
\rho_0)^2   .
\end{equation}
The vortex kinetic
energy terms ${\cal L}_{kin}$ are given explicitly by ${\cal H}_{kin}$ in Eqs.~(\ref{vorthop}-\ref{rothop})
except with $a_j^t \rightarrow a_j$.

The fermion action $S_f = \int_\tau \sum_{\bf r} {\cal L}_f$ is
given by,
\begin{equation}
\label{eq:Lf}
{\cal L}_f =  f_{{\bf r} \sigma}^\dagger (\partial_0 - i \beta_0)
f^{\vphantom\dagger}_{{\bf r} \sigma} + {\cal H}_f ,
\end{equation}
with ${\cal H}_f$ given in Eq.~(\ref{eq:Hf}).  The two sectors are
coupled together by the electric field in the electron hopping term, and
by the Chern-Simons action $S_{cs} = \int_\tau \sum_{{\bf r} = {\sf r} +
  {\bf w}} {\cal L}_{cs}$ with
\begin{eqnarray}
{\cal L}_{cs} & = & {i \over  \pi}  \beta_0({\bf r}) \epsilon_{ij} \partial_i \alpha_j({\sf r}) \nonumber \\
&& +  {i \over \pi} \beta_i({\bf r}) \epsilon_{ij} \big[
\partial_j \alpha_0({\sf r} + \hat{x}_i ) - \partial_0 \alpha_j
({\sf r} + \hat{x}_i ) \big] .
\label{Lcs}
\end{eqnarray}
Notice that in the absence of the electron hopping term ($t_e =
0$), the ``electric field'' $e_j$ enters quadratically in the
action, and can then be integrated out to give ${\cal L}_a
\rightarrow \tilde{\cal L}_a$, with
\begin{equation}
\tilde{\cal L}_{a} = {1 \over 2u_v }(\partial_0 a_j - \partial_j a_0)^2 +
v_0^2 (\epsilon_{ij} \partial_i a_j - \pi \rho_0)^2   .
\end{equation}

Finally, it is useful in some circumstances to treat the external gauge
field by making the shift
$\overline{e}_i \rightarrow \overline{e}_i+A_i/\pi$, which removes all
coupling of $A_i$ to the fermions, and furthermore leaves $A_i$ {\sl linearly
coupled} to a charge ``3-current'' of the usual form,
$S_A = \int_\tau \sum_{r} {\cal L}_A$,
with
\begin{equation}
\label{LA}
{\cal L}_A = i A_\mu({\bf r}) J_\mu ({\bf r}) .
\end{equation}
Here $J_\mu$ is the charge 3-current given explicitly by,
\begin{equation}
\label{eq:Jzero}
J_0({\bf r}) = {1 \over \pi} \epsilon_{ij} \partial_i a_j({\sf r})  ,
\end{equation}
\begin{equation}
\label{eq:Jj}
J_i({\bf r}) = {1 \over  \pi}  \epsilon_{ij} [ \partial_j a_0({\sf r} +
\hat{x}_i ) - \partial_0 a_j({\sf r} +
\hat{x}_i ) ]  ,
\end{equation}
with ${\bf r} = {\sf r} + {\bf w}$.  Notice that the 3-current is
conserved as required: $\partial_0 J_0({\bf r}) + \partial_i J_i({\bf r}
- \hat{x}_i ) =0$.  This form is useful for a variety of calculations,
particularly within the purely bosonic RL model discussed in
Secs.~\ref{sec:roton-liquid}-\ref{sec:inst-roton-liqu}, but less so in
some RFL calculations best done in the ``electron gauge'' (see below),
which is incompatible with the above shift of $\overline{e}_i$.  

As can be seen
from the equations of motion obtained from, $\delta {\cal
L}/\delta \alpha =0$ and $\delta {\cal L} / \delta \beta =0$, the
effect of the Chern-Simons term is to attach $\pi$ flux in
$\alpha$ ($\beta$) to the spinon (vortex) world-lines;
\begin{equation}
\epsilon_{\mu \nu \lambda}  \partial_\nu \alpha_\lambda = \pi
J^s_\mu ;  \hskip0.5cm \epsilon_{\mu \nu \lambda} \partial_\nu
\beta_\lambda = \pi J^v_\mu ,
\end{equation}
where $J^s_\mu$ and $J^v_\mu$ are the spinon and vortex
three-currents.  Here $\mu,\nu,\lambda = 0,x,y$ run over the three
space-time coordinates.

Finally we comment on the nature of the gauge symmetries of the
full action $S$ in the Lagrangian representation.  In particular,
associated with the three gauge fields $a_\mu,\alpha_\mu$ and
$\beta_\mu$ are three {\it independent} space-time gauge
symmetries. Specifically, these are:
\begin{enumerate}
\item
  \begin{eqnarray}
    \label{eq:gauge1}
    \theta_{\sf r} & \rightarrow & \theta_{\sf r} + \Theta_{\sf r}, \nonumber \\
    a_\mu({\sf r}) & \rightarrow & a_\mu({\sf r}) + \partial_\mu \Theta_{\sf
    r},
  \end{eqnarray}
\item
  \begin{eqnarray}
    \label{eq:gauge2}
    \theta_{\sf r} & \rightarrow & \theta_{\sf r} + \chi_{\sf r}, \nonumber \\
    \alpha_\mu({\sf r}) & \rightarrow & \alpha_\mu({\sf r}) - \partial_\mu \chi_{\sf
    r},
  \end{eqnarray}
\item
  \begin{eqnarray}
    \label{eq:gauge3}
    f_{{\bf r}\sigma} & \rightarrow & f_{{\bf r}\sigma}e^{i\Lambda_{\bf
        r}}, \nonumber \\
    \beta_\mu({\bf r}) & \rightarrow & \beta_\mu ({\bf r}) +
\partial_\mu \Lambda_{\bf r} ,
  \end{eqnarray}
\end{enumerate}
with $\Theta_{\sf r}, \chi_{\sf r}$ and $\Lambda_{\bf r}$
arbitrary functions of space and imaginary time.  Because of the
gauge invariance of $S$ under these three distinct
transformations, we are free to fix gauges {\it independently} for
the three gauge fields.

In addition to these three gauge symmetries, the full Lagrangian
has two {\it global} symmetries.  By construction,
the spinon Lagrangian ${\cal L}_s$ conserves the total
spin, and since the electrical 3-current in Eqs.~(\ref{eq:Jzero}-\ref{eq:Jj}) satisfies
a continuity equation, $\partial_\mu J_\mu =0$,
the total electrical charge $Q = \sum_{\bf r} J_0({\bf r})$ is
also conserved.

A particularly convenient gauge choice for the gauge field $\beta_\mu$
is,
\begin{equation}
\beta^l_i({\bf r}) = - \pi \overline{e}^l_i({\bf r}) ,
\end{equation}
with $\overline{e}_i$ defined in terms of the ``electric field''
$e_j({\sf r})$ in Eq.~(\ref{eq:defineebar}).  Remarkably, in this
gauge the electron creation operator equals the spinon creation
operator. To see this, it is convenient to shift $\alpha_0
\rightarrow \alpha_0 + a_0$ and then integrate out $a_0$, which
constrains $\vec\nabla\times\vec\beta = \pi\vec\nabla\cdot
\vec{e}$, or equivalently $\beta_i^t = -\pi \overline{e}^t_i$.
Together with the above gauge choice this implies that $\beta_i
\equiv - \pi \overline{e}_i$, so that from Eq.~(\ref{electronU1})
one has $c_{{\bf r}\sigma} = f_{{\bf r}\sigma}$. We refer to this
as the ``electron gauge''. The possibility to choose a gauge
within the Lagrangian formulation of the $U(1)$ vortex-spinon
theory which makes $f_{{\bf r}\sigma}$ an electron operator,
suggest that it should be possible to re-formulate the
vortex-spinon {\it Hamiltonian} entirely in terms of vortices and
electrons.  This is indeed the case, as we demonstrate briefly in
Appendix \ref{ap:elecform}.

\section{The Roton Liquid}
\label{sec:roton-liquid}

We first focus on the bosonic charge sector of the theory, ignoring
entirely the fermions.  Specifically, in the full Euclidean action in
Eq.~(\ref{Sfullb}) we retain {\it only} the charge action, $S_c$, and
the coupling to the external electromagnetic field, $S_A$.  We take the
fermionic density as a non-fluctuating constant.  As discussed in
Sec.~\ref{sec:u1-vortex-spinon-1}, this constant should be determined by
energetics.  We assume here that the largest energy scales in the
problem are those of the vortices, i.e. $\kappa_r,u_v$ etc.  In this
case, one expects that vortex kinetic energy (rotonic or otherwise) is
minimized when the vortices experience zero average magnetic flux.  We
therefore choose the fermionic density {\sl equal} to the total charge
density, setting ${1 \over \pi} \epsilon_{ij} \partial_i \alpha_j =
\rho_0$ and also putting $\alpha_0=0$.  Note that this choice is
essential to recovering an ordinary Fermi liquid state (see
Sec.~\ref{sec:vortex-hopping},App.~\ref{sec:fermiliquid}) and hence is also
natural in this sense.  We remark that, while we will continue to assume
the average fermion density is equal to $\rho_0$ in the bulk of this
paper, we will return to another possibility -- and its physical regime
of relevance -- in the discussion section.

It is also convenient to isolate the
fluctuations in the charge density by defining,
\begin{equation}
a_j = a_j^b + \tilde{a}_j   ,
\end{equation}
with ``background'' density ${1 \over \pi} \epsilon_{ij} \partial_i a^b_j
= \rho_0 $.
We can then take $\alpha_j = a^b_j$,
so that $a_j - \alpha_j = \tilde{a}_j$.
Of the three vortex kinetic energy processes which enter in ${\cal L}_{kin}$,
we hereafter and in the rest of the paper drop entirely the vortex pair hopping process
in Eq.~(\ref{vortpairhop})
putting $t_{2v}=0$.  For now we also set the single vortex hopping processes
to zero, putting $t_v=0$ in Eq.~(\ref{vorthop}), but will return to their effects
in Section IV.  Of interest here is the new roton hopping process,
${\cal L}_{r} \equiv {\cal H}_{r}$ in Eq.~(\ref{rothop}), which can be conveniently recast in
the form:
\begin{equation}
{\cal L}_{r} = - \kappa_r {\cal C}[\tilde{a}] \cos[\Delta_{xy} \theta -
 {1 \over 2}(\partial_x \tilde{a}_y +\partial_y \tilde{a}_x)]  ,
\end{equation}
with
\begin{equation}
{\cal C}[\tilde{a}] = \cos(\epsilon_{ij}
\partial_i \tilde{a}_j/2) .
\label{rotamp}
\end{equation}
More generally, with spinon fluctuations included
one has ${\cal C} = \cos[\epsilon_{ij} \partial_i (a_j - \alpha_j)/2]$.
Clearly, ${\cal C}$ is maximal -- and hence the rotonic kinetic energy
most negative -- for
$\epsilon_{ij}\partial_i\tilde{a}_j=0$, which is true on average for
this choice of fermion density.

We next choose the gauge $\vec{\nabla}\cdot\vec{\tilde{a}} =0$, and
integrate over $a_0$ (with $u_0 \rightarrow 0$).
Having dropped the vortex hopping processes, the remaining charge Lagrangian is then given by,
\begin{equation}
{\cal L}_c = {\cal L}_{pl} + {\cal L}_{\theta}  ,
\end{equation}
with
\begin{equation}
{\cal L}_{pl} = {1 \over 2u_v} [ (\partial_0 \tilde{a}_j)^2 + v_0^2  (\epsilon_{ij} \partial_i \tilde{a}_j)^2 ]  ,
\label{Lplas}
\end{equation}
\begin{equation}
{\cal L}_{\theta} = {1 \over 2u_v} (\partial_j \partial_0 \theta)^2 -
\kappa_r{\cal C}[\tilde{a}] \cos[\Delta_{xy}\theta -
 {1 \over 2}(\partial_1 \tilde{a}_2+\partial_2 \tilde{a}_1)]  .
\label{Ltheta}
\end{equation}
To analyze the phases of this model
it is instructive to represent the Lagrangian ${\cal L}_\theta$
in Hamiltonian
form by re-introducing the vortex number operator, $\hat{N}_{\sf
  r}$:
\begin{eqnarray}
  \hat{H}_\theta & = & {u_v \over 2} \sum_{{\sf r},{\sf r}^\prime}
  \hat{N}_{\sf r}
  \hat{N}_{{\sf r}^\prime} V({\sf r} - {\sf r}^\prime) \nonumber \\ &&
  \hspace{-0.3in}- \kappa_r   \sum_{\sf r} {\cal C}[\tilde{a}]
 \cos(\Delta_{xy}\hat{\theta}_{\sf
    r}-\frac{1}{2}(\partial_x \tilde{a}_1+\partial_y\tilde{a}_x))  .
\label{Htheta}
\end{eqnarray}
Again $V({\sf r})$ is Fourier
transform of the discrete inverse Laplacian operator, $V({\bf k})
\equiv 1/{\cal K}^2({\bf k})$, with ${\cal K}^2({\bf k}) = \sum_j
2(1-\cos k_j)$.

The first term in Eq.~(\ref{Htheta}) describes a logarithmically
interacting gas of (integer strength) vortices moving on
the dual 2d square lattice.  When $\kappa_r =0$, this model
will undergo a finite temperature Kosterlitz-Thouless transition\cite{2dXY} from
a high temperature ``vortex plasma'' into the low temperature ``vortex
dielectric''.  This corresponds, of course, to a transition into
a superconducting phase.  With $\kappa_r=0$ the Kosterlitz-Thouless
transition temperature will be set by the vortex interaction strength,
$u_v$.  But upon increasing the strength of the roton hopping,
one expects the transition temperature to be suppressed, and for
$\kappa_r \gg u_v$ to be driven all the way to zero.
Thus, at zero temperature, upon increasing the single dimensionless
ratio, $\kappa_r/u_v$, one expects a quantum phase transition
out of the superconducting phase and into a new phase (see Fig. 1) -
the ``Roton Liquid''.

\begin{figure}
\begin{center}
\vskip-2mm
\hspace*{0mm}
\centerline{\fig{2.8in}{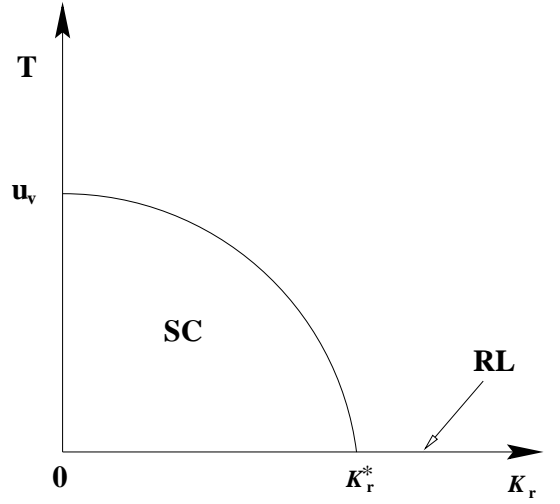}}
\vskip-2mm
\caption{Schematic phase diagram of the Hamiltonian in the charge sector, $H_c = H_\theta + H_{pl}$,
with zero vortex hopping strength, $t_v=0$.  When the roton hopping vanishes,
$\kappa_r=0$, $H_c$ describes a classical
logarithmically interacting vortex gas, and
has a superconductor-to-normal transition
at a Kosterlitz-Thouless temperature, $T_{KT} \approx u_v$.
With increasing roton hopping, $T_{KT}$ decreases being driven to zero
at $\kappa_r^* \approx u_v$ where there is a quantum phase transition
from the superconductor into the Roton liquid phase.
}
\label{fig:RLphasediag}
\end{center}
\end{figure}

\subsection{Harmonic theory and Excitations}
\label{sec:harm-theory-excit}

To access the properties of the Roton Liquid (RL), we consider
$\kappa_r \gg u_v$, where it is presumably valid to expand
the cosine terms in Eq.~(\ref{Htheta}) for small argument,
giving $\hat{H}_\theta = \hat{H}_{rot} + ...$, with
\begin{eqnarray}
\label{eq:Hrot}
\hat{H}_{rot} & = & {u_v \over 2} \sum_{{\sf r},{\sf r}^\prime} \hat{N}_{\sf r}
\hat{N}_{{\sf r}^\prime} V({\sf r} - {\sf r}^\prime) +
\frac{\kappa_r}{8} \sum_{\sf r} (\epsilon_{ij}\partial_i
\tilde{a}_j)^2  \nonumber \\
& & + \frac{\kappa_r}{2}
  \sum_{\sf r} [\Delta_{xy}\hat{\theta}_{\sf
    r}- \frac{1}{2} (\partial_x \tilde{a}_y+\partial_y \tilde{a}_x)]^2  .
\end{eqnarray}
With this expansion, it is no longer legitimate to
restrict $\theta$ to the interval $[0,2\pi]$.   Consistency then dictates that
the eigenvalues of the vortex number operator no longer be restricted
to integers, but allowed to take on any real value from $[- \infty, \infty]$.

The full Roton Liquid Hamiltonian, $\hat{H}_{RL} =\hat{H}_{pl}+\hat{H}_{\rm rot}$ is
quadratic and can be readily diagonalized.  This is most conveniently
done by returning to the Lagrangian framework, described now by
\begin{eqnarray}
  {\cal L}_{RL} & = & {1 \over 2u_v} [ (\partial_0 \tilde{a}_j)^2 +
  \tilde{v}_0^2  (\epsilon_{ij} \partial_i \tilde{a}_j)^2
  ] \nonumber \\
  &  & \hspace{-0.5in}+
  {1 \over 2u_v} (\partial_j \partial_0 \theta)^2 +{ \kappa_r \over 2}
  [\Delta_{xy} \theta -
  {1 \over 2}(\partial_x \tilde{a}_y+\partial_y \tilde{a}_x)]^2   ,
\label{LRL}
\end{eqnarray}
with $\tilde{v}_0=\sqrt{v_0^2 + \kappa_r u_v/4}$.
To proceed to describe the normal modes of this quadratic Lagrangian,
we define Fourier transforms
\begin{eqnarray}
  O({\bf r},\tau) & = & \int_{{\bf k},\omega_n} e^{i {\bf k}\cdot{\bf r} -i\omega_n\tau} O({\bf k},\omega_n), \\
  {\sf O}({\sf r},\tau) & = & \int_{{\bf k},\omega_n} e^{i {\bf k}\cdot{\sf r}-i\omega_n\tau} {\sf O}({\bf k},\omega_n),
\end{eqnarray}
for fields $O,{\sf O}$ on the original and dual lattices,
respectively.  Here integration $\int_{\bf k} \equiv \int d^2{\bf
  k}/(2\pi)^2$ is taken over the Brillouin zone $|k_1|,|k_2| < \pi$ and
$\int_{\omega_n} \equiv \int_{- \infty}^\infty d\omega_n/(2\pi)$
defines the integration measure (at zero temperature) for the
Matsubara frequency $\omega_n$.  At non-zero temperature, one simply
replaces $\int_{\omega_n} \rightarrow \beta^{-1}\sum_{\omega_n}$, with
$\omega_n=2\pi n/\beta$.  It is moreover convenient to define
\begin{equation}
  {\cal K}_j({\bf k}) = -i(e^{ik_j}-1),
\end{equation}
so that upon Fourier transformation, the discrete derivatives behave
intuitively,
\begin{equation}
  \partial_j \rightarrow_{FT} i {\cal K}_j({\bf k}),
\end{equation}
and of course $\partial_0 \rightarrow -i\omega_n$ as usual.  We also
introduce the transverse gauge field:
\begin{equation}
\tilde{a}_i({\bf k}) =\epsilon_{ij} {  i{\cal K}^*_j({\bf k})  \over {\cal
    K}({\bf k}) } a({\bf k})    ,
\end{equation}
with ${\cal K}^2({\bf k})  = \sum_j |{\cal K}_j({\bf k})|^2$
and $|{\cal K}_j({\bf k})| = 2 |\sin(k_j/2)|$.

To now diagonalize ${\cal L}_{RL}$, it is convenient to define
a real two-component field
$\Upsilon_a$ via
\begin{eqnarray}
  a({\bf k},\omega_n) & = & \sqrt{u_v} e^{i{\bf k}\cdot{\bf w}} \Upsilon_1({\bf
    k},\omega_n), \\
  \theta({\bf k},\omega_n) & = & \frac{\sqrt{u_v}}{{\cal K}({\bf
      k}) } \Upsilon_2({\bf k},\omega_n).
\end{eqnarray}
Then the action, $S_{RL}=\sum_{\sf r}\int_\tau {\cal L}_{RL}$ is
\begin{eqnarray}
  S_{RL} & = & \frac{1}{2}
\int_{{\bf k},\omega_n}  \hspace{-0.2in} \Upsilon_\alpha({\bf k},\omega_n) G^{-1}_{\alpha\beta}({\bf k},\omega_n)
  \Upsilon_\beta({\bf -k},-\omega_n),
  \label{eq:upsilons}
\end{eqnarray}
with
\begin{equation}
  \label{eq:gf}
  G_{\alpha\beta} =
  \frac{G^0 \delta_{\alpha\beta} + G^z \sigma^z_{\alpha\beta} + G^x
    \sigma^x_{\alpha\beta}}{(\omega_n^2+\omega_{pl}^2)(\omega_n^2+\omega_{rot}^2)} ,   \end{equation}
where $\vec{\sigma}$ is the usual vector of Pauli matrices.
Here we have defined,
\begin{eqnarray}
  G^0 & = & \omega_n^2 + \frac{1}{2} v_+^2 {\cal K}^2, \\
  G^z & = & -\frac{1}{2} v_+^2 {\cal K}^2 + v_1^2 \frac{|{\cal
      K}_x{\cal K}_y|^2}{{\cal K}^2}, \\
  G^x & = & \frac{v_1^2}{2} (|{\cal K}_x|^2 -
    |{\cal K}_y|^2)\frac{\tilde{\cal K}_x\tilde{\cal K}_y}{{\cal K}^2},
\label{eq:gf1}
\end{eqnarray}
with $v_1=\sqrt{\kappa_r u_v}$ and $\tilde{\cal K}_j = 2\sin (k_j/2)$.  The poles in
$G_{\alpha\beta}$ at $\omega=i\omega_n=\pm \omega_{pl}, \pm \omega_{rot}$
describe two types of collective modes.

The first excitation is a plasmon with a
renormalized dispersion,
\begin{equation}
  \omega_{pl}^2({\bf k}) = \frac{1}{2} \left[ v_{+}^2 {\cal K}^2 +
    \sqrt{ v_{-}^4 {\cal K}^4 + \tilde{v}_0^2 v_1^2 (|{\cal K}_x|^2
- |{\cal K}_y|^2 )^2 }\right],
\end{equation}
with velocities,
\begin{equation}
  v_{\pm}=\sqrt{\tilde{v}_0^2 \pm v_1^2/4} .
\end{equation}
The plasmon frequency vanishes at the center of the Brillouin zone,
${\bf k}=0$, and in the absence of long-ranged Coulomb interactions
disperses linearly $\omega_{pl} = v_{pl} |{\bf k}|$ at
small wavevectors, $k_j \rightarrow 0$.
But the associated plasmon velocity, $v_{pl}(\phi)$,
depends upon the ratio, $k_y/k_x = \tan(\phi)$.
In particular, along the zone diagonals with $k_y= \pm k_x$
the velocity is minimal and unaffected by the vortices with $v_{pl} = \tilde{v}_0$, whereas
it takes it's maximum
value, $v_+$, along the $k_x$ or $k_y$ axes.

This upward shift in the plasmon frequency is due to a
``level repulsion'' with the second collective mode - the gapless roton,
which disperses as,
\begin{equation}
  \omega_{rot}^2({\bf k}) = \frac{1}{2} \left[ v_+^2 {\cal K}^2 -
   \sqrt{ v_{-}^4 {\cal K}^4 + \tilde{v}_0^2 v_1^2 (|{\cal K}_x|^2
- |{\cal K}_y|^2 )^2 }\right]  .
\end{equation}
For $|k_x| \ll 1$ {\sl and fixed $k_y$} the roton dispersion {\it vanishes},
$\omega_{rot} \sim v_{rot} |k_x|$, with
\begin{equation}
  v_{rot} = \frac{\tilde{v}_0}{v_{+}} v_1  .
\end{equation}
Remarkably, the Roton Liquid phase supports
a gapless ``Bose surface'' of roton excitations,
along the $k_x=0$ and $k_y=0$ axes.  These roton excitations
describe {\it gapless and transverse} current fluctuations,
which are obviously not present in a conventional bosonic superfluid.

With long-range Coulomb interactions present one would have simply,
$v_0^2 \rightarrow v_0^2({\bf k}) \sim {1 \over |{\bf k}|}$, giving
the familiar 2d plasmon dispersion, $\omega_{pl} \sim \sqrt{|{\bf
    k}|}$.  In addition, the roton velocity $v_{rot}$ becomes
dependent upon $k_y$.  We note in passing that the roton velocity is
in either case determined not only from the dynamics of $\theta$ but
also that of $\tilde{a}$, as is evident from its dependence upon
$\tilde{v}_0/v_{+}$.  It will be sometimes instructive in the
following to consider the simple limit $v_0\sim\tilde{v}_0 \rightarrow
\infty$, in which the spatial fluctuations of $\tilde{a}$ vanish and
the roton mode is entirely decoupled from $\tilde{a}$.

\subsection{No Meissner effect in roton liquid}
\label{sec:no-meissner-effect}

We now employ the Gaussian theory to examine some of the electrical
properties of the Roton liquid phase.
Consider first the response of the RL phase to
an applied magnetic field.
In the presence of a magnetic field, $B = \epsilon_{ij} \partial_i A_j$, there is an additional term that
one must add to the Lagrangian, which
from Eq.~(\ref{LA}) takes the form:
\begin{equation}
{\cal L}_A = {i \over \pi} a_0 B   .
\end{equation}
If $a_0$ is integrated out from ${\cal L}_c$ in Eq.~(\ref{Lc}) with this
additional term present, the Hamiltonian, $H_\theta(\hat{N}_{\sf r},
\hat{\theta}_{\sf r})$ in Eq.~(\ref{Htheta}) becomes simply,
$H_\theta(\hat{N}_{\sf r}- {1 \over \pi} B, \hat{\theta}_{\sf r})$.  As
expected, the vortex density will become non-zero in the presence of the
magnetic field.  Since the vortex number operator in $H_\theta$ has
integer eigenvalues, it is not generally possible to shift away the
applied B-field.  But in the Roton Liquid phase (at zero temperature)
where the cosine term can be expanded to quadratic order (as in
$H_{rot}$), the vortex number operator has a continuous spectra, and one
can formally eliminate the B-field by shifting, $\hat{N}_{\sf r}
\rightarrow \hat{N}_{\sf r} + {1 \over \pi} B$, for all ${\sf r}$.
Since the ground state energy of the RL phase is thus independent of the
applied B-field, both the magnetization and the magnetic susceptibility,
$\chi = \partial M/\partial B$ vanish.  Unlike in a superconductor,
where $\chi = -{1 \over 4\pi}$, there is {\it no} Meissner effect in the
Roton Liquid (strictly speaking, there is never a Meissner effect in a
single two dimensional layer, but one can consider an infinite stack of
electrically decoupled but magnetically coupled layers, which would
exhibit a Meissner effect when the layers are true 2d superconductors,
but not when they are RLs).  Physically, since the RL phase supports
gapless roton excitations, the state cannot screen out an applied
magnetic field.

\subsection{Off-Diagonal Quasi-Long-Range Order}
\label{sec:quasi-diagonal-long}

We next consider the Cooper pair propagator in the Roton Liquid,
\begin{equation}
  G^{cp}({\bf r}_1 - {\bf r}_2,\tau_1-\tau_2)  = \langle
  B_{{\bf r}_1}(\tau_1)
  B^\dagger_{{\bf r}_2}(\tau_2) \rangle ,
\end{equation}
with the Cooper pair destruction operator $B_{\bf r}$ given in
Eq.~(\ref{pairop}) as an infinite product of exponentials, $e^{2\pi i
  e_j}$, running along the string.  The propagator for the $d$-wave pair
field, $B_{{\bf r},{\bf r}+\hat{\bf x}_j}$, 
\begin{eqnarray}
  \label{eq:Gcpij}
    G^{cp}_{ij}({\bf r},\tau) & = & 
  \left\langle B^{\vphantom\dagger}_{{\bf 0},\hat{\bf
        x}_i}(0)B^\dagger_{{\bf r},{\bf 
    r}+\hat{\bf x}_j}(\tau)\right\rangle,
\end{eqnarray}
behaves similarly, and will
be discussed at the end of this sub-section.

\subsubsection{Equal time correlator}

We consider at first the equal time correlator, with $\tau_1=\tau_2$.
The path independence of the string rests on the condition,
$(\vec{\nabla}\cdot\vec{e})({\sf r}) = N_{\sf r}$, with {\it integer}
vortex number, $N_{\sf r}$.  Unfortunately, within the tractable
harmonic approximation valid for most quantities in the roton liquid
phase (with cosine terms in the roton hopping expanded to quadratic
order), the condition of integer vortex number is {\it not} satisfied,
and the results for $G^{cp}({\bf r},0)$ depend upon the choice of
string.  We believe that the correct behavior can be extracted by taking
the string running along the straightest and ``shortest'' (using the
``city block'' metric $|x_1-x_2|+|y_1-y_2|$) path between the two
points, ${\bf r}_1$ and ${\bf r}_2$.  As we shall see, the Cooper pair
propagator calculated in this way has an anisotropic spatial power-law
decay.  Preliminary calculations suggest that, once perturbative
corrections to the harmonic approximation (using the formalism
established in Sec.~\ref{sec:superc-inst}) are taken into account (even
if they are irrelevant in the renormalization group sense), simple
variations in the string do not modify the power-law decay of
$G^{cp}({\bf r},0)$, but only change the (non-universal)
prefactor.\cite{BFunpub}\ 

We take ${\bf r}_1 - {\bf r}_2= X\hat{x}+Y\hat{y}$, and ${\bf r}_2
= {\bf w}$, with integer $X,Y\geq 0$. Then, upon expressing the
correlator $G^{cp}(X,Y,0)$ as an imaginary time path integral, one
obtains an extra term in the Euclidean action, $S=\int_\tau
\sum_{\sf r} {\cal
  L}$ in Eq.~(\ref{Sfullb}), with
\begin{equation}
  {\cal L} \rightarrow {\cal L} + i e_j({\sf r}
  ,\tau) {\cal J}_j({\sf r},\tau)  .
\end{equation}
The c-number ``source'' field is given as,
\begin{equation}
{\cal J}_j({\sf r},\tau) = 2 \pi \delta(\tau)\big[
\delta_{j2}\sum_{x'=0}^{X-1} \delta_{x,x'+1}\delta_{y,0}-\delta_{j1}
\sum_{y'=0}^{Y-1} \delta_{x,X}\delta_{y,y'} \big] .
\end{equation}
The Fourier transform is simply,
\begin{equation}
  {\cal J}_j({\bf k},\omega_n) = 2\pi \big[\delta_{j2}
  \frac{1-e^{-ik_x X}}{i{\cal K}_x} - \delta_{j1} \frac{e^{-ik_x
      X}(1-e^{-ik_y Y})}{i{\cal K}_y}\big].
\end{equation}
Integrating out the electric field $e_j$ in the gauge
$\vec{\nabla}\cdot \vec{a}=0$ and decomposing the source
fields into transverse and longitudinal parts, ${\cal J}_t = i
\epsilon_{ij} {\cal K}_i {\cal J}_j/{\cal K}$, ${\cal J}_l=i {\cal
  K}_i^* {\cal J}_i/{\cal K}$, one obtains the result
\begin{eqnarray}
  \label{eq:stringups}
  G^{cp}(X,Y,0) & = & \Big\langle \exp\big\{ -\int_{{\bf
      k},\omega_n}\big[
  \frac{i\omega_n}{\sqrt{u_v}}( {\cal J}_t \Upsilon_1-{\cal J}_l
  \Upsilon_2 ) \nonumber \\ && + \frac{1}{2u_v} ({\cal J}_t^2+{\cal
    J}_l^2) \big] \big\}
  \Big\rangle_{\Upsilon} ,
\end{eqnarray}
where the Gaussian average over $\Upsilon$ is to be taken with respect
to $S_{RL}$ in Eq.~(\ref{eq:upsilons}).  Performing this Gaussian
integral, one obtains
\begin{eqnarray}
  \label{eq:Gcpthree}
  G^{cp}(X,Y,0) & = & \exp[ -\Gamma_l -\Gamma_t
  -\Gamma_{lt}],
\end{eqnarray}
where
\begin{eqnarray}
  \label{eq:Gammacps}
  \Gamma_l & = & \int_{{\bf k}\omega_n} \frac{|{\cal J}_l|^2
    (1-\omega_n^2 G_{22})}{2u_v}, \\
  \Gamma_t & = & \int_{{\bf k}\omega_n} \frac{|{\cal J}_t|^2
    (1-\omega_n^2 G_{11})}{2u_v}, \\
  \Gamma_{lt} & = & -\int_{{\bf k}\omega_n} \frac{{\cal J}_l{\cal J}_t^*
    \omega_n^2 G_{12}}{u_v},
\end{eqnarray}
with $G_{ij}$ given in Eq.~(\ref{eq:gf}).

Investigation of $\Gamma_t$ and $\Gamma_{lt}$ shows that the
corresponding integrands are non-singular at small $k_x$ or $k_y$, and
hence go to finite limits for large $|X|$ and/or large $|Y|$.
They will thus affect only the amplitude of the Cooper pair propagator at
large distances, and we henceforth neglect them.  Singular behavior at
long distances {\sl does} arise from $\Gamma_l$, in line with the
intuition that it is vortex fluctuations which disrupt the
superconducting phase, since ${\cal J}_l$ couples to the longitudinal
electric field, which through $\vec\nabla\cdot\vec{e} = N$ describes
the vorticity.  To evaluate $\Gamma_l$, we first perform the frequency
integration to obtain
\begin{eqnarray}
  \label{eq:gammal1}
  \Gamma_l & = & \int_{\bf k} |{\cal J}_l|^2 \frac{\tilde{v}_0 v_1
    |{\cal K}_x {\cal K}_y| + v_1^2|{\cal K}_x {\cal K}_y|^2/{\cal
      K}^2}{4u_v (\omega_{pl}+\omega_{rot})} .
\end{eqnarray}
Next, we explicitly express the square of the longitudinal string,
\begin{eqnarray}
  \label{eq:stringl}
  |{\cal J}_l|^2 & = & \frac{(2\pi)^2}{{\cal K}^2} \Big[
  \left|\frac{{\cal K}_x}{{\cal K}_y}\right|^2 |1-e^{-ik_y Y}|^2 +
  \left|\frac{{\cal K}_y}{{\cal K}_x}\right|^2 |1-e^{-ik_x X}|^2
  \nonumber \\
  & & + 2{\rm Re} \{ (1-e^{-ik_x X})(1-e^{-ik_y Y}) \} \Big].
\end{eqnarray}
The first two terms in Eq.~(\ref{eq:stringl}) are singular for small
$k_y$, $k_x$ respectively for very large $Y$,$X$, leading to
a logarithmic dependence when inserted in Eq.~(\ref{eq:gammal1}).  The
final term in Eq.~(\ref{eq:stringl}), by contrast, is singular only
for {\sl both} $k_x$,$k_y$ small, and this singularity, inserted into
Eq.~(\ref{eq:gammal1}), is weak and integrable.  Extracting the
logarithmic parts, one finds
\begin{equation}
  \label{eq:gammal2}
  \Gamma_l \sim \Delta_c (\ln |X| + \ln|Y|),
\end{equation}
for $|X|,|Y| \gg 1$, with
\begin{equation}
\Delta_c = 2\pi {\tilde{v}_0 \over v_+ } \sqrt{ {\kappa}_r \over
u_v }  .
\label{Cooperexp}
\end{equation}
Hence we have
\begin{equation}
  \label{eq:ODQLRO}
  G^{cp}(X,Y,0) \sim \frac{{\rm const}}{|X|^{\Delta_c}|Y|^{\Delta_c}}.
\end{equation}
This establishes
that the Roton liquid phase has off-diagonal quasi-long-ranged
order (ODQLRO) at zero temperature.

\subsubsection{Unequal time correlator}

We now consider the Cooper pair propagator at unequal times.
Unfortunately, it is difficult to produce a simple and general
calculation for arbitrary spatial and time separations.  In
particular, clearly, by square symmetry, we expect $G^{cp}(X,Y,\tau) =
G^{cp}(Y,X,\tau)$.  Any choice of strings, however, necessarily
creates an asymmetry between the two spatial directions.  As we have
been unable to resolve this dilemma, we instead focus on the simple
case in which the pair is created and annihilated on a single row of
the lattice, i.e. $G^{cp}(X,0,\tau)$.  We will see that this
correlator decays as a power law both in space and time.

To proceed, we take ${\bf r}_1 - {\bf r}_2= X\hat{x}$, ${\bf r}_2 =
{\bf w}$, $\tau_1=\tau$, $\tau_2=0$, with $X,Y\geq 0$.  With this
choice, the string in Fourier space becomes
\begin{equation}
  \label{eq:stringtau}
  {\cal J}_j({\bf k},\omega_n) = 2\pi \delta_{j2}
  \frac{1-e^{-ik_x X+i\omega_n\tau}}{i{\cal K}_x}.
\end{equation}
Repeating the same manipulations as above, one again obtains (with
negligible contributions from the transverse part of the string)
\begin{equation}
  \label{eq:cptau}
  G^{cp}(X,0,\tau) \sim \exp[ -\tilde\Gamma_l ],
\end{equation}
with
\begin{equation}
  \label{eq:gammalt}
  \tilde\Gamma_l \sim \frac{\Delta_c}{2} \ln ( X^2 + v_{rot}^2 \tau^2 ),
\end{equation}
where for simplicity we have taken $v_{rot}$ independent of $k_y$ (for
a non-trivial $v(k_y)$, the logarithm is simply averaged uniformly
over the $k_y$ axis).  This gives
\begin{equation}
  \label{eq:Gctau}
  G_c(X,0,\tau) \sim \frac{\rm const.}{(X^2+v_{rot}^2 \tau^2)^{\Delta_c/2}}.
\end{equation}
Note that this power-law form implies a power-law local tunneling
density of states for Cooper pairs, $\rho_{cp}(\epsilon) \sim
\epsilon^{\Delta_c-1}$.

We conjecture that the full
correlator satisfies a simple scaling form with ``z=1'' scaling:
\begin{equation}
  \label{eq:scaling}
  G^{cp}(X,Y,\tau) \sim \frac{{\rm
      const}}{|X|^{\Delta_c}|Y|^{\Delta_c}} {\cal
    G}(\frac{X}{v_{rot}\tau},\frac{Y}{v_{rot}\tau}).
\end{equation}
Combining our two calculations above implies ${\cal G}(\chi,0) =
(\chi^2/(\chi^2+1))^{\Delta_c/2}$.

\subsubsection{ODQLRO of the $d$-wave pair field}
\label{sec:odqlro-d-wave}

We now briefly discuss the analogous ODQLRO of the $d$-wave pair field
described by $G^{cp}_{ij}({\bf r},\tau)$ in Eq.~(\ref{eq:Gcpij}).  This
quantity is plagued by the same string ambiguities as the local Cooper
pair propagator, but to a larger degree, since it involves {\sl two}
separate strings emanating from the two sites shared by the initial bond
and ending at the two sites shared by the final bond.  Although we are
confident $G^{cp}_{ij}({\bf r},\tau)$ has a power-law form consistent
with ODQLRO, we are unable to determine the precise nature of these
correlations with reliability.  For instance, consider the equal time
pair field correlator for two bonds along the $x$-axis,
$G^{cp}_{11}({\bf r},0)$.  For two bonds in the same row, ${\bf
  r}=X\hat{\bf x}_1$, the strings can be chosen to line all in the same
row, and since the logarithmic divergence controlling the ODQLRO arises
from small $k_x$ in this case, the power law exponent is unchanged, i.e.
$G^{cp}_{11}(X,0,0) \sim {\rm Const}/|X|^{\Delta_c}$, with $\Delta_c$
given above.  We believe that, since this choice of string is by far the
most natural, this is probably the correct result.  If, instead, we
choose to separate the two pair fields along a single column, ${\bf
  r}=Y\hat{\bf x}_2$, then the two strings involved cannot be taken
entirely atop one another.  Different choices for the strings then give
different results.  For instance, making the symmetric choice of two
parallel strings (each of strength $\pi$ rather than $2\pi$) gives a
decay exponent reduced from $\Delta_c$ to $\Delta_c/2$ in the $Y$
direction, while choosing the strings to overlap everywhere except the
two ends reproduces the previous exponent $\Delta_c$ without any
reduction.  Since we are unable to reliably resolve this ambiguity, we
are unable to determine the exact form of the $d$-wave pair field
correlator.  Instead, we will take the pragmatic approach of
approximating the correlations by those of the local pair field,
$G^{cp}_{ij}({\bf r},\tau) \approx G^{cp}({\bf r},\tau)$.  It should be
understood that the decay exponent $\Delta_c$ may need to be
renormalized and/or the correlator corrected slightly to obtain detailed
results for a specific model.

\subsection{Conductivity in the harmonic theory}
\label{sec:cond-harm-theory}

Given the above result of ODQLRO, it is natural to expect a very
large and perhaps infinite conductivity in the Roton liquid.
Indeed, neglecting the effects of vortex hopping, ${\cal L}_v$ in
Eq.~(\ref{vorthop}), one can readily see (e.g. from
Eq.~(\ref{Htheta})) that the total vortex number on each row and
each column of the 2d lattice is separately conserved.  Thus it is
impossible to set up a vortex flow, and hence by the Josephson
relation to generate an electric field.  Using the quadratic roton
liquid Lagrangian ${\cal L}_{RL}$, this expectation can be
directly confirmed.  In particular, we can integrate out {\sl all}
dynamical fields ($\theta,a$) from the Lagrangian leaving only the
external gauge field $A^\mu$, and thereby extract the polarization
tensor and conductivity.  This is most conveniently carried out in
the gauge $A_0=0$, and assuming no magnetic field $\partial_x
A_y-\partial_y A_x=0$.  In this case, one may write
\begin{equation}
 S_A = \int_{{\bf k},\omega_n} \frac{-i\omega_n {\cal K}_j({\bf
     k})}{\pi{\cal K}({\bf k})}e^{-i{\bf k}\cdot{\bf w}} a({\bf k},\omega_n)
 A_j(-{\bf k},-\omega_n).
\end{equation}
Integrating out $a$ and $\theta$ using the Green's function $G_{11}$
in Eq.~(\ref{eq:gf}) , one obtains, in the limit $|{\bf k}| \rightarrow
0$, the effective action
\begin{equation}
  S^{0}_A = \frac{1}{2} \int_{{\bf k},\omega_n} \Pi^0_{ij}({\bf
    k},\omega_n) A_i({\bf k},\omega_n) A_j(-{\bf k},-\omega_n),
\end{equation}
with
\begin{equation}
  \Pi^0_{ij}({\bf k},\omega_n) \sim \frac{u_v}{\pi^2}\frac{k_i k_j}{k^2} ,
\end{equation}
as $|{\bf k}| \rightarrow 0$.  The dependence of $\Pi^0_{ij}$ on the
orientation of ${\bf k}$ is due to the fact that we have assumed
$B=0$, which according to Faraday's law requires ${\bf \nabla}\times
{\bf E} = -\partial_t {\bf B}=0$.  Hence we must choose ${\bf k}$
parallel to the electric field ${\bf A}={\bf E}/(-i\omega)$.  Thus we
extract an {\sl isotropic} conductivity tensor $\sigma^0_{ij} =
\delta_{ij} \sigma^0$, with
\begin{equation}
  \sigma^0(\omega) = \frac{u_v}{\pi^2} \frac{1}{-i\omega},
\end{equation}
characteristic of a system with no dissipation.

The above conclusion for the quadratic RL Lagrangian is, however,
modified by the vortex hopping terms.  As we detail in the next section,
the effects of a small vortex hopping term depend sensitively on the
parameters that enter in the harmonic theory of the roton liquid - in
particular the dimensionless ratio $u_v/\kappa_r$.  There are two
regimes.  When this ratio is larger than a critical value, vortex
hopping is ``relevant'' and grows at low energies destabilizing the
roton liquid phase.  On the other hand, for small enough $u_v/\kappa_r$
the vortex hopping strength scales to zero and the roton liquid phase is
stable.  In this latter case, the effects of vortex hopping on physical
quantities can be treated perturbatively.  In particular, we find that
the conductivity in the roton liquid diverges as a power law in the low
frequency and low temperature limit.

\section{Instabilities of the Roton Liquid}
\label{sec:inst-roton-liqu}

We first consider the instabilities of the roton liquid due to the
presence of a vortex hopping term, and examine the effects of such
processes on the electrical transport.  In the subsequent subsection
we consider the legitimacy of the harmonic expansion required to
obtain the quadratic RL Lagrangian.  This is achieved by performing a
''plaquette duality'' transformation\cite{EBL}, where it is possible to address
this issue perturbatively.  Again we find two regimes depending on the
parameters in the quadratic roton liquid Lagrangian.  A stable regime
for small $u_v/\kappa_r$ wherein the harmonic roton liquid description
is valid, and an instability towards a superconducting phase when this
ratio is large.

\subsection{Vortex hopping}
\label{sec:vortex-hopping}

The analysis of the stability of the RL to vortex hopping is
mathematically nearly identical to the stability analysis of the
Exciton Bose Liquid (EBL) of Ref.~[\onlinecite{EBL}] with respect to
boson hopping.  Taking over those methods, we note that the vortex
hopping operator exhibits {\sl one-dimensional power-law
  correlations},
\begin{equation}
  \label{eq:hoppower}
 \langle e^{i\partial_y\theta_{\sf r}(0)} e^{-i\partial_y\theta_{{\sf
        r}+{\bf r}}(\tau)} \rangle_{RL} = \delta_{y,0} {\cal R}(x,\tau) ,
\end{equation}
with a power law form at large space-time separations,
\begin{equation}
\label{rotgreen}
{\cal R}(x,\tau) =
  \frac{c}{(x^2+v_{rot}^2 \tau^2)^{\Delta_v}},
\end{equation}
where ${\bf r}=x{\bf\hat{x}}+y{\bf\hat{y}}$
and $c$ is a dimensionless constant.
Here, ${\cal R}(x,\tau)$ is essentially the single roton
Greens function,
describing the space-time propagation of a roton with a dipole
oriented
along the $\hat{y}$ axis.
When calculated using the RL Lagrangian, one finds
\begin{equation}
\label{DeltavRL}
  \Delta_v = \frac{1}{4\pi} \sqrt{\frac{u_v}{\kappa_r}} \frac{v_+}{\tilde{v}_0}.
\end{equation}
Simple calculations show that this power-law behavior is not modified
by including the fluctuating $a_y$ field in the vortex hopping
operator, $\hat{T}_j({\sf r},\tau) = e^{i(\partial_y\theta_{\sf r} +
  a_y({\sf r}))}$, i.e.
\begin{equation}
  \label{eq:hoppower2}
  \langle \hat{T}_y({\sf r},0) \hat{T}^*_y({\sf r}+{\bf r},\tau)
   \rangle_{RL} \sim
  \frac{\delta_{y,0}}{(x^2+v_{rot}^2 \tau^2)^{\Delta_v}},
\end{equation}
with only a change in the prefactor. Notice that the exponent
$\Delta_v$ characterizing the power law decay of the roton
propagator is inversely proportional to the analogous exponent
$\Delta_c$ in Eq.~(\ref{Cooperexp}) which gives the power law
decay of the Cooper pair propagator.  Indeed, for the Roton liquid
Lagrangian studied here we find the simple identity, $\Delta_v
\Delta_c = 1/2$.  But with inclusion of other terms in the
original Hamiltonian such as the spinons or further neighbor roton
hopping terms, this equality will be modified.

The arguments of Ref.~[\onlinecite{EBL}] imply that the vortex hopping
term is then relevant for $\Delta_v <2$.  In this regime, the vortex
hopping strength grows large when scaling to low energies, and one
expects the vortices to condense at zero temperature.  In this case it is legitimate to
expand the cosine term in Eq.~(\ref{vorthop})
and one generates a ``dual Meissner
effect'' where the gauge fields that are minimally coupled to the
vortices become massive.  In the presence of spinons this leads to a
mass term of the form, ${\cal L}_v \sim {t_v \over 2} (a_j -
\alpha_j)^2$, which confines one unit of electrical charge to each
spinon presumably driving one into a Fermi liquid phase.  If we ignore
fluctuations in the spinon density, or drop the spinons entirely
retaining a theory of Cooper pairs, the resulting phase is a charge
ordered bosonic insulator.  The nature of the charge ordering will in
this case depend sensitively on the commensurability of the Cooper
pair density ($\rho_0/2$) with the underlying square lattice.  In the
simplest commensurate case with one Cooper pair per site ($\rho_0 =
2$), a featureless Mott insulating state obtains.

\subsubsection{Electrical Resistance in the
 roton liquid}
\label{sec:electr-resist-roton}

When $\Delta_v>2$, on the other hand, vortex hopping is ''irrelevant'',
and the effects of the hopping on physical quantities can be treated
perturbatively.  Despite its irrelevance, we expect the vortex hopping
to strongly modify the Gaussian result for the conductivity by
introducing dissipation.  To understand how this occurs, it is
instructive to first consider the simple limit alluded to earlier in
which $v_0 \rightarrow \infty$.  In this limit, the longitudinal density
fluctuations described by $\tilde{a}$ at non-zero wavevector are
suppressed, and the roton mode is purely captured by the $\theta$ field.
The zero wavevector (but non-zero frequency) piece of $a$, however,
remains non-zero in this limit and can be used to calculate the
conductivity in an RPA fashion.  In particular, we take into account the
roton fluctuations and their associated dissipation induced by vortex
hopping by calculating the effective action for $\tilde{a},A$ upon
integrating out $\theta$ to second order (the lowest non-trivial
contribution) in $t_v$.  Starting then with the Lagrangian, ${\cal
  L}_{RL} + {\cal L}_A + {\cal L}_v$, expanding to second order in the
vortex hopping action $S_v$, and integrating over the $\theta$ field and
the gauge field $a({\bf k} \ne 0)$ with $v_0 \rightarrow \infty$ gives,
\begin{eqnarray}
  S_{a,A}^{\rm eff} & = & \sum_{{\bf r}={\sf r}+{\bf w}}\int_\tau
  [\frac{1}{2u_v} 
  (\partial_0\tilde{a}_j)^2 -\frac{i}{\pi}\epsilon_{ij} \tilde{a}_i({\sf r})
  \partial_0 A_j({\bf r}-{\bf\hat{x}_j})] \nonumber \\
& & + S^{(2)}_{a,A},
\end{eqnarray}
where
\begin{equation}
  S^{(2)}_{a,A} = -\frac{t_v^2}{2} \sum_{{\sf r,r'}}\int_{\tau,\tau'}
  \left\langle \cos(\partial_i \theta - \tilde{a}_i)_{{\sf r}\tau}
    \cos(\partial_j \theta-\tilde{a}_j)_{{\sf r'}\tau'} \right\rangle_\theta,
\end{equation}
where $\langle\cdot\rangle_\theta$ indicates the average with respect
to the Gaussian action for $\theta$.  From Eq.~(\ref{eq:hoppower}),
one can carry out this average to obtain,
\begin{eqnarray}
  S^{(2)}_{a,A} & \sim & -\frac{t_v^2}{4} \bigg\{
  \sum_{{\sf r},x}\int_{\tau,\tau'}
 {\cal R}(x,\tau - \tau^\prime) \nonumber \\
  & \times &  \cos[\tilde{a}_y({\sf r},\tau)-\tilde{a}_y({\sf
    r}+x{\bf\hat{x}},\tau')] + (x \leftrightarrow y)\bigg\} ,
\end{eqnarray}
with the roton propagator ${\cal R}(x,\tau)$ given as in Eq.~(\ref{rotgreen}),
except with $\Delta_v \rightarrow \sqrt{u_v/\kappa_r}/(4\pi)$ in this limit.
Following the usual RPA strategy, we expand $S^{(2)}$ to quadratic
order in $\tilde{a}$ to obtain,
\begin{equation}
\label{RPArot}
  S_{a,A}^{(2)} =   \frac{t_v^2}{4} \int_{{\bf
  k},\omega_n} \tilde{{\cal R}}(\omega_n) 
 |\tilde{a}_j(\omega_n)|^2,
\end{equation}
with the definition, $\tilde{{\cal R}}(\omega_n) = {\cal R}(0)
- {\cal R}(\omega_n)$ and,
\begin{equation}
\label{Protgreen}
{\cal R}(\omega_n) \equiv  {\cal R}(k_x=0,\omega_n) .
\end{equation}
The ${\bf k} = 0$ limit
is valid when
$v_0\rightarrow\infty$.
At low frequencies one has,
\begin{equation}
{\cal R}(\omega_n) =   - {\cal C}_\gamma  \frac{| \omega_n /v_{rot} |^{1 + \gamma} }{v_{rot} \sin \frac{\pi}{2}
(1 + \gamma) }  + ... ,
\label{rotpol}
\end{equation}
with ${\cal C}_\gamma > 0$ a dimensionless constant.  Here we have retained
explicitly the leading singular frequency dependence and
dropped analytic terms consisting of even powers of $\omega_n$.
The exponent $\gamma$ is defined as,
\begin{equation}
\gamma = 2 \Delta_v - 3 .
\end{equation}
and in the stable regime of the RL phase, $\gamma >1$.

Finally, upon
integrating out $\tilde{a}_j$, one obtains a renormalized
electromagnetic response tensor
\begin{equation}
  \Pi_{ij} =  \frac{\omega_n^2}{u_v^{-1} (\pi \omega_n)^2 +
    \frac{\pi^2 t_v^2}{2} \tilde{{\cal R}}(\omega_n) } \frac{k_i
    k_j}{k^2}.
\label{eq:piRPA}
\end{equation}

It is now straightforward to extract the conductivity by analytic
continuation,
\begin{equation}
  \sigma(\omega) = \left.\frac{\Pi_{xx}(k_x\rightarrow
      0,k_y=0,\omega_n)}{\omega_n} \right|_{i\omega_n\rightarrow
    \omega+i\delta}.
\end{equation}
One obtains an appealing Drude form:
\begin{equation}
  \label{eq:drude}
  \sigma(\omega) = \frac{1}{-i\omega ( \pi^2/u_v) +
i \frac{\pi^2 t_v^2}{2}  \tilde{{\cal R}}_{ret}(\omega) /\omega  } ,
\end{equation}
with the retarded propagator obtained from analytic continuation:
\begin{equation}
{\cal R}_{ret}(\omega)= {\cal R}(\omega_n) |_{i\omega_n\rightarrow
    \omega+i\delta}.
\end{equation}
The non-analytic frequency dependence of ${\cal R}_{ret} (\omega)$
contributes to the dissipative (real) part of the resistance
(per square), $R(\omega) \equiv {\rm Re} \sigma^{-1}(\omega)$,
which is quadratic in the vortex hopping amplitude, $t_v^2$:
\begin{equation}
\label{eq:resist}
 R(\omega) =   \frac{\pi^2 t_v^2} {2\omega} Im {\cal R}_{ret}(\omega) =
{\cal C}_\gamma \frac{ \pi^2 t_v^2}{2v_{rot}^2} \left[ \frac{|\omega|}{v_{rot}}\right]^\gamma .   \end{equation}
Note that at the point for which vortex hopping is just marginal,
$\gamma =1$, the resistance becomes linear in frequency ${\rm Re}
\sigma(\omega) \sim 1/\omega$.

We can readily extend this result to finite temperatures,
by using the finite temperature roton propagator:
\begin{equation}
{\cal R}(\tau) = c_\gamma \left[
 \frac{ \pi/v_{rot}\beta } { sin(\pi \tau/\beta) } \right]^{\gamma + 2}  .
\end{equation}
The retarded roton propagator follows upon analytic continuation,
and can be most readily extracted by using the identity:
\begin{equation}
\label{eq:identity}
Im {\cal R}_{ret}(\omega) = \sinh(\beta \omega/2) \int^\infty_{-\infty}
dt e^{i \omega t} {\cal R}(\tau \rightarrow \frac{\beta}{2} + it ) .
\end{equation}
Upon combining Eq.~(\ref{eq:resist}-\ref{eq:identity}) we thereby obtain
a general expression for the finite temperature and frequency
(dissipative) resistance in the roton liquid phase:
\begin{equation}
R(\omega,T) = c_\gamma \frac{(\pi t_v)^2}{v_{rot}^2}
\left[ \frac{\pi T}{v_{rot}}\right]^\gamma
\tilde{R}_\gamma(\omega/\pi T)   ,\label{eq:RLresist}
\end{equation}
with a universal crossover scaling function,
\begin{equation}
\tilde{R}_\gamma(X) = \frac{2^\gamma}{\Gamma(2+\gamma)} |\Gamma(1 + \frac{\gamma
+iX}{2})|^2  \frac{\sinh(\pi X/2)}{X}  ,
\end{equation}
interpolating between the d.c. resistance at finite temperature
and
the $T=0$ a.c. behavior.
Since $\tilde{R}_\gamma(X \rightarrow 0)$ is finite, the d.c. resistance varies as a power law in temperature: $R(T) \sim T^\gamma$.
At the boundary of the
RL phase with $\gamma =1$, a linear temperature dependence is predicted.
At large argument,
\begin{equation}
\tilde{R}_\gamma(X \rightarrow \infty) = \frac{\pi}{ \Gamma(2 + \gamma)}
X^\gamma   ,
\end{equation}
so that the resistance crosses over smoothly to
the zero temperature form, $R(\omega, T=0) \sim |\omega|^\gamma$.

This RPA treatment has the appeal that it produces the natural physical
result that the effect of the weak (irrelevant) vortex hopping is to
generate a small {\sl resistivity} $\sim t_v^2$.  Formally, it is
correct for $v_0=\infty$ because the RPA reproduces the exact
perturbative result for the electromagnetic response tensor to
$O(t_v^2)$ in this case.  Unfortunately, when the spatial fluctuations
of $\tilde{a}_j$ are not negligible, i.e. for $v_0 < \infty$, even the
$O(t_v^2)$ term is not obtained correctly.  More generally, the
fluctuations of $\tilde{a}_j$ and $\theta$ must be treated on the same
footing.  Therefore in the general case we instead integrate out {\sl
  both} fields and obtain more directly the correction to $\Pi_{ij}$ to
$O(t_v^2)$.  The calculations are described in
appendix~\ref{sec:appendix} .  This does not yield the appealing
``Drude'' form in Eq.~(\ref{eq:drude}) but instead the Taylor expansion
of Eq.~(\ref{eq:piRPA}) to $O(t_v^2)$,
\begin{equation}
  \label{eq:polar3}
  \Pi_{ij}^{(2)} \sim - \frac{t_v^2 u_v^2}{2v_{rot}^{2\Delta_v -1}} |\omega_n|^{2\Delta_v-4} \frac{k_i
    k_j}{k^2} ,
\end{equation}
except with the scaling dimension $\Delta_v$, given explicitly in
Eq.~(\ref{DeltavRL}), now fully renormalized by the plasmon
fluctuations.  Provided the vortex hopping is irrelevant, this is
sufficient to recover properly the low-frequency behavior of the
resistivity, Eq.~(\ref{eq:resist}) to $O(t_v^2)$.  In particular, 
formally inverting the perturbative result for $\sigma(\omega,T)$ to
$O(t_v^2)$, we infer the appropriate d.c. dissipative
resistance $R(T) \sim t_v^2 T^{\gamma}$, with $\gamma =
2\Delta_v -3$.

\subsection{``Charge Hopping''}
\label{sec:charge-hopping}

In this subsection we examine the legitimacy of the ``spin wave''
expansion employed in Sec. IIIA to obtain the Gaussian Roton
liquid Lagrangian. This is most readily achieved by passing to a
dual representation, which exchanges vortex operators for new
``charge'' operators.  This procedure is a quantum analog of the
mapping from the classical 2d $XY$-model to a sine-Gordon
representation\cite{2dXY}, the latter suited to examine
corrections to the low temperature ``spin wave'' expansion. As we
shall find, there are parameter regimes where the ``charge''
hopping terms are irrelevant and the RL phase is stable.  But
outside of these regimes, the ``charge'' quasiparticles become
mobile at low energies and condense - driving an instability into
a conventional superconducting phase. Throughout this subsection
we will drop the vortex hopping term, $H_v$, focusing on the
parameter regimes of the Roton Liquid phase where it is irrelevant
(i.e. $\Delta_v > 2$).

\subsubsection{Plaquette Duality}
\label{sec:plaquette-duality}

To this end we now employ the ``plaquette duality''
transformation, originally introduced in Ref.~[\onlinecite{EBL}]
in the context of the Exciton-Bose-Liquid phase.  Consider
the charge sector
of the theory, with Hamiltonian $H_c = H_{pl} + H_{\theta} + H_v$.
This Hamiltonian
is a function of the vortex phase field and number operators,
$\theta_{\sf r}, N_{\sf r}$ (living on the sites of the dual lattice),
as well as the (transverse) gauge field
$a^t_j$ and it's conjugate transverse ``electric field'', $e^t_j$.
The plaquette duality exchanges $\theta_{\sf r}$ and $N_{\sf r}$
for a new set of canonically conjugate fields  which live on the sites of the
original 2d square lattice.  The two new fields, denoted $\tilde{\phi}_{\bf r}$ and
$\tilde{n}_{\bf r}$, are defined
via the relations,
\begin{equation}
  \pi N_{{\bf r} + {\bf w}} \equiv \Delta_{xy} \tilde{\phi}_{\bf r}   ,
\end{equation}
\begin{equation}
  \pi \tilde{n}_{\bf r} \equiv \Delta_{xy} \theta_{{\bf r} - {\bf w}}   .
\end{equation}
Although $\tilde{\phi}_{\bf r}$ and $\tilde{n}_{\bf r}$ are conjugate fields satisfying,
\begin{equation}
\label{nphicomm}
[\tilde{n}_{\bf r},\tilde{\phi}_{{\bf r}^\prime} ]= i \delta_{{\bf r} {\bf r}^\prime}  ,
\end{equation}
they can not strictly be interpreted as phase and number operators
since the eigenvalues of $\tilde{\phi}_{\bf r} = \pi m$ for arbitrary
integer $m$, whereas $\pi \tilde{n}_{\bf r}$ is $2 \pi$ periodic.
It is important that $\tilde{\phi}_{\bf r}$ and $\tilde{n}_{\bf r}$
not be confused with the ``chargon'' phase and number operators introduced in Section II
which were denoted $\phi_{\bf r},n_{\bf r}$ -
{\it without} the tildes.  As we discuss below, $e^{i\tilde{\phi}_{\bf r}}$
does in fact create a charge-like excitation, but it is {\it not} the chargon.

Under the change of variables, $H_\theta(\theta,N) \rightarrow
H_\phi(\tilde{\phi},\tilde{n})$,
\begin{equation}
H_\phi = H_u  - \kappa_r \sum_{\sf r} \cos[\pi \tilde{n}_{{\sf r} + {\bf w}} -
{1 \over 2} (\partial_x \tilde{a}_y + \partial_y \tilde{a}_x)]   ,
\end{equation}
with the definition,
\begin{equation}
H_u = {u_v \over 2 \pi^2  }  \sum_{{\bf r} {\bf r}^\prime}
\Delta_{xy} \tilde{\phi}_{\bf r} \Delta_{xy} \tilde{\phi}_{{\bf  r}^\prime} V({\bf r} - {\bf r}^\prime) .
\end{equation}

In the Roton Liquid phase the cosine term in $H_\phi$ is expanded
to quadratic order, and $H_\phi + H_{pl} \rightarrow H_{RL}$.
To be consistent, both $\tilde{n}$ and $\tilde{\phi}$ must then be allowed to take on
any real value, and it is convenient to pass to a Lagrangian written
just in terms of $\tilde{\phi}$:
\begin{eqnarray}
  L_{RL}  & = & L_{pl} + H_u + {1 \over 2 }  \sum_{\bf r}
  { (\partial_0 \tilde{\phi}_{\bf r} )^2    \over \pi^2 \kappa_r}
\nonumber \\
 & & + \frac{i}{2\pi}
  \sum_{\sf r} \partial_0 \tilde{\phi}_{{\sf r} + {\bf w}}
  (\partial_x \tilde{a}_y +
  \partial_y \tilde{a}_x )  .
\label{LRLdual}
\end{eqnarray}
In this dual form the Roton Liquid Lagrangian depends
(quadratically) on the field $\tilde{\phi}_{\bf r}$, which
lives on the sites of the direct lattice,
and the (transverse) gauge field, $\tilde{a}_j({\sf r})$,
defined on the links of the dual lattice.

\subsubsection{Superconducting Instabilities}
\label{sec:superc-inst}

This dual formulation is ideal for studying the legitimacy
of the ``spin-wave'' expansion needed to obtain the quadratic
RL Lagrangian.  The crucial effect of the spin-wave expansion
was in softening the integer constraint on
the eigenvalues of $\tilde{\phi}/\pi$,
allowing $\tilde{\phi}$ to take on all real values in $L_{RL}$.
It is, however, possible to mimic the effects
of this constraint by adding a potential term to $L_{RL}$ of the form,
\begin{equation}
L_\lambda  = - \lambda \sum_{\bf r} \cos(2 \tilde{\phi}_{\bf r})  .
\label{cospert}
\end{equation}
When $\lambda \rightarrow \infty$ the integer constraint is enforced,
whereas the RL phase corresponds to $\lambda =0$.
Stability of the RL phase can be studied by treating $\lambda$ as a {\it small}
perturbation to the quadratic RL Lagrangian.
But as discussed in Ref.~[\onlinecite{EBL}], one should also consider other
perturbations to $L_{RL}$ which might be even more relevant.
Generally, one can add any local operator involving
$\tilde{\phi}_{\bf r}$ at a set of nearby spatial points
which is $2 \pi$ periodic in $2 \tilde{\phi}$, and satisfies all
the discrete lattice symmetries (i.e.. translations, rotations
and parity).  For example, terms of the form $\cos(2 \ell \tilde{\phi})$
for arbitrary integer $\ell$ are allowed, although these
will generically become less relevant with increasing $\ell$.
as we shall see,
for our ``minimal'' model of the RL phase, the most relevant
perturbation is of the form,
\begin{equation}
L_t = - t_c \sum_{\bf r} \sum_{j=1,2} \cos(2\partial_j \tilde{\phi}_{\bf r} )  .
\label{chhoppert}
\end{equation}

Before studying the perturbative stability to such operators, we
try to get some physical intuition for the meaning of the operator
$e^{i\tilde{\phi}}$.  From the commutation relations in
Eq.~(\ref{nphicomm}), the operator $e^{i\tilde{\phi}_{\bf r}}$ increase
$\tilde{n}_{\bf r}$ by one, and creates some sort of quasiparticle
excitation on the spatial site ${\bf r}$. Since the perturbation
in Eq.~(\ref{cospert}) changes the number $\tilde{n}_{\bf r}$ by $\pm 2$,
the total number of these quasiparticles, $\tilde{n}_{tot} = \sum_{\bf r}
\tilde{n}_{\bf r}$ is {\it not} conserved, but the ``complex charge'', $Q_c
= e^{i\pi \tilde{n}_{tot}}$ is conserved.   The perturbation in
Eq.~(\ref{chhoppert}) can then be interpreted as a ``charge
hopping'' process. To get some feel for the nature of the
quasiparticle, it is instructive to introduce an external magnetic
field, $B = \epsilon_{ij}
\partial_i A_j$, which enters into $H_u$ above via the
substitution, $\Delta_{xy} \phi_{\bf r} \rightarrow \Delta_{xy}
\phi_{\bf r} - B_{{\bf r} + {\bf w}}$. A spatially
uniform field, $B$, can readily be removed from the Gaussian RL
Lagrangian by letting
\begin{equation}
\tilde{\phi}_{\bf r} \rightarrow \tilde{\phi}_{\bf r} + B x y .
\end{equation}
But then $B$ appears in the cosine perturbations, in ``almost'' a
minimal coupling form.  For example, one has the combination
$\partial_x \tilde{\phi} - 2A_x$ where we have chosen the gauge $A_x = -B
y/2$ and $A_y =Bx/2$, suggesting that $e^{i\tilde{\phi}}$ carries the
Cooper pair charge. But the $y-$derivative enters as $\partial_y
\tilde{\phi} + 2A_y$ - with the {\it wrong} sign.  Thus, this quasiparticle
is {\it not} a conventional electrically charged particle.
Nevertheless, as we show below, condensation of the quasiparticle
with $\langle e^{i\tilde{\phi}} \rangle \ne 0$ does drive the RL phase
into a superconducting state.

To evaluate the relevance of the various perturbations
in the RL phase, requires diagonalizing the associated action, $S_{RL}$.
In momentum space one has,
\begin{equation}
  S_{RL} =  \int_{k_\mu} [ \frac{D_{\phi \phi}}{2} |\tilde{\phi}|^2
  + \frac{D_{aa}}{2} |a|^2 + D_{a \phi} a(k_\mu) \tilde{\phi}(-k_\mu)  ],
\end{equation}
\begin{equation}
 D_{\phi \phi} = {1 \over \pi^2 \kappa_r } [ \omega_n^2  + { u_v \kappa_r
|{\cal K}_x {\cal K}_y |^2  \over {\cal K}^2  }  ] ,
\end{equation}
\begin{equation}
 D_{aa} = {1 \over u_v } [ \omega_n^2  + \tilde{v}_0^2 {\cal K}^2 ]  ,
\end{equation}
\begin{equation}
 D_{a \phi} = { \omega_n ({\cal K}_x^2 - {\cal K}_y^2) e^{i{\bf k} \cdot {\bf w}}  \over 2  \pi {\cal K} }  ,
\end{equation}
with $k_\mu = ({\bf k}, \omega_n)$.
Evaluation of the two-point function of $e^{2i\tilde{\phi}}$ then gives,
\begin{equation}
\langle e^{i 2\tilde{\phi}_{\bf r}(\tau)} e^{-i2\tilde{\phi}_{\bf 0}(0)} \rangle_{RL}
\sim  {  \delta_{{\bf r},{\bf 0}}   |\tau|^{-2 \Delta_c } } ,
\label{eq:e2phi}
\end{equation}
with scaling dimension,
\begin{equation}
  \Delta_c = 2 \pi \sqrt{ \kappa_r \over u_v } { \tilde{v}_0 \over v_+ } .
\label{Deltac}
\end{equation}
We note in passing that this power-law form of the ``charge''
operator ($e^{2i\tilde{\phi}}$) two-point function in Eq.~(\ref{eq:e2phi})
differs from $\exp(-\ln^2 \tau)$ behavior of the corresponding
object in the Exciton Bose liquid of Ref.~\onlinecite{EBL}, due to
the long-range logarithmic interactions between vortices.
Stability of the RL phase requires $\Delta_c > 1$. Evaluation of
the two-point function for the ``charge'' hopping operator,
$e^{2i\partial_y \tilde{\phi}}$, gives,
\begin{equation}
\langle e^{i 2 \partial_y \tilde{\phi}_{\bf r}(\tau)} e^{-i2
\partial_y \tilde{\phi}_{\bf 0}(0)} \rangle_{RL}
\sim  {  \delta_{y,0} \over  (x^2 + v_{rot}^2 \tau^2 )^{\Delta_c} }  ,
\label{eq:chargehopping}
\end{equation}
with ${\bf r} = x{\bf\hat{x}} + y {\bf\hat{y}}$, and with
the {\it same} scaling dimension $\Delta_c$ as above.
However, since this two-point function decays algebraically in
two (rather than one) space-time dimensions, the perturbation
$t_c$ is relevant for $\Delta_c < 2$.

When $\Delta_c <2$, the ``charge'' hopping process will grow at low
energies, and will destabilize the roton liquid phase.  Not
surprisingly, the resulting quantum ground state is
superconducting. Indeed, the exponent $\Delta_c$ in Eq.~(\ref{Deltac}) above
is in fact {\it identical} to the exponent characterizing the
power law decay of the Cooper pair propagator in Eq.~(\ref{Cooperexp}),
so that for small $\Delta_c$ the Roton liquid phase is already ``almost'' superconducting.
Moreover, when the vortex core
energy greatly exceeds the roton hopping strength, $u_v \gg
\kappa_r$, the Hamiltonian $H_\theta$ in Eq.~(\ref{Htheta}) is
deep within it's superconducting phase.  This limit precisely
corresponds to $\Delta_c \ll 1$, the limit where the ``charge''
hopping is strongly relevant.  More directly, when the ``charge''
hopping strength $t_c$ grows large, the field $\tilde{\phi}$ gets trapped
at the minimum of the cosine potentials in
Eqs.~(\ref{cospert},\ref{chhoppert}), and it is legitimate to
expand the cosine potentials to quadratic order.  Once massive,
the expectation value $\langle e^{i\tilde{\phi}} \rangle \ne 0$ - and the
charge quasiparticle has condensed.  Moreover, setting $\tilde{\phi}=0$ in
Eq.~(\ref{chhoppert}) in the presence of an applied magnetic field
will generate a term of the form,
\begin{equation}
  {\cal L}_t \sim {t_c  \over 2} \vec{A}^2   ,
\end{equation}
indicative of a Meissner response.

In the resulting superconducting phase, the rotons  - gapless in the
RL phase - become gapped.
This follows upon expanding the cosine term in Eq.~(\ref{cospert}),
${\cal L}_\lambda = 2 \lambda \tilde{\phi}^2$,
which leaves the roton liquid Lagrangian quadratic, and allow one to readily extract the
modified roton dispersion.  For $k_x \rightarrow 0$ at fixed $k_y$, this gives:
\begin{equation}
\omega_{rot}({\bf k}) = \sqrt { v_{rot}^2 k_x^2 + m_{rot}^2   }  ,
\end{equation}
with the ``roton mass gap'' given by, $m_{rot} = 2\pi \sqrt{\kappa_r \lambda}$.

Since the product $\Delta_c \Delta_v = {1 \over 2}$, it is not
possible to have both the vortex hopping and the ``charge'' hopping
terms simultaneously irrelevant.

\section{The Roton Fermi Liquid Phase}
\label{sec:roton-spinon-liquid}

We now put the fermions back into the description of the Roton liquid.
We first consider setting the explicit pairing term in the fermion
Hamiltonian ${\cal H}_f$ in Eq.~(\ref{eq:Hf}) to zero: $\Delta = 0$.  As
in Sec.~\ref{sec:roton-liquid}, we will assume for the most part (with
the exception of Sec.~\ref{sec:phase-diagram} in the discussion) that
the equilibrium fermion density is equal to the charge density $\rho_0$.
This choice naturally minimizes Coulomb energy and vortex kinetic
energy, as discussed therein.  As we shall see, in this way we will
arrive at a description of a novel non-Fermi liquid phase - the Roton
Fermi liquid - which supports a gapless Fermi surface of quasiparticles
coexisting with a gapless set of roton modes. We then reintroduce a
non-zero pairing term, and study the perturbative effects of $\Delta$.
We argue that, when the scaling exponent that describes the decay of the
off-diagonal order in the Roton liquid is large enough, $\Delta_c>
\Delta_c^* > 3/2$, the explicit pairing term is perturbatively
irrelevant, and the RFL phase with a full gapless Fermi surface is
stable.

Even in the absence of the explicit pairing term -- which couples
the fermionic and vortex degrees of freedom in a highly non-linear
manner -- the rotons and quasiparticles interact through
(three-)current-current interactions ``mediated'' by the
$\beta_\mu$ and $e_i$ fields.  Although these interactions are
long-ranged for individual vortices, they are not for rotons,
which carry no net vorticity. Moreover, due to phase space
restrictions we find that the residual short-ranged interactions
asymptotically decouple at low energies.  The resulting RFL phase
supports both gapless charge and spin excitations with no broken
spatial or internal symmetries, just as in a conventional Fermi
liquid.  But, due to the vortex sector of the theory, the RFL
phase is demonstrably a non-Fermi liquid, with a gapless ``Bose
surface'' of Rotons and with ODQLRO in the Cooper pair field but
no Meissner effect (see below).  Moreover, the quasiparticles at
the RFL Fermi surface are sharp (in the sense of the electron
spectral function), but electrical currents are carried by the
(quasi-)condensate.  Even with impurities present the resistivity
vanishes as a power law of temperature in the RFL.  The power law
exponent, $\gamma$, varies continuously, but is greater than or
equal to one.  For $\gamma <1$, the zero temperature RFL phase is
unstable to a quantum confinement transition, which presumably
drives the system into a conventional Fermi liquid phase.

To describe the RFL, we start with the general Lagrangian,
Eqs.~(\ref{Sfullb}-\ref{eq:Lf}), and first make the same approximation
as in the RL of expanding the roton hopping term to quadratic order.
That is, to leading order, ${\cal L}_{kin} \approx {\cal L}_r^0$, with
\begin{eqnarray}
  {\cal L}_r^0   =  {\kappa_r \over 2} [ \partial_{xy} \theta & -
  & {1 \over 2} \lbrace \partial_x (a_y - \alpha_y) + \partial_y(a_x - \alpha_x)
  \rbrace ]^2 \nonumber \\
  & & + {\kappa_r \over 8} [\epsilon_{ij} \partial_i (a_j - \alpha_j)]^2.
\end{eqnarray}
As before for the RL, this approximation will be corrected
perturbatively by ``charge'' and vortex hopping terms, which will {\sl
  not} be expanded.

Turning to the fermionic sector, we assume for the moment that the
power-law decay (ODQLRO) of the Cooper pair field is sufficiently
rapid that the pair field term $\Delta_j$ can be neglected.  We
argue later this is correct for $\Delta_c> \Delta_c^* > 3/2$.
This gives a non-anomalous fermionic Lagrange density $L_f$, which
we further presume is well-described by Fermi liquid theory, again
checking the correctness of this assumption perturbatively in the
couplings to $\beta_\mu$ and $e_i$.   Hence we replace ${\cal L}_f
\approx {\cal L}_f^0 + {\cal
  L}_f^1$, with (working for
simplicity at zero temperature)
\begin{eqnarray}
  \label{eq:Lf0}
  {\cal L}_f^0 & = & f^\dagger_{{\bf r}\sigma}(\partial_0 - \mu)
  f^{\vphantom\dagger}_{{\bf r}\sigma} -t \sum_j f^\dagger_{{\bf r} +
    {\bf \hat{x}}_j \sigma} f^{\vphantom\dagger}_{{\bf r} \sigma},  \\
  \label{eq:Lf1}
  {\cal L}_f^1 & = & -i\beta_0({\bf r}) f_{{\bf r}\sigma}^\dagger
  f^{\vphantom\dagger}_{{\bf r}\sigma} \\ &&  - i\sum_j
  [t\beta_j({\bf r}) +
  t_e (\pi\overline{e}_j({\bf
    r})+A_j({\bf r}))] f^\dagger_{{\bf r} +
    {\bf \hat{x}}_j \sigma} f^{\vphantom\dagger}_{{\bf r} \sigma}
  \nonumber \\
  & & \hspace{-0.4in} + \sum_j [\frac{t_s}{2} \beta_j({\bf r})^2 +
  \frac{t_e}{2}(\beta_j({\bf r})+ \pi\overline{e}_j({\bf
    r})+A_j({\bf r}))^2]
  f^\dagger_{{\bf r} +
    {\bf \hat{x}}_j \sigma} f^{\vphantom\dagger}_{{\bf r} \sigma} . \nonumber
\end{eqnarray}
Note that, to leading order, the fermionic dispersion is
controlled by the sum of the two hopping amplitudes, $t=t_s+t_e$.
Here we have included explicitly the physical external vector
potential, $A_j({\bf r})$, in the electron hopping term.  Some
care needs to be taken when treating $A_j({\bf r})$.  The above
form is correct provided $A_j({\bf r})$ is also coupled into $e_j$
in the quadratic Hamiltonian in the ``canonical'' fashion $\pi
\overline{e}_j \rightarrow \pi \overline{e}_j + A_j$, in the
roton/plasmon portion of the Hamiltonian. It is {\sl not} correct
if this canonical coupling is removed by shifting
$\overline{e}_j$, which is the procedure needed to generate the
$A_\mu J_\mu$ coupling in Eq.~(\ref{LA}).  If the latter form of
the Lagrangian is used, the vector potential should be removed
from the electron hopping term.  Either choice is correct if used
consistently.

The full Lagrangian that we then use to access the RFL phase is
given by,
\begin{equation}
{\cal L}_{RFL} = {\cal L}_a + {\cal L}_0 + {\cal L}_r^0 + {\cal
L}_{cs} + {\cal L}_f^0 + {\cal L}_f^1  , \label{LRFL}
\end{equation}
with the definition,
\begin{equation}
{\cal L}_0 = {1 \over 2 u_0} (\partial_0 \theta - a_0 +
\alpha_0)^2 .
\end{equation}
The interaction terms between the Fermions and the fields
$\beta_j$ and $e_j$ in ${\cal L}_f^1$ will be treated in the
random phase approximation. Doing so, one arrives at a tractable,
if horrendously algebraically complicated Lagrangian describing
the RFL, which is quadratic in the fields
$\theta,a_\mu,e_j,\alpha_\mu$ and $\beta_\mu$. This makes
calculations of nearly any physical quantity possible in the RFL.

Before turning to these properties, we verify (in the remainder of this
section) the above claims that the coupling of fermions and vortices
does not destabilize the RFL -- i.e. it neither modifies the form of the
low energy roton excitations at the ``Bose surface'' nor the fermionic
quasiparticles at the Fermi surface.

\subsection{Quasiparticle Scattering by Rotons}
\label{sec:selectr-scatt-rotons}

In this subsection, we show that the coupling of the electronic
quasiparticles to the vortices does not destroy the Fermi surface.
To do so, we will integrate out the vortex degrees of freedom to
arrive at effective interactions amongst the quasiparticles.  This
procedure is somewhat gauge dependent.  To provide a useful
framework for the calculation of the electron spectral function in
the following section, we will choose the gauge $\vec{\nabla}\cdot
\vec{\beta} = \pi \epsilon_{ij}\partial_i e_j$, in which the
$f,f^\dagger$ operators create fermionic quasiparticles with
non-vanishing overlap with the bare electrons, without the need
for any additional string operators. This is essentially
equivalent to working in the electron formulation of Appendix
\ref{ap:elecform}.

Since this gauge choice explicitly involves
$\epsilon_{ij}\partial_i e_j$, we must employ a path integral
representation in which the transverse component of the electric
field, $e^t_j$, has not been integrated out. It is further
convenient to fix the two remaining gauge choices according to
$\vec{\nabla}\cdot\vec{a}=\alpha_0=0$, and to integrate out the
field $a_0$ in the $u_0 \rightarrow 0$ limit. The full RFL
Lagrangian density (before imposing the constraint $\beta^l_j =
\pi e^t_j$) then takes the form,
\begin{equation}
  \label{eq:laget0}
  {\cal L}_{RFL} = {\cal L}_{vort} + {\cal L}^0_f(\beta_\mu) + {\cal
  L}_f^1(\beta_\mu)  ,
\end{equation}
with
\begin{eqnarray}
  \label{eq:laget}
  {\cal L}_{vort} & = & \frac{u_v}{2} (e^t_j)^2 + i e^t_j \partial_0 a^t_j +
  \frac{v_0^2}{2u_v} (\epsilon_{ij}\partial_i a^t_j - \pi
  \overline{\rho})^2 \\
  & & + \frac{i}{\pi}\epsilon_{ij} \alpha_i (\partial_j \beta_0 -
  \partial_0 \beta_j) + {\cal L}_\theta(\alpha_j-a^t_j) ,
  \nonumber
\end{eqnarray}
where
\begin{eqnarray}
  \label{eq:Lthetaet}
  {\cal L}_\theta(\alpha_j) & = &   \frac{1}{2u_v} (\partial_0 \partial_i
  \theta)^2 + \frac{\kappa_r}{8}(\epsilon_{ij}\partial_i \alpha_j)^2
  \nonumber \\
  & + & \frac{\kappa_r}{2}[\Delta_{xy}\theta +
  \frac{1}{2}(\partial_x\alpha_y+\partial_y\alpha_x)]^2 .
\end{eqnarray}

To assess the perturbative effects of the vortices upon the electronic
quasiparticles, we wish to integrate out the vortices perturbatively
in the coupling of the gauge field $\beta_\mu$ to the fermions.  For
this we require the correlation functions of the $\beta_\mu$ fields
(neglecting the couplings inside ${\cal L}_f$, and in particular to
lowest order just $\langle \beta_\mu \beta_\nu \rangle$).  To obtain
the latter, we add a source term to the Lagrangian,
\begin{equation}
  \label{eq:betasource}
  {\cal L}_{vort} \rightarrow {\cal L}_{vort} +i( \lambda_0 i\beta_0 +
  \vec{\lambda}\cdot\vec{\beta}).
\end{equation}
Here we have included an extra factor of $i$ with $\beta_0$ to
compensate for the factor of $i$ present in the coupling of
$\beta_0$ to the fermion density.  Upon fully integrating out the
$e^t, \theta,a^t,\beta_\mu,\alpha_j$ fields, the coefficient of
$\lambda_\mu \lambda_\nu$ in the effective action will give (half)
the desired correlator.  We perform the integration in two stages.
First, imposing the constraint $\beta^l = \pi e^t$, we eliminate
$e^t$, shift $\alpha_j \rightarrow \alpha_j - (\partial_j \theta -
a^t_j)$, and integrate out the $\theta$, $a^t_j$, and $\beta_\mu$
fields.  One obtains ${\cal L}_{vort} \rightarrow \tilde{\cal
  L}_{vort}(\alpha_j,\lambda_\mu)$, with
\begin{eqnarray}
  \label{eq:lvorttilde}
  \tilde{\cal L}_{vort} & = & \frac{v_0^2}{2u_v}
  (\epsilon_{ij}\partial_i \alpha_j +i\pi \tilde{\lambda}_0)^2 +
  \frac{\kappa_r}{4} [(\partial_x \alpha_y)^2 + (\partial_y
  \alpha_x)^2] \nonumber \\
  & & + \frac{1}{2u_v} (\partial_0 \alpha_i - \pi \epsilon_{ij}
  \lambda_j)^2 -t_v \cos(\alpha_j ),
\end{eqnarray}
with $\tilde\lambda_0 = \lambda_0 -i\overline{\rho}$.

In Eq.~(\ref{eq:lvorttilde}) we have added back in the vortex
hopping term, ${\cal L}_v  = -t_v \cos(\alpha_j)$, neglected in
the RFL Lagrangian.  In the RFL phase, the vortex hopping is
irrelevant, and scales to zero at low energies.  If we put
$t_v=0$, the remaining integrations over $\alpha_j$ are Gaussian
and can be readily performed. We will return to the effects of
non-zero vortex hopping upon the fermions in
Sec.~\ref{subsec:resist}. The final effective action then takes
the form,
\begin{equation}
  \label{eq:Svorteff}
  S_{eff}(\lambda_\mu) = - \frac{1}{2}\int_{{\bf k},\omega_n}\!
  U^{(0)}_{\mu\nu}({\bf k},\omega_n) \lambda_\mu({\bf k},\omega_n)
  \lambda_\nu(-{\bf k},-\omega_n).
\end{equation}
Here
\begin{equation}
  \label{eq:Umunu}
  U^{(0)}_{\mu\nu}({\bf k},\omega_n) = \frac{\pi^2}{u_v(\omega_n^2 +
    \omega_{pl}^2)(\omega_n^2+\omega_{rot}^2)} u_{\mu\nu}({\bf k},\omega_n)
\end{equation}
specifies the $\beta_\mu$ propagator: $\langle
\beta_0\beta_0\rangle_0=U^{(0)}_{00}$ , $\langle
\beta_i\beta_j\rangle_0=-U^{(0)}_{ij}$,$\langle
i\beta_0\beta_j\rangle_0=-U^{(0)}_{0j}$, where the superscript
zero reminds us that this is the result to zeroth order in
$t_v=0$, and we have the definitions,
\begin{eqnarray}
  \label{eq:lambdamunu}
  u_{00} & = & v_0^2[\omega_n^4 + \frac{v_1^2 {\cal K}^2}{2} \omega_n^2 +
  \frac{v_1^4}{4} |{\cal K}_x {\cal K}_y|^2], \\
  u_{xx} & = & -v_1^2[\tilde{v}_0^2 |{\cal K}_x {\cal K}_y|^2 + v_+^2
  |{\cal K}_x|^2 \omega_n^2], \\
  u_{yy} & = & -v_1^2[\tilde{v}_0^2 |{\cal K}_x {\cal K}_y|^2 + v_+^2
  |{\cal K}_y|^2 \omega_n^2], \\
  u_{xy} & = & -v_0^2 {\cal K}_x {\cal K}_y \omega_n^2, \\
  u_{0j} & = & -v_0^2 {\cal K}_x i\omega_n [2\omega_n^2 + v_1^2 ({\cal
    K}^2 - {\cal K}_j^2)].\label{eq:lambda0j}
\end{eqnarray}

In the d.c. limit, these interactions simplify considerably, and one
obtains the simple results
\begin{eqnarray}
  \label{eq:Udc}
  U^{(0)}_{00}({\bf k},\omega_n=0) & = & \frac{\pi^2 \kappa_r}{4}
  \left(\frac{v_0}{\tilde{v}_0}\right)^2, \\
  U^{(0)}_{ij}({\bf k},\omega_n=0) & = & -\frac{\pi^2}{u_v} \delta_{ij},
\end{eqnarray}
and $U^{(0)}_{0i}({\bf k},\omega_n=0) = 0$.

Since $i\beta_0$ and $\beta$ couple to the fermion density and
current respectively, ${\cal S}_{eff}$ mediates an effective
frequency and wavevector dependent interaction between fermions.
Since, in the d.c. limit, $-U^{(0)}_{00}({\bf k},\omega_n=0)$ and
$U^{(0)}_{ii}({\bf k},\omega_n=0)$ are finite and wavevector
independent, they describe local (on-site) repulsive quasiparticle
density-density and attractive current-current interactions,
respectively.

Generally, a repulsive density-density interaction between fermions
will lead to Fermi liquid corrections in the quasiparticle dynamics
and thermodynamics, but will not destroy the Fermi surface.  On the
other hand, the current-current interaction for near-neighbor
quasiparticle hopping can be rewritten in terms of antiferromagnetic,
near-neighbor repulsive, and pairing interactions:
\begin{equation}
  H_J = J \sum_{{\bf r}, {\bf r}^\prime} \left[{\bf S}_{\bf r} \cdot
  {\bf S}_{{\bf r}^\prime} + \frac{1}{4} n^f_{\bf r} n^f_{\bf r'}
  + 2\overline{\Delta}^f_{\bf r} \Delta^f_{\bf r'} \right] ,
\end{equation}
up to a shift of the fermion chemical potential, with ${\bf S} \equiv
f^\dagger \frac{\vec\sigma}{2} f^{\vphantom\dagger}$, (with $\vec\sigma$
the vector of Pauli matrices), $n^f = f^\dagger f^{\vphantom\dagger}$,
$\Delta^f = f_\uparrow f_\downarrow$ and $\overline{\Delta}^f =
f^\dagger_\downarrow f^\dagger_\uparrow$.  Here, $J \sim t_s^2 /u_v$ is
inversely proportional to the vortex core energy. One expects that the
antiferromagnetic interaction can lead to an a Cooper instability in the
$d$-wave (or extended $s-$wave) channel.  The repulsive pairing
interaction interaction dis-favors an $s$-wave Cooper instability.
Hence it seems possible that for small $u_v$, for which this $J$ is
large, a spontaneous quasiparticle pairing instability may occur, most
probably of $d$-wave symmetry.  Another possibility, which appears very
natural for {\sl extremely} small doping $x\rightarrow 0^+$ (and $u_v
\rightarrow 0^+$), is that the antiferromagnetic interaction drives an
antiferromagnetic spin-density-wave instability.  We will discuss both
possibilities in the discussion section.

\subsection{Roton Scattering by Quasiparticles}
\label{sec:roton-scatt-selectr}

Having established that the Fermi surface remains intact (apart
from a possible BCS-type pairing instability) in the presence of
gapless rotons, we need to see how the fermions feed back and
effect the roton modes.  To this end, we will integrate out all
fields except $\theta$, treating the fermions within the random
phase approximation (RPA), to study the effects upon the roton
dispersion.  The RPA is complicated by the two distinct
amplitudes, $t_s$ and $t_e$, describing spinon and electron
hopping processes, the latter coupling to the electric field $e_j$
as well as the $\beta_j$ gauge field.  We begin with the RFL
Lagrangian in Eq.~(\ref{LRFL}).  It is convenient to first shift
$\alpha_\mu \rightarrow \alpha_\mu + a_\mu$ then integrate out
$a_0$ and take $u_0 \rightarrow 0$, which constrains
$\vec\nabla\times\vec\beta = \pi\vec\nabla\cdot \vec{e}$ and
$\alpha_0 = -\partial_0 \theta$. Choosing in addition
$\vec\nabla\cdot\vec\beta =- \pi \vec\nabla\times\vec{e}$, we
essentially return to the electron formulation, with $\beta_i=-\pi
\overline{e}_i$.  This choice is convenient, since the $\beta_i$
and $\pi\overline{e}_i$ fields present in the electron hopping
term precisely cancel, and only the $-\pi \overline{e}_i$ appears
``like a gauge field'' in the spinon hopping term.  The fermionic
quasiparticles can then be integrated out in the random phase
approximation (RPA).  One finds $S_f \rightarrow S_{RPA}$, with
\begin{eqnarray}
  S_{RPA}(\beta_\mu) & = & {1 \over 2} \int_{{\bf k},\omega_n}\!\!\!
  [ \Pi_{00} |\beta_0({\bf k},\omega_n)|^2 + \frac{\pi^2
    t_s^2}{t^2}\Pi_{11} |e_l({\bf k},\omega_n)|^2 \nonumber \\
  & & + \pi^2
  t_s(1\!-\!\frac{t_s}{t}\!) T_{nn} |e({\bf k},\omega_n)|^2 ] ,\label{eq:sRPA}
\end{eqnarray}
where $T_{nn} = \langle f^\dagger_{{\bf r}\sigma}
f^{\vphantom\dagger}_{{\bf r+\hat{x}}\sigma}\rangle$ is the nearest
neighbor kinetic energy, and $t=t_s+t_e$.  Here $\Pi_{00}$ and
$\Pi_{11}$ are respectively the density-density and current-current
response kernels (polarization bubbles + diamagnetic contribution in the
case of $\Pi_{11}$) that would obtain for the Fermi sea were a single
gauge field minimally coupled to the fermions.  These depend upon the band
structure.  We will not require particular expressions for these
quantities, but will use the fact that both $\Pi_{00}({\bf k},\omega_n)$
and $\Pi_{11}({\bf k},\omega_n)$ are finite and generally non-vanishing
at {\sl fixed wavevector} $|{\bf k}|>0$ and $\omega_n=0$.  Since we will
focus upon the low energy rotons, which occur near the principle axes in
the Brillouin zone, we have dropped terms in Eq.~(\ref{eq:sRPA}) that
are small for $\omega_n \sim k_x \ll k_y$ and $\omega_n \sim k_y \ll
k_x$ (e.g.  due to the parity symmetry, $x \rightarrow -x$, of the 2d
square lattice, $\Pi_{01}(k_x=0,k_y,\omega_n)$ vanishes).  We also note
that Eq.~(\ref{eq:sRPA}) does not have the usual RPA form even in this
limit, due to the non-minimal coupling form of the fermion Lagrangian
(minimal coupling is restored only for $t_s\rightarrow t$).

It may be helpful to keep in mind the forms for a circular Fermi
surface (e.g. valid at small electron densities), where, at small
wavevectors and frequencies, one has
\begin{equation}
  \Pi_{00}({\bf k}, \omega_n) \sim m^* ,
\end{equation}
where $m^* \sim t^{-1}$ is the effective mass and
\begin{equation}
  \Pi_{11} ({\bf k}, \omega_n) \sim {1 \over m^*} [ {\bf k}^2 +
  {|\omega|  \over  v_F k  } ] ,
\end{equation}
which is valid for $|\omega | < v_F k $.  We stress, however, that our
results do not depend upon these particular forms.

Since $S_{RPA}$ is quadratic, one can perform the remaining
integrals  straightforwardly.  We choose
$\vec{\nabla}\cdot\vec{a}=\vec{\nabla}\cdot\vec{\alpha}=0$, and
integrate out $a$, $\beta_0$ and $e$, to obtain
\begin{eqnarray}
  \label{eq:atlag}
  S & = &  \int_{{\bf k},\omega_n}\!\!\! \big\{
  \frac{\omega_n^2}{2\tilde{u}_v} \alpha^2 + \frac{\kappa_r}{8}\frac{4v_0^2 +
    \tilde{v}_1^2}{\tilde{v}_1^2}\alpha^2 + \frac{\omega_n^2 {\cal
      K}^2}{\overline{u}_v} \theta^2 \nonumber \\
  & &
  + \frac{\kappa_r}{2}|{\cal K}_x{\cal K}_y \theta + \frac{({\cal
    K}_x^2 - {\cal K}_y^2)}{2{\cal K}} \alpha|^2 \big\},
\end{eqnarray}
where $\tilde{u}_v = u_v + \pi^2 t_s (1-t_s/t) T_{nn}$,
$\overline{u}_v = \tilde{u}_v + \pi^2 t_s^2/t^2 \Pi_{11}$, and
$\tilde{v}_1^2 = v_1^2 + \kappa_r \pi^2 \Pi_0 v_0^2$.  Without loss of
generality, we focus upon the branch of rotons with $\omega_n \sim k_x
\ll k_y \sim O(1)$, for which the first term in Eq.~(\ref{eq:atlag})
is negligible, and the remaining $\alpha$ integral can be carried out
to obtain finally the effective action
\begin{equation}
  S_{rot} = \frac{1}{2}\int_{{\bf k},\omega_n}^\prime  \frac{\tilde{\cal
    K}_y^2}{\overline{u}_v} \left[ \omega_n^2 + v_{rot}^2(k_y)
  k_x^2\right] |\theta|^2, \label{eq:Srotsec5}
\end{equation}
with
\begin{equation}
  v_{rot}^2(k_y) = v_1^2 \frac{\overline{u}_v}{u_v}
{  v_0^2 + {1 \over 4} \tilde{v}_1^2    \over   v_0^2 + {1 \over 2}
  \tilde{v}_1^2  }  .\label{eq:vrotfull}
\end{equation}

We have thereby shown that even a gapless Fermi sea  {\it
  does not} lead to a qualitative change in the gapless Bose-surface
of rotons.  Due to the $k_y$-dependence of $\Pi_{00}$ and $\Pi_{11}$
(implicit in $\tilde{v}_1$ and $\overline{u}_v$), the roton velocity
is now seen to depend upon $k_y$.  Additional ``direct''
quasiparticle-quasiparticle interactions (not mediated by the rotons)
would in any case similarly renormalize $v_{rot}$.  But the location
of the Bose surface and the qualitative low energy dispersion of the
roton modes are seen to be unaffected by the fermions.  Together with
the earlier demonstration of the stability of the Fermi
surface, this result establishes the RFL phase as a stable 2d
non-Fermi liquid with {\it both} gapless charge and spin excitations.

\section{Instabilities of the Roton Fermi Liquid}
\label{sec:inst-roton-spin}

As for the RL, the RFL has potential instabilities to superconducting
and Fermi Liquid states, driven by effects/terms neglected in the
previous subsection.  We address each in turn now.

\subsection{Landau Fermi Liquid Instability}
\label{sec:fermi-liqu-inst}

The arguments of Sec.~\ref{sec:vortex-hopping} for determining the
relevance or irrelevance of the neglected vortex hopping term
within the simpler RL continue to hold for the full RFL, provided
the proper renormalized roton liquid parameters are employed.  In
particular, the vortex hopping term continues to be described by a
$1+1$-dimensional scaling dimension $\Delta_v$, which is, however,
renormalized by the statistical interactions with the fermions, to
wit:
\begin{equation}
  \label{eq:deltavrenorm}
  \Delta_v = \frac{1}{4\pi} \int_{-\pi}^{\pi} \! \frac{dk_y}{2\pi}\,
  \frac{\overline{u}_v(k_y)}{v_{rot}(k_y)},
\end{equation}
with $v_{rot}(k_y)$ given from Eq.~(\ref{eq:vrotfull}).  The same
arguments thus continue to apply, and the vortex hopping term is
irrelevant for $\Delta_v>2$.  For $\Delta_v<2$, we expect an
instability to a state with proliferated vortices.  The dual
Meissner effect for this vortex condensate confines particles with
non-zero gauge charge, as discussed in
Sec.~\ref{sec:vortex-hopping}.  Coming from the RFL, the natural
expectation is then that the system becomes a Fermi liquid.  This
hypothesis is fleshed out in Appendix \ref{sec:fermiliquid}.

\subsection{Superconducting Instabilities}
\label{sec:superc-inst-1}

As for the RL, the RFL can also be unstable to a superconducting
state.  However, the inclusion of the fermionic quasiparticles opens new
routes to superconductivity from the RFL.  We explore each of
these in turn.

\subsubsection{``Charge'' hopping}
\label{sec:charge-hopping-1}

As for the vortex hopping considered above, the arguments for the
relevance of the ``charge'' hopping in Sec.~\ref{sec:superc-inst} for
the RL continue to apply with only a renormalization of the roton
liquid parameters.  In particular, the scaling dimension defined from
the charge hopping, Eq.~(\ref{eq:chargehopping}), is modified to
\begin{equation}
  \label{eq:deltacrenorm}
  \Delta_c = \pi \int_{-\pi}^\pi \! \frac{dk_y}{2\pi} |{\cal K}_y(k_y)|^2
  \frac{v_{rot}(k_y)}{\overline{u}_v(k_y)}.
\end{equation}
With this modification, the charge hopping processes become
relevant for $\Delta_c<2$, as before.  It should be noted that,
including the renormalizations due to scattering by fermions,
$\Delta_c \Delta_s \neq 1/2$, owing to the $k_y$ dependence of the
$v_{rot}$ and $\overline{u}_v$.

As we established in Sec.~\ref{sec:superc-inst} by employing the
plaquette duality transformation, when the charge hopping
processes are relevant the rotons become gapped out and the system
exhibits a Meissner effect.  Here, we briefly comment on the
corresponding fate of the fermionic quasiparticles, which are
gapless across the Fermi surface in the RFL phase. The relevance
of the charge hopping in the plaquette dualized representation,
indicates that it is not legitimate to expand the cosine term that
enters into the roton hopping Lagrangian, ${\cal L}_r$.  Rather,
in the original vortex representation, the state corresponds to a
``vortex vacuum", and the properties of the state can be accessed
by taking all of the vortex hopping processes small,
$t_v,t_{2v},\kappa_r \ll u_v$. At zeroth order in the vortex
kinetic terms, the full Hamiltonian of the $U(1)$ vortex-spinon
field theory in Eq.~(\ref{HamU1vs}) is independent of $\alpha_j$,
so that $\beta_j$ commutes with $\hat{H}$ and naively can be taken
as c-numbers. A simple choice consistent with the condition
$N_{\sf r}=0$ is $\beta_j=0$, and in the vortex vacuum one also
has that $e^\ell_j =0$.  In this case, the full fermionic
Hamiltonian describing the quasiparticles in Eq.~(\ref{eq:Hf})
reduces to the Bogoliubov-deGennes form, apart from the coupling
to $e^t_j$ which describes the Doppler shift couplings to the
super-flow.  In particular, the string operator which enters into
the explicit pairing term equals unity, ${\cal S}^2_{\bf r}=1$, so
that the fermionic quasiparticles are paired. We denote by
$|0\rangle_f$ the ground state of $\hat{H}_f$ with
$\beta_j=e_j=0$, which is easily found by filling the
Bogoliubov-deGennes levels below the Fermi energy. Our naive
ground state is then
\begin{equation}
  \label{eq:u1naive}
  |0\rangle_0 = |0\rangle_f \otimes |\beta_j=0\rangle \otimes |N_{\sf
  r}=0\rangle.
\end{equation}
In the U(1) sector, unfortunately, we must take somewhat more
care, since the condition $\beta_j=0$ is in fact inconsistent with
the gauge constraint ${\cal G}_f(\Lambda_{\bf r})=1$.  We can,
however, project into the U(1) subspace using the operator
\begin{equation}
  \label{eq:projector}
  \hat{P}=\prod_{\bf r} \delta(f^\dagger_{{\bf r} \sigma}
  f^{\vphantom\dagger}_{{\bf r} \sigma} - {1 \over \pi}
  \epsilon_{ij} \partial_i \alpha_j({\bf r} - {\bf w})).
\end{equation}
Since $[\hat{P},\hat{H}]=[\hat{P},{\cal G}_v(\chi_{\sf r})]=0$,
the state
\begin{equation}
  \label{eq:projgs}
  |0\rangle = \hat{P}|0\rangle_0
\end{equation}
satisfies all gauge constraints and remains an eigenstate of $H$
(with zero kinetic terms). This establishes that the resulting
superconducting state is an ordinary BCS-type superconductor with
paired electrons.

\subsubsection{BCS instability}
\label{sec:selectr-bcs-inst}

As we saw in Section V, the coupling of the spinons to the gapless
rotons in the RFL phase leads to Heisenberg exchange and pair
field interactions between the fermions with strength $J \sim
t_s^2/u_v$.  In the physically interesting limit $t \approx t_s
\gg t_e$, the quasiparticles move primarily as spinons, i.e.
without any associated electrical charge, and hence do not
experience a long-ranged Coulomb repulsion.  Thus one may imagine
that the above interactions could lead to a BCS pairing
instability at ``high'' energies (still below the quasiparticle
Fermi energy to make the Fermi liquid description appropriate, but
quantitatively large compared to e.g. more conventional BCS
critical temperatures) of order $J$.  Here we explore the
properties of the resulting phase which emerges when the fermions
pair spontaneously and condense.  To this end, we focus on the
fluctuations of the phase, $\Phi$ of the fermion pair field,
$\langle f_{\uparrow} f_{\downarrow} \rangle = \Delta_f
e^{i\Phi}$.  Keeping $\Delta_f$ fixed, and working once more in
the electron  gauge $\beta_i = -\pi \overline{e}_i$, we integrate
out the fermions entering in ${\cal L}_{RFL}$ to generate an
effective action for $\Phi, \beta_0$, and $\overline{e}_i$.
Specifically, one obtains ${\cal L}^0_f + {\cal L}_f^1 \rightarrow
{\cal
  L}_\Phi$, with
\begin{eqnarray}
{\cal L}_\Phi & = & \frac{g_f}{2} (\partial_0 \Phi - 2 \beta_0)^2 + \frac{
  \rho_s^f t_s}{2}(\partial_i\Phi + 2\pi
\overline{e}_i)^2
 \\ & & \hspace{-0.5in} + \frac{\rho_s^f t_e}{2}(\partial_i \Phi - 2A_i)^2
-\alpha\frac{ k_B T}{t\Delta_f} (t \partial_i \Phi + 2 \pi
t_s\overline{e}_i - 2 t_e A_i)^2. \nonumber
\label{eq:LPhi}
\end{eqnarray}
Here $g_f$ is of order the density of states at the Fermi surface, $A_i$
is (time-independent for simplicity) external vector potential, and
$\rho_s^f$ is the fermion pair (``superfluid'') density.  The final term
represents the low but non-zero temperature corrections to the
superfluid density appropriate to the case of d-wave pairing.  Here
$\alpha$ is a non-universal constant of order one representing the
effects of Fermi liquid corrections, or equivalently, high-energy
renormalizations of the ``Doppler shift'' coupling constant.

The excitations and response functions of the system can then be
obtained from the RFL Lagrangian by shifting $\alpha_\mu
\rightarrow \alpha_\mu + a_\mu$ and then integrating out $a_0,
a_j$ and $\alpha_0$.  Having made the replacement, ${\cal L}_f
\rightarrow {\cal L}_\Phi$, one thereby obtains ${\cal L}_{RFL}
\rightarrow {\cal L}_{eff}$, with an effective Lagrangian given by
${\cal L}_{eff} =\hat{\cal L}_{rot}+{\cal L}_\Phi$, where
\begin{eqnarray}
  \label{eq:Lrothat}
  \hat{\cal L}_{rot} & = & \frac{u_v}{2} (e_i+ \frac{1}{\pi}
  \epsilon_{ij}A_j)^2 + i \partial_0 e_i (\partial_i\theta+\alpha_i)
  \nonumber \\
  & & +
  \frac{u_v}{2\pi^2 v_0^2}\beta_0^2 + \frac{i}{\pi} \beta_0
  \epsilon_{ij} \partial_i \alpha_j \\
  & & + \frac{\kappa_r}{8}(\epsilon_{ij}\partial_i \alpha_j)^2 +
  \frac{\kappa_r}{2}[ \Delta_{xy}\theta + \frac{1}{2}(\partial_x
    \alpha_y + \partial_y \alpha_x)]^2. \nonumber
\end{eqnarray}

First, it is instructive to consider the transverse electromagnetic
response, in particular consider a static, transverse external gauge
field, $\partial_0 A_i = \vec\nabla\cdot\vec{A}=0$.  Since the external
gauge field is at zero frequency, $\alpha_j,\beta_0$, $\theta$, and $e_t$ are
decoupled from $A_i$ and $e_l$ in this limit, and we may thus neglect all but
the first term in Eq.~\ref{eq:Lrothat}.  In addition, $\Phi$ is
decoupled from $e_l$ ($\overline{e}_t$) as well, so we may simplify the
effective Lagrangian to

\begin{eqnarray}
  \label{eq:efflagmeissner}
{\cal L}_{eff}=
  \frac{u_v}{2}({e_l}+ \frac{A_t}{\pi })^2 +
  2 t_e \rho_s^f A_t^2+  2\pi^2t_s\rho_s^f e_l^2
\nonumber \\  -\alpha\frac{ k_B T}{t\Delta_f} (2 \pi
t_s e_l - 2 t_e A_t)^2.
\end{eqnarray}

Integrating over $e_l$ then gives the superfluid stiffness $K_s$ as the
coefficient of $2 A_t^2$ (corresponding to the pair-field phase
stiffness) in the effective action:
\begin{eqnarray}
  \label{eq:rhos}
  K_s & & =  \\
  & & \frac{\rho_s^f(t u_v + 4\pi^2 t_e t_s \rho_s^f
    )}{u_v + 4\pi^2 t_s\rho_s^f} -
     \frac{( t u_v + 4\pi^2 t_e t_s \rho_s^f )^2}{(u_v + 4\pi^2
         t_s\rho_s^f )^2} \frac{2\alpha k_B T}{t\Delta_f}\nonumber
\end{eqnarray}
to linear order in temperature.  Note that, for $\rho_s^f\neq 0$, the
system is a true charge superfluid, displaying a Meissner effect.

We would next like to establish the fate of the roton excitations.
To do these, we let $A_i=0$, specialize to $T=0$, and consider the
appropriate limit $\omega_n \sim k_x \ll k_y\sim O(1)$,
integrating out fields to obtain an effective action for $\theta$
alone.  Integrating out $\Phi$, one obtains in this limit
\begin{eqnarray}
  \label{eq:Phiintegral}
  {\cal L}_\Phi & \rightarrow & 2 g_f \beta_0^2 + 2\pi^2 \rho_s^f t_s
  e_l^2,
\end{eqnarray}
using the above conditions on $\omega_n$ and ${\bf k}$, and $\partial_0
\beta_0 \ll \epsilon_{ij}\partial_i e_j$, which follows in this limit.
Further taking the gauge $\vec\nabla\cdot\vec\alpha=0$, one sees that
$e_t$ couples only to $\partial_0\alpha$ in Eq.~(\ref{eq:Lrothat}), and
thus generates only negligible terms at low frequencies and may be
dropped.  We therefore need only the effective action for
$\alpha=\alpha_t$, $\theta$, $\beta_0$ and $e_l$, $S_{eff}=\int_{{\bf
    k}\omega_n} s_{eff}$, which in this limit becomes
\begin{eqnarray}
  \label{eq:seffatbe}
  s_{eff} & & = \frac{u_v + 4\pi^2 \rho_s^f t_s}{2} e_l^2 +
  \frac{u_v+4\pi^2 v_0^2 g_f}{2\pi^2 v_0^2}\beta_0^2 - |{\cal
    K}_y|\omega_n e_l \theta \nonumber \\ && \hspace{-0.3in} + \frac{i|{\cal
      K}_y|}{\pi}\beta_0 \alpha
   +
  \frac{\kappa_r}{8}{\cal K}_y^2 \alpha^2 + \frac{\kappa_r {\cal
      K}_y^2}{2}|k_x \theta + \frac{{\rm sign}(k_y)}{2} \alpha|^2.
\end{eqnarray}
Performing the integration over $\alpha$, $\beta_0$ and $e_l$, one
obtains the final effective action for $\theta$:
\begin{eqnarray}
  \label{eq:Seffthetafinal}
  s_{eff}(\theta) & = & \frac{1}{2}\frac{{\cal K}_y^2}{u_v+4\pi^2
    \rho_s^f t_s}\left(\omega_n^2 + \tilde{v}_{rot}^2
    k_x^2\right)|\theta|^2,
\end{eqnarray}
where
\begin{eqnarray}
  \label{eq:vrotss}
  \tilde{v}_{rot}^2 = v_1^2 \frac{u_v+4\pi^2 \rho_s^f t_s}{u_v}
  \frac{u_v(v_0^2+\frac{v_1^2}{4}) + \pi^2 g_f v_0^2 v_1^2}{u_v(v_0^2 +
    \frac{v_1^2}{2})
    + 2\pi^2 g_f v_0^2 v_1^2}.
\end{eqnarray}
Thus, despite the superconductivity induced by quasiparticle pairing,
the gapless roton excitations survive, with some quantitative
modifications to their velocity and correlations.

As we shall explore further in the concluding sections, in the limit of
a very small ``bare'' vortex core energy appropriate in the under-doped
limit, $u_v \sim x \rightarrow 0$, there is a large energy scale for
pairing, $J \sim t_s^2 /u_v$.  It is then natural to assume $t\rho_s^f
\approx t(1-x)
\gg u_v$.  If we presume that the fermionic kinetic energy is
predominantly due to ``spinon hopping'', $t\approx t_s \gg t_e$, then
one has
\begin{equation}
  \label{eq:Ksapprox}
  K_s \approx \frac{u_v}{4\pi^2} + t_e\rho_s^f - (\frac{u_v}{4\pi^2} +
  t_e\rho_s^f)^2 \frac{2\alpha k_{\scriptscriptstyle B}T}{t\Delta_f
    (\rho_s^f)^2}.
\end{equation}
Thus, despite the large bare superfluid stiffness coming from the BCS
pairing of the quasiparticles, the renormalized stiffness is small, set
by the bare vortex core energy, $K_s = {u_v \over 4\pi^2} \sim x$ or the
electron hopping $t_e$ (presumed small).  It is the renormalized
stiffness which determines the vortex core energy, and sets the scale
for the finite temperature Kosterlitz-Thouless superconducting
transition, $T_{KT} \sim u_v \sim x$.  In this way, one can understand
the large separation in energy scales between the pseudo-gap line at
scale $J$ and the superconducting transition temperature, $T_{KT} \sim
x$.  Unfortunately (since it is in apparent conflict with the small
amount of experimental data for this quantity), along with this small
superfluid stiffness, one obtains a small linear temperature derivative,
$\left.\partial K_s/\partial T\right|_{T=0}$, due to the same mechanism.
This is similar to results for the U(1) gauge theory for the $t-J$
model.

To complete the analysis, one should consider the effects of the
heretofore neglected explicit pairing term in this novel ``rotonic''
superconductor.  In particular, one may imagine that, once the fermionic
pair field has condensed, it may induce true off-diagonal long-range
order in the rotonic sector through the proximity effect.  Naively,
ODLRO appears incompatible with gapless rotons, so one may expect the
explicit pairing term to induce a gap in the roton spectrum.  While this
is possible, it is easy to see that it is not inevitable.  To see this,
note that, in the rotonic superconductor, the fermionic pair-field in
the explicit pairing term may be replaced simply by its mean-field value,
$\Delta_1 c_{{\bf r}+\hat{\bf 
        x}_1 \sigma} \epsilon_{\sigma\sigma'}c_{{\bf r}\sigma'} =
    \Delta_2 c_{{\bf r}+\hat{\bf 
        x}_2 \sigma} \epsilon_{\sigma\sigma'}c_{{\bf
        r}\sigma'}=\Delta_f$.  This leads, in the roton sector, to an
    additional term, 
\begin{eqnarray}
  \label{eq:Hdelta}
  H_\Delta & = & - \Delta_f \sum_{j\bf r} (B_{{\bf r},{\bf r}+\hat{\bf
      x}_j}^\dagger + {\rm h.c.}).
\end{eqnarray}
Indeed, Eq.~(\ref{eq:Hdelta}), since it embodies a linear coupling to
the bosonic pair field, $B_{{\bf r},{\bf r}+\hat{\bf x}_j}$, will induce
ODLRO in the $B_{\bf r}$ operators.  However, this need not itself be a
mechanism to induce a roton gap.  Noting that from
Sec.~\ref{sec:quasi-diagonal-long}, the boson pair field correlators are
power-law in form, we expect that for $\Delta_c$ sufficiently large,
perturbation theory in $\Delta_f$ will be regular and convergent, and
the gapless rotons will be preserved.  Due to the anisotropic structure
of these correlators, we cannot reliably determine the relevance or
irrelevance of $\Delta_f$ by simple power-counting arguments.  We note
that for $\Delta_c>3/2$, the induced ODLRO is expected to be linear in
$\Delta_f$, i.e. the bosonic pair-field susceptibility is finite.
However, the criterion for irrelevance is probably more stringent, e.g.
$\Delta_c>3$.  To determine the precise value $\Delta_c^*$ such that for
$\Delta_c>\Delta_c^*$, the explicit pairing term remains irrelevant
after condensation of quasiparticle pairs, would require a more careful
analysis of the structure of the perturbation theory in $\Delta_f$.  We
leave this for braver souls, and content ourselves with the observation
that there is a range of stability to this perturbation.  We note that,
however, since the RL itself even without the pairing term is always
unstable to either vortex or ``charge'' hopping, the rotonic
superconductor, with true gapless rotons, exists at best as an
intermediate energy scale crossover.  Nevertheless, provided the
inevitable roton gap is small (i.e. for small $\Delta_f,t_c, t_v$), we
expect the gap onset will only slightly modify the values for the superfluid
stiffness and its linear temperature derivative determined above.

\subsubsection{Explicit pairing term}
\label{sec:stab-expl-spin}

The final potential instability of the RFL we consider is from the
explicit (spinon) pairing term $\Delta$.  Recall from
Sec.~\ref{sec:u1-vortex-spinon-1}, 
\begin{eqnarray}
  \label{eq:explicit}
  {\cal H}_\Delta & = &   \sum_j e^{i\beta_j({\bf r})} \Delta_j
    [{\cal S}_{\bf r}]^2 f_{{\bf r}+\hat{\bf
        x}_j \sigma} \epsilon_{\sigma\sigma'}f_{{\bf r}\sigma'}+ {\rm
      h.c.}\\
    & = & \sum_j \Delta_j B_{{\bf r},{\bf r}+\hat{\bf x}_j}^\dagger
    c_{{\bf r}+\hat{\bf 
        x}_j \sigma} \epsilon_{\sigma\sigma'}c_{{\bf r}\sigma'}+ {\rm
      h.c.},\label{eq:explicitelec}
\end{eqnarray}
where the last line is written in electron variables, or equivalently in
the ``electron gauge'' with $\beta_j=-\pi\overline{e}_j$.  This
representation is convenient because it eliminates all unphysical gauge
fluctuations, which, although they do not appear in any physical
properties, may enter inadvertently through approximations.  Had we
considered instead a Hamiltonian with $s$-wave pairing, we would have
had
\begin{eqnarray}
  \label{eq:Hswave}
  {\cal H}_\Delta^{s-{\rm wave}} & = & \Delta^s [{\cal S}_{\bf r}]^2
  f_{{\bf r}\sigma}\epsilon_{\sigma\sigma'}f_{{\bf r}\sigma'} + {\rm
    h.c.} \nonumber \\
  & = & \Delta^s B_{\bf r}^\dagger c_{{\bf
      r}\sigma}\epsilon_{\sigma\sigma'}c_{{\bf r}\sigma'} + {\rm 
    h.c.}
\end{eqnarray}
Since we have already determined the correlations of the boson pair
field operator $B_{\bf r}$ (which exhibits ODQLRO), in
Sec.~\ref{sec:quasi-diagonal-long}, all the operators appearing in
Eq.~(\ref{eq:Hswave}) have well understood properties at this stage, and
so we will discuss this case for simplicity.  For the physically more
interesting scenario of $d$-wave pairing, we require instead the
behavior of the bond pair field correlations.  As discussed in
Sec.~\ref{sec:odqlro-d-wave}, this behavior is qualitatively identical
to that of the local pair field, with a possible renormalization of
$\Delta_c$.  Hence we believe that the results we obtain for local
($s$-wave) pairing -- in particular that the pairing term is irrelevant
for sufficiently large $\Delta_c$ -- carry over to the $d$-wave case,
provided a possible (and for the moment unknown) $O(1)$ modification of
$\Delta_c$ is made in the relations.

We would like to determine the ``relevance'' of ${\cal H}_\Delta$ in the
renormalization group (RG) sense, i.e. whether its presence destabilizes
the low energy properties of the RFL.  Unfortunately, due to the
extremely anisotropic nature of the rotonic spectrum, and the very
different nature of the low energy electronic quasiparticle states at
the Fermi surface, we do not know how to formulate a proper RG
transformation.  However, we do note that correlations of ${\cal
  H}_\Delta$ decay as power laws in space and imaginary time in the
theory with $\Delta=0$, since then these correlations factor into
$G^{cp}$ (with power-law behavior described in
Sec.~\ref{sec:quasi-diagonal-long}) and the fermion pair-field
correlator, which has the power law form characteristic of a Fermi
liquid.  Clearly, for sufficiently large $\Delta_c$, the Cooper pair
propagator $G^{cp}({\bf r},\tau)$ will decay sufficiently rapidly that
perturbation theory in $\Delta$ does not generate any (primitive)
singularities.  However, determining the critical $\Delta_c^*$ above
which this occurs is beyond the scope of the current study.  A simple
argument clearly bounds $\Delta_c^*>3/2$.  In particular, one may
attempt to integrate out the vortices perturbatively in ${\cal
  H}_\Delta$ in a cumulant expansion.  The first non-trivial term in
this expansion occurs at second order, and generates an attractive
fermion pair-field to pair-field interaction whose vertex is
simply the Fourier transform of $G^{cp}({\bf r},\tau)$.  For
$\Delta_c<3/2$, a simple scaling analysis indicates that this
Fourier transform is divergent at ${\bf q}=\omega_n=0$.  Thus for
such values of $\Delta_c$, this attractive Cooper channel
interaction will overwhelm any other repulsive interaction that
might be present at low energies, leading to a Cooper instability
and pairing.  This analysis, however, neglects higher cumulant
terms which are certainly present in integrating out the vortices,
and which are presumably crucial in determining the ultimate
limits of stability of the RFL.  Nevertheless, and this is all we
shall require at present, it is clear that $\Delta_c^*$ exists and
is not infinite, so that a non-vanishing region of stability also
exists. When $\Delta_c < \Delta_c^*$ the explicit pairing term
will be ``relevant'', destabilizing the RFL phase, presumably
driving it into a conventional superconducting phase with paired
electrons and gapped rotons.  Given our lack of knowledge of
$\Delta_c^*$, the additional uncertainty in the decay exponent of ODQLRO
is not particularly damaging.

\section{Properties of the RFL phase}

\label{sec:RFLproperties}

Having established the existence of the RFL phase, we now discuss
some of it's properties.  Since the rotons are effectively
decoupled from the fermions at low energies, much of the physics
of the RFL phase follows directly from the results we established
for the charge sector in Sections III and IV.  Specifically, one
expects {\it three} different gapless excitations in the RFL phase
- a gapless longitudinal plasmon, a Bose surface of gapless rotons
and a set of particle-hole excitations across the Fermi surface.
The {\it spin} physics in the RFL phase is then qualitatively
similar to that of a conventional normal metal.  Also like a
metal, the RFL has ODQLRO in the Cooper pair field.  Unlike an
ordinary metal, however, this power law off-diagonal order exists
with two distinct unrelated powers, one originating from the
two-particle excitations of the Fermi sea, and the other from the
gapless rotons. Because the rotons and electronic quasiparticles
exist as nearly independent excitations in the RFL, one would
expect a sharp electron spectral function despite the critical
rotons. While this is true, as we demonstrate below in detail, the
quasiparticles do scatter appreciably and in a singular manner
from the gapless rotons at particular ``hot spots'' on the Fermi
surface.  Nevertheless, even at these hot spots, the decay rate
vanishes more rapidly than the electron's energy.

Despite the existence of such long lived electron-like quasiparticle
excitations, the electrical transport in the RFL phase is strikingly
non-Drude like, as we now demonstrate.  Specifically, in the presence of
impurities the electrical conductivity at low temperatures is dominated
by the quasi-condensate in the charge sector, diverging at low
temperatures as $\sigma \sim T^{-\gamma}$ with $\gamma >1$.  But as we
shall see, the {\it Hall conductivity} behaves very differently, being
dominated by the electron-like quasiparticles.

\subsection{Electrical Conductivity}
\label{subsec:resist}
Here we evaluate the electrical resistance in the RFL phase.  As in
Section IVA, it will be necessary to include the effects of the vortex
hopping term, even though vortex hopping is technically ``irrelevant''
when $\gamma >1$.  The RFL phase exhibits the same ``emergent
symmetry'' as in the Roton liquid -- the number of vortices on every
row and column being independently conserved -- so inclusion of vortex
hopping is necessary to generate {\it any} dissipative resistance
whatsoever.  In order to access the
Hall conductivity we apply a uniform external magnetic field.  We
choose a gauge with $A_0=0$, and set,
\begin{equation}
A_j({\sf r},\tau) = A^B_j({\sf r}) + \tilde{A}_j(\tau)  ,
\end{equation}
with $\epsilon_{ij} \partial_i A^B_j = B$ the external magnetic field
and $\tilde{A}_j$ a time-dependent source term used to extract the
conductivity tensor.  As might be expected it is important to include
the effects of elastic scattering from impurities. 

It is convenient to
employ, as in Sec.~\ref{sec:selectr-bcs-inst}, in which the fermions
have been integrated out, their effects felt only through an
additional contribution to the effective action.  In particular, we
choose,  as in Sec.~\ref{sec:selectr-bcs-inst}, the electron gauge, and
write
\begin{equation}
  \label{eq:actionforRFLsigma}
  S_{eff} = \int_\tau \sum_{\sf r} \left[ \hat{\cal L}_{rot} - t_v
    \cos(\partial_i \theta+\alpha_i)\right] + S_f(e_i,\beta_0,\tilde{A}_i),
\end{equation}
with $\hat{L}_{rot}$ given in Eq.~(\ref{eq:Lrothat}). Here $S_f$
represents the fermionic terms in the action.  We proceed by
shifting $\overline{e}_j \rightarrow \overline{e}_j -A_j^B/\pi$,
which takes $A_j \rightarrow \tilde{A}_j$ in $\hat{L}_{rot}$
without any further changes since $A_j^B$ is time independent.
This has the effect for the fermions of making the magnetic field
appear uniformly in both electron and spinon hopping terms.
Integrating out the fermions then  perturbatively in
$\overline{e}_i$, $\beta_0$, and $\tilde{A}_i$ effectively
replaces $S_f \rightarrow \tilde{S}_f = \int_{{\bf k}\omega_n}
s_f$, where
\begin{eqnarray}
  \label{eq:SfRFLsigma}
  s_f & = & \frac{\Pi_{00}}{2} |\beta_0({\bf k},\omega_n)|^2 + \frac{\pi^2 t_s}{2} T_{nn}|\overline{e}_j|^2 + \frac{t_e}{2}
  T_{nn} |\tilde{A}_j|^2 \nonumber \\
  &   & \hspace{-0.2in} +
  \frac{\tilde{\Pi}_{ij}}{2t^2}(\pi t_s \overline{e}_i + t_e
  \tilde{A}_i)_{{\bf k},\omega_n}(\pi t_s\overline{e}_j +t_e
  \tilde{A}_j)_{-({\bf k}\omega_n)}.
\end{eqnarray}
In Eq.~(\ref{eq:SfRFLsigma}), both $\Pi_{00}$ and $\tilde\Pi_{ij}$ are
functions of ${\bf k}$ and $\omega_n$ (and $B$ through $A_j^B$ which
appears in the electron hopping term), and we have neglected a
cross-term $\Pi_{0i}$ between $\beta_0$ and the spatial gauge fields,
which is negligible in all the limits of interest.  We will require the
behavior of $\tilde{\Pi}_{ij}$ in two regimes.  Since the external
fields are spatially uniform, we will need $\tilde{\Pi}({\bf
  k=0},\omega_n)$ at low frequencies
\begin{eqnarray}
  \label{eq:Piuniform}
  \tilde{\Pi}_{ij}({\bf k=0},\omega_n) \approx \sigma_{xx}^f |\omega_n|
  \delta_{ij} + \sigma_{xy}^f \omega_n \epsilon_{ij},
\end{eqnarray}
where $\sigma_{ij}^f$ is the conductivity tensor for the fermions.
Roton fluctuations are dominated in contrast by $\omega_n \sim k_x \ll
k_y \sim O(1)$ (or the same with $k_x \leftrightarrow k_y$).  In this
limit, $\tilde{\Pi}_{ij}$ becomes a non-trivial function of the $O(1)$
component of the wavevector, but has the useful property (due to square
reflection symmetry) of decoupling in the longitudinal and transverse
basis,
\begin{eqnarray}
  \label{eq:Pilt}
  \tilde\Pi_{ij}({\bf k},\omega_n) \approx \tilde{\Pi}_t \left( \delta_{ij}
    - \frac{{\cal K}_i^{\vphantom{*}}{\cal K}_j^*}{{\cal K}^2}\right) +
  \tilde\Pi_l 
  \frac{{\cal K}_i^{\vphantom{*}}{\cal K}_j^*}{{\cal K}^2} . 
\end{eqnarray}
In this limit, we note that $\tilde\Pi_t$ is related to $\Pi_{11}$ of
Sec.~\ref{sec:roton-scatt-selectr}\ by $\Pi_{11} = \tilde{\Pi}_t+
T_{nn}$.

As discussed above, the conductivity is finite only once the vortex
hopping is included to break the row/column symmetries of the RFL.
Hence to extract the conductivity, we must compute the effective
action for $\tilde{A}_j$ to $O(t_v^2)$, which gives the first
non-trivial correction.  This requires only Gaussian integrals, with
no further approximations.  However, for ease of presentation, it is
convenient, analogously to Sec.~\ref{sec:electr-resist-roton}, to take the
simplifying limit $u_v/v_0^2 + \Pi_{00} \ll 1/\kappa_r$.  In this
limit the fluctuations of $\beta_0$ are extremely strong, which in
turn strongly suppresses the fluctuations of $\alpha^t({\bf
  k},\omega_n)$ except at ${\bf k=0}$.  Furthermore, choosing
the gauge $\vec\nabla\cdot\vec\alpha=0$, we may thereby take
$\alpha_i({\sf r},\tau)$ to be a function of $\tau$ only.  Doing so,
we may drop all spatial derivatives of $\alpha_j$ in $\hat{\cal
  L}_{rot}$.  Furthermore, since fluctuations of $\alpha_j$ are only
temporal, it can be accurately treated in an RPA fashion.  

To carry out the RPA calculation, we first perform the integral over
$e_i$, which gives an effective action in terms of the remaining
$\theta$, $\alpha_i$, and $\tilde{A}_i$ fields,
$S_{eff}\rightarrow S'_{eff}=S'_1(\tilde{A}_i)  
+ S'_2(\alpha_i,\tilde{A}_i)+ S'_3(\theta, \alpha_i)$, with 
\begin{widetext}
  \begin{eqnarray}
    \label{eq:speff}
    S'_1 & = & L^2\int_{\omega_n} \frac{1}{2}\big\{ 
    (\eta t_s + t_e)T_{nn}\delta_{ij} 
    + \frac{1}{t^2}\left(t_e^2 +
      \eta^2 t_s^2+ 2 \eta t_e t_s \right)\tilde{\Pi}_{ij}({\bf
      k=0},\omega_n) \big\} 
    \tilde{A}_i(-\omega_n)\tilde{A}_j(\omega_n), \\
    \label{eq:speff2}
    S'_2 & = & L^2 \int_{\omega_n} \left\{
      \frac{\eta \omega_n^2}{2u_v}\left[\delta_{ij} +\frac{\eta \pi^2
          t_s^2}{u_v t^2} \tilde{\Pi}_{ij}({\bf
          k=0},\omega_n)\right]\alpha_i(-\omega_n)\alpha_j(\omega_n)
      -\frac{\eta 
        \omega_n}{\pi} \epsilon_{ij} \alpha_i(-\omega_n)
      \tilde{A}_j(\omega_n) \right\}  \\ 
    \label{eq:speff3}
    S'_3 & =&\int_{{\bf k}\omega_n}  \frac{1}{2}\left[\frac{\omega_n^2
        {\cal K}^2}{\overline{u}_v}+ 
      \kappa_r|{\cal K}_x{\cal K}_y|^2 \right]|\theta({\bf
      k},\omega_n)|^2 - t_v \sum_{\sf 
      r}\int_\tau \cos(\partial_i \theta+\alpha_i), 
  \end{eqnarray}
\end{widetext}
where $\eta = u_v/(u_v+\pi^2 t_s T_{nn})$, and $\overline{u}_v = u_v +
\pi^2 t_s T_{nn} + \pi^2 t_s^2 \tilde{\Pi}_t/t^2 = \tilde{u}_v + \pi^2
t_s^2 \Pi_{11}/t^2$ as in Sec.~\ref{sec:roton-scatt-selectr}.  In
Eqs.~(\ref{eq:speff}),~(\ref{eq:speff2}), $L^2$ is the system volume
(number of sites in the square lattice), arising since
$\alpha_i,\tilde{A}_i$ are spatially constant. In
obtaining Eqs.~(\ref{eq:speff}-~\ref{eq:speff2}), we expanded to linear
order in $\tilde\Pi_{ij}({\bf k=0},\omega_n)$ since we are interested
ultimately in the low-frequency limit, and $\tilde{\Pi}_{ij}$ is linear
in frequency.  In Eq.~(\ref{eq:speff3}), we used the fact that $\theta$
coupled to only to the longitudinal part of $e_i$ and hence only to
$\tilde\Pi_t$.  Note that the quadratic part of Eq.~(\ref{eq:speff3})
reproduces precisely Eqs.~(\ref{eq:Srotsec5}-~\ref{eq:vrotfull}), in the
limit $v_0 \gg \tilde{v}_1$, as assumed herein.  

We now can integrate out $\theta$ to $O(t_v^2)$.  This proceeds
identically as in Sec.~\ref{sec:electr-resist-roton}, with $a_j$
replaced by $-\alpha_j$, and with the renormalized roton liquid
parameters $u_v \rightarrow \overline{u}_v$.  Hence we obtain the
correction $S'_3(\theta,\alpha_i) \rightarrow S''_3(\alpha_j)$, with 
\begin{eqnarray}
  \label{eq:S3pp}
  S''_3 & = & \frac{t_v^2}{4} L^2 \int_{\omega_n} \tilde{\cal R}(\omega_n)
  |\alpha_j(\omega_n)|^2, 
\end{eqnarray}
with $\tilde{\cal R}(\omega_n) \sim
-|\omega_n|^{1+\gamma}$ as in Eq.~(\ref{rotpol}) of
Sec.~\ref{sec:electr-resist-roton}, but $\gamma=2\Delta_v-3$ with the
renormalized $\Delta_v$ given in Eq.~(\ref{eq:deltavrenorm}).  

With this replacement, the remaining quadratic integral over $\alpha_i$
can be easily performed to determine the physical response kernel,
$S'_1+S'_2+S''_3 \rightarrow S_{resp}(\tilde{A}_j)$, with
\begin{eqnarray}
  \label{eq:Sresp}
  S_{resp} & = & L^2 \int_{\omega_n} \frac{1}{2} \tilde{A}_i(-\omega_n)
  \Pi^{RFL}_{ij}(i\omega_n) \tilde{A}_j(\omega_n),
\end{eqnarray}
where
\begin{eqnarray}
  \label{eq:PiRFL}
  \Pi^{RFL}_{ij}(i\omega_n) & \approx & \tilde{\Pi}_{ij}^f +
  \tilde{\Pi}_{ij}^{rot}, 
\end{eqnarray}
and
\begin{eqnarray}
  \label{eq:pif}
  \tilde{\Pi}_{ij}^f & = & t_e T_{nn}\delta_{ij} +
  \frac{t_e^2+2\eta t_e t_s}{t^2} \tilde\Pi_{ij}({\bf k\! =\!0},\omega_n) ,
  \\
  \tilde{\Pi}_{ij}^{rot} & = &     \left[\frac{u_v}{\pi^2} - \frac{u_v^2 t_v^2
      \tilde{\cal R}(\omega_n)}{2\pi^2 \omega_n^2}\right]\delta_{ij}.
\end{eqnarray}
The conductivity is obtained from the Kubo formula as
\begin{equation}
  \label{eq:kubo}
  \sigma_{ij}(\omega) = \left[\frac{\Pi_{ij}^{RFL}(i\omega_n)-t_e
  T_{nn}\delta_{ij}}{\omega_n} \right]_{i\omega_n\rightarrow
  \omega+i\delta}.
\end{equation}
As in Sec.~\ref{sec:electr-resist-roton}, we see that the rotonic
contribution to the conductivity
is not perturbative (at low frequencies) in $t_v$, so to capture the
expected behavior, we replace it by,
\begin{equation}
  \label{eq:pirotdrude}
  \sigma^{rot}_{ij} =
  \left[\frac{\tilde{\Pi}_{ij}^{rot}}{\omega_n}\right]_{i\omega_n\rightarrow \omega+i\delta}
  \rightarrow \frac{\delta_{ij}}{-i\omega \frac{\pi^2}{u_v} + i
  \frac{\pi^2 t_v^2}{2} \tilde{\cal R}_{ret}(\omega)/\omega }    .
\end{equation}
This gives
\begin{equation}
  \label{eq:condadd}
  \sigma_{ij}(\omega,T) = \sigma_{ij}^{rot}(\omega,T)+\frac{t_e^2+2\eta t_e
  t_s}{t^2}\sigma_{ij}^f(\omega, T)  .
\end{equation}

Notice that from Eq.~(\ref{eq:condadd}), the conductivity is the
sum of separate fermion and roton contributions, and that the only
effects of the fermions upon the rotonic piece is to modify the
exponent $\gamma$ (or $\Delta_v$) implicit in $\tilde{\cal
R}(\omega)$.  Moreover, the fermionic contribution vanishes for
$t_e=0$, as expected on physical grounds since the spinon hopping
term does not transport charge.

Consider first the dissipative diagonal d.c. (sheet) resistance,
$R(T) = Re[\sigma^{-1}_{xx}]$, which at low temperatures is
completely dominated by the rotonic contribution, given as in
Eq.~(\ref{eq:RLresist}),
\begin{equation}
  \label{eq:RFLcond}
R(T) = c_\gamma \frac{(\pi t_v)^2}{v_{rot}^2} \left[ \frac{\pi
T}{v_{rot}}\right]^\gamma  ,
\end{equation}
except with $v_{rot}$ and the exponent $\gamma$ renormalized by
the interactions with the fermions. We thereby arrive at the
important conclusion that the resistance in the RFL phase varies
as $R(T) \sim T^\gamma$, with $\gamma >1$.  When $\gamma < 1 $ the
RFL phase is unstable to confinement into a Fermi liquid phase,
and right at the quantum confinement transition the resistance is
linear in temperature.

The Hall conductivity and hence Hall angle on the other hand, are
dominated by the fermionic quasiparticles. Specifically,
$\sigma_{xy} \approx \sigma_{xy}^f$, and in the d.c. limit the
fermionic Hall conductivity can be approximately obtained from a
Drude expression, $\sigma_{xy}^f \approx \omega_c \tau_f (n_f e^2
\tau_f)/m \sim \omega_c \tau_f^2$, where $\omega_c = eB/m$ is the
cyclotron frequency. The fermionic scattering time $\tau_f$ has a
variety of contributions, from elastic impurity scattering to
interactions with rotons, considered in the next section, and
hence may (or may not) have temperature dependence of its own - in
contrast to the diagonal conductivity,  $\sigma_{xx} \sim
T^{-\gamma}$, which diverges as $T \rightarrow 0$ due to the
rotonic contribution. These considerations suggest that the
cotangent of the Hall angle, defined in terms of the resistivity
tensor $\rho_{ij}$ as $\cot(\Theta_H) \equiv \rho_{xx} /
\rho_{xy}$, will vary as,
\begin{eqnarray}
  \label{eq:thetaH}
  \cot(\Theta_H) & \sim & \frac{\sigma_{xx}}{\sigma_{xy}} \sim
  \frac{T^{-\gamma}}{\omega_c \tau_f^2} \sim \frac{1}{\omega_c
    T^{\gamma}\tau_f^2} .
\end{eqnarray}

The complete absence of a roton contribution to the Hall conductivity
$\sigma_{xy}^{RFL}$  is a consequence of the
magnetic field independence of the roton liquid Lagrangian, ${\cal
  L}_{RL}$, either in it's original or dual forms, Eqs.~(\ref{LRL}) and
(\ref{LRLdual}), respectively.  But as discussed in Sec. IVB, the dual
representation of ${\cal L}_{RL}$ allows for additional terms involving
cosines of the dual field $\phi$, which are present due to the
underlying discreteness of the vortex number operator.  While being
irrelevant in the Roton liquid phase, these neglected terms {\it do
  depend} on the external magnetic field and if retained will lead to a
non-vanishing roton contribution to the Hall conductivity.  But
this contribution is expected to vanish rapidly at low
temperatures.  If on the other hand, these terms are {\it
relevant} and drive a superconducting instability at low
temperature, they will likely contribute signifigantly to the Hall
conductivity.  A careful analysis of the Hall response {\it above}
the superconducting transition temperature in this situation will
be left for future work.

\subsection{Electron Lifetime}

Although the electron spectral function is expected to be sharp at zero
temperature right on the Fermi surface in the ideal RFL ``fixed point''
theory, at finite energies (or temperatures) one expects the electrons
to scatter off the rotons through interactions neglected in the RFL,
causing a decay.  Here, we consider two contributions to this electron
decay rate at lowest order in the perturbations to the RFL.  We consider
two scattering mechanisms.  First, scattering due to coupling of the
quasiparticle current to boson currents through the $\beta_\mu$ and
$e_i$ terms present in the fermionic hamiltonian.  Because singular
bosonic current fluctuations are primarily induced by vortex hopping,
non-trivial contributions to the fermion lifetime through this mechanism
occur first at $O(t_v^2)$.  Second, we consider scattering of
quasiparticles due to ``superconducting fluctuations'', i.e. fermion
decay mediated by the ``Jospephson coupling'' $\Delta$ to the bosonic
pair field.  Since we expect $\Delta \ll t_v$ on physical grounds (see
Sec.~\ref{sec:effective-parameters}), this $O(\Delta^2)$ contribution is
naively much smaller than the former one.  However, depending upon the
parameters of the RFL, this need not be the case.  This point will be
returned to in the discussion.

As discussed above, to compute the spectral function it is
simplest to work in the gauge $\beta^\ell=\pi e^t$, in which we
may assume the electron operator $c_{{\bf r}\sigma} \sim f_{{\bf
r}\sigma}$, and hence study
\begin{equation}
{\cal G}_f ({\bf r}, \tau) = - \langle T_\tau f_{{\bf r} \sigma} (\tau)
f^\dagger_{{\bf 0} \sigma}(0)  \rangle   .
\end{equation}
In other gauges, it would be necessary to include string factors as
indicated in Eq.~(\ref{electronU1}).

Neglecting the fluctuating $\beta_\mu$ fields, one has ${\cal
  G}_0({\bf k},\omega_n) = (i\omega_n - \epsilon_{\bf k} )^{-1}$, with
dispersion, $\epsilon_{\bf k} = -2t \cos(k_j) - \mu_s$.
Both fluctuations of the $\beta_\mu$ gauge fields, which mediate retarded
interactions amongst the quasiparticles, as well as the explicit
interactions in the pairing term, will induce a self
energy $\Sigma({\bf k},\omega_n)$, defined by
\begin{equation}
  \label{eq:selfenergydef}
  {\cal G}_0({\bf k},\omega_n) = \frac{1}{i\omega_n -\epsilon_{\bf
      k}-\Sigma({\bf k},\omega_n)}.
\end{equation}
We will compute the lowest order corrections to the
electron self energy, obtaining a single-particle lifetime from the
imaginary part of its retarded continuation, $\Sigma_{ret}({\bf
  k},\omega) = \Sigma({\bf k},\omega_n \rightarrow -i\omega+ 0^+)$ in
the usual way.  At the level of this discussion, the two types of
interactions act in parallel to scatter quasiparticles, so
\begin{eqnarray}
  \label{eq:addsigmas}
  \Sigma({\bf k},\omega) & = &   \Sigma_{cf}({\bf k},\omega) +
  \Sigma_{sf}({\bf k},\omega).  
\end{eqnarray}
We will focus upon the imaginary part $\Sigma''({\bf k},\omega)$, which
describes broadening of the electron spectral function, $1/\tau_f(T) =
\Sigma''({\bf k},0;T)$.  We have
\begin{eqnarray}
  \label{eq:matthieson}
  {\tau_f}^{-1} & = &   (\tau_f^{cf})^{-1}+ (\tau_f^{sf})^{-1}.
\end{eqnarray}
In principle this single-particle ``lifetime'' is
distinct from the momentum scattering rate which is relevant in
discussions of transport quantities.  However, we will use the behavior
obtained for $1/\tau_f$ as a crude guide to the quasiparticle transport
as well, leaving a more careful treatment for future study.

\subsubsection{Scattering mediated by vortex hopping-enhanced current
  fluctuations} 
\label{sec:scatt-medi-vort}

Taking into account quadratic fluctuations of the $\beta_\mu$ fields,
the leading self energy correction due to current fluctuations is
\begin{eqnarray}
\Sigma_{cf}({\bf k}, i\omega_n) & = & \sum_{\mu\nu} \int_{{\bf q},
  \omega_n^\prime} \!\!\!\!
v_{F\mu} v_{F\nu} {\cal G}_0({\bf k}-{\bf q},\omega_n - \omega_n^\prime)
U_{\mu\nu}( {\bf q}, \omega_n^\prime) \nonumber \\
& & \hspace{-0.75in}= \sum_{\mu\nu} \int_{{\bf q}, \omega_n^\prime}
v_{F\mu} v_{F\nu} \frac{1}{i(\omega_n-\omega_n^\prime)-\epsilon_{{\bf
    k-q}}}U_{\mu\nu}( {\bf q}, \omega_n^\prime),
\end{eqnarray}
with the definitions, $v_{Fj} = \partial \epsilon_{\bf k} /
\partial k_j$, $v_{F0} = 1$.  Here we have written the self-energy
in terms of the full $\beta-$gauge-field correlator, $U_{\mu\nu}$
to all orders in $t_v$, rather than restricting to $t_v=0$ as we
did in obtaining Eq.~(\ref{eq:Umunu}).

Introducing a spectral representation,
\begin{equation}
  U_{\mu\nu}({\bf k}, \omega_n) = \int^\infty_{-\infty} { d \omega \over \pi}
{ U_{\mu\nu}^{\prime\prime}({\bf q}, \omega)  \over  \omega -  i\omega_n  }  ,
\end{equation}
with $U_{\mu\nu}^{\prime \prime}(\omega) = Im
U_{\mu\nu}^{ret}(\omega) =  Im U_{\mu\nu}(\omega_n \rightarrow
-i\omega+ 0^+)$, allows one to analytically continue to obtain the
(retarded) self energy,
\begin{equation}
\Sigma_{ret}({\bf k}, \omega) = \Sigma({\bf k}, \omega_n)|_{i \omega_n
  \rightarrow \omega + i0^+ }   .
\end{equation}
For positive frequencies, $\omega > 0$, the
imaginary part, $\Sigma^{\prime \prime} = Im \Sigma_{ret}$ is given by,
\begin{eqnarray}
\label{eq:Imself}
\Sigma_{cf}^{\prime \prime}({\bf k}, \omega) & = & \\
& & \hspace{-0.5in}\sum_{\mu\nu} \int_{\bf q} v_{F\mu} v_{F\nu}
U_{\mu\nu}^{\prime \prime}({\bf q},\omega - \epsilon_{{\bf k} +
{\bf q}} )  \Theta(\epsilon_{{\bf k} + {\bf q}}) \Theta(\omega -
\epsilon_{{\bf k} + {\bf q}}).\nonumber
\end{eqnarray}

The self-energy can now be evaluated by considering progressive orders
in $t_v$.  To zeroth order, the expressions for $U_{\mu\nu}^{(0)}({\bf
  k},\omega_n)$ may be taken from
Eqs.~(\ref{eq:Umunu}-\ref{eq:lambda0j}).  We note that, because they
come from the quadratic RL Lagrangian, they contain only simple poles.
Furthermore, they are explicitly real functions of $i\omega_n$, so
that, upon analytic continuation, their retarded correlator has zero
imaginary part.  Since $U_{\mu\nu}^{(0)\prime\prime}=0$, one thus
has $\Sigma_{cf}^{(0)\prime\prime}({\bf k},\omega)=0$.

Thus there is no broadening of the quasiparticle peak at zeroth
order in the vortex hopping.  The first non-trivial contribution
to the quasiparticle lifetime occurs though the $O(t_v^2)$
corrections to the gauge field propagator
$U_{\mu\nu}=U^{(0)}_{\mu\nu}+t_v^2 U^{(2)}_{\mu\nu}+\cdots$.  This
correction is obtained by integrating out $\alpha_j$ to second
order in $t_v$ starting from Eq.~(\ref{eq:lvorttilde}) to obtain
$S_{eff}(\lambda_\mu)$ to $O(t_v^2)$ using the cumulant expansion.
We shift $\alpha_i({\sf r},\tau) \rightarrow \alpha_i({\sf
r},\tau) - \upsilon_i({\sf r},\tau)$, to eliminate the linear
terms in $\alpha_i$ in the Lagrangian, with $\upsilon_i$ linear in
$\lambda_\mu$ given explicitly by,
\begin{widetext}

\begin{eqnarray}
  \label{eq:alphashift}
  \upsilon_x({\bf k},\omega_n) & = & -\frac{\pi}{(\omega_n^2 +
    \omega_{pl}^2)(\omega_n^2+\omega_{rot}^2)} [ v_0^2 {\cal K}_x{\cal
    K}_y (v_0^2 {\cal K}_x \lambda_0 +i\omega_n \lambda_x) -
  (\omega_n^2 + v_+^2 |{\cal K}_x|^2)(v_0^2 {\cal K}_y \lambda_0
  +i\omega_n \lambda_y)], \\
  \upsilon_y({\bf k},\omega_n) & = & \frac{\pi}{(\omega_n^2 +
    \omega_{pl}^2)(\omega_n^2+\omega_{rot}^2)} [ v_0^2 {\cal K}_x{\cal
    K}_y (v_0^2 {\cal K}_y \lambda_0 + i\omega_n \lambda_y) -
  (\omega_n^2 + v_+^2 |{\cal K}_y|^2)(v_0^2 {\cal K}_x \lambda_0 +
  i\omega_n \lambda_x)].\nonumber
\end{eqnarray}
After this shift the correction to the effective action can be
formally written,
\begin{eqnarray}
  \label{eq:deltaSeff}
  S_{eff}^{(2)} & = & - \frac{t_v^2}{2}\sum_{ij}\sum_{{\sf
      r,r'}}\int_{\tau,\tau'} \big\langle \cos (\alpha_i({\sf r},\tau)
  - \upsilon_i({\sf r},\tau)) \cos (\alpha_j({\sf r},\tau)
  - \upsilon_j({\sf r},\tau)) \big\rangle_\alpha,
\end{eqnarray}
where the subscript $\alpha$ indicates a Gaussian average over
$\alpha_i$ with respect to the quadratic terms in
Eq.~(\ref{eq:lvorttilde}).  Since $\upsilon_i$ is linear in
$\lambda_\mu$, this correction is not quadratic in $\lambda_\mu$,
indicating that the $\beta_\mu$ fluctuations are not Gaussian.  To
evaluate the leading order self-energy correction, however, we require
only the two-point function of $\beta_\mu$, and hence may expand
to quadratic order in $\upsilon_i$.  One finds
\begin{eqnarray}
  \label{eq:u2a}
  S_{eff}^{(2)} & = & -\frac{t_v^2}{4} \int_{{\bf k}\omega_n} \left(
  |\upsilon_x({\bf k},\omega_n)|^2 \tilde{\cal R}(k_y,\omega_n) +
  |\upsilon_y({\bf k},\omega_n)|^2 \tilde{\cal R}(k_x,\omega_n)\right).
\end{eqnarray}
Here $\tilde{\cal R}(k,\omega_n)$
is the two-point function of the vortex hopping operator considered in
Sec.~\ref{sec:vortex-hopping}, i.e.
\begin{eqnarray}
  \label{eq:vorthopalpha}
  \tilde{\cal R}(k,\omega_n) & = & \sum_x \int_\tau (1-\cos(k x + \omega_n
  \tau)) \left\langle e^{i\alpha_y(x,y,\tau)} e^{-i\alpha_y(0,y,0)}
  \right\rangle_\alpha \nonumber \\
  & \sim & \frac{A(\Delta_v)}{\sin \pi(\Delta_v-1)}  (\omega_n^2 +
  v_{rot}^2 k^2)^{\Delta_v-1},
\end{eqnarray}
\end{widetext}
where the latter behavior gives the leading non-analytic term for
small $\omega_n, k$, with $A(\Delta_v)$ a constant prefactor.  An
analytic quadratic term is also present (and larger than the given
form for $\Delta_v>2$), but does not contribute to the imaginary part
of the self-energy for the same reasons described above for the
$t_v=0$ terms.

Next, we must analytically continue to obtain the
$U^{(2)\prime\prime}_{\mu\nu}({\bf k},\omega)$.  From
Eqs.~(\ref{eq:alphashift}), one immediately sees that
$\upsilon_i^{ret}({\bf k},\omega)$ is purely real.  Hence, the
imaginary part comes entirely from the analytic continuation of
${\tilde R}(k_i,\omega_n)$.  One has
\begin{equation}
  \label{eq:rimag}
  {\cal R}''(k_i,\omega) \sim  A(\Delta_v) (\omega^2-v_{rot}^2
  k_i^2)^{\Delta_v-1} \Theta(|\omega|-v_{rot}|k_i|){\rm sgn}(\omega).
\end{equation}
$ $From inspection of Eq.~(\ref{eq:u2a}), one thereby sees the imaginary
part of the retarded gauge field correlator, $U^{(2)''}_{\mu\nu}({\bf
  k},\omega)$ , is the sum of two terms, which are non-zero only for
$|\omega|>v_{rot}|k_x|$ or $\omega>v_{rot}|k_y|$, respectively.
Thus the momentum integrals in the expression for $\Sigma''_{cf}$ in
Eq.~(\ref{eq:Imself}), are constrained not only by
$0<\epsilon_{\bf
  k+q}<\omega$ but also either $\epsilon_{\bf k+q}<
\omega-v_{rot}|q_x|$ or $\epsilon_{\bf k+q}< \omega-v_{rot}|q_y|$, for
the two terms, respectively.

We focus on the self-energy for momenta exactly on the Fermi
surface, $|{\bf k}|=k_F$.  In this case, the constraints clearly
require small ${\bf q}$ for small $\omega$.  For such small
wavevectors, one may approximate $\epsilon_{\bf k+q} \approx
\vec{v}_F\cdot\vec{q}+ q^2/2m^*$.  For a generic point on the
Fermi surface (taken for simplicity in the quadrant with
$k_x,k_y>0$ , for which $\vec{v}_F$ makes an angle $\theta_F \neq
0,\pi/2$ to the $x$ axis, the $q^2$ term is negligible, and one
has $\epsilon_{\bf k+q} \approx v_F (q_x \cos\theta_F + q_y
\sin\theta_F)$.  Applying the above constraints, one finds that
both $q_x$ and $q_y$ are integrated over a small bounded region in
which both are $O(\omega)$.  Thus for such generic points on the
Fermi surface, we should consider the limit of
$U^{(2)\prime\prime}_{\mu\nu}({\bf q},\omega-\epsilon_{\bf k+q})$
in which all arguments are $O(\omega)$.  In this limit, inspection
of Eqs.~(\ref{eq:alphashift},\ref{eq:vorthopalpha}) shows that the
correlator satisfies a scaling form,
\begin{equation}
  \label{eq:genericscaling}
  U^{(2)\prime\prime}_{\mu\nu}({\bf q},\omega-\epsilon_{\bf k+q}) \sim
  \omega^{2(\Delta_v-2)} {\cal U}(q_x/\omega, q_y/\omega),
\end{equation}
with a well-behaved limit as ${\bf q}\rightarrow 0$.  Inserting this
into Eq.~(\ref{eq:Imself}) and rescaling the momentum integrals by
$\omega$, one sees that $\Sigma''_{cf}({\bf k}_F,\omega) \sim
\omega^{2(\Delta_v-1)}$ for $\theta_F \neq 0,\pi$.  Since
$\Delta_v\geq 2$ in the stable region of the RFL, this dependence is
always as weak or weaker than the ordinary $\omega^2$ scattering rate
due to Coulomb interactions in a Fermi liquid.

The special cases when $\theta_F =0,\pi/2$ require separate
consideration.  For these values, the Fermi velocity is along one
of the principal axes of the square lattice, and the linear
approximation for $\epsilon_{\bf k+q}$ is inadequate.  Taking for
concreteness $\theta_F=0$, we have instead $\epsilon_{\bf k+q}
\approx v_F q_x + q_y^2/2m^*$, and it is not obviously consistent
to neglect the $q_y^2$ term.  For the first term (involving
$\tilde{R}''(q_y,\omega-\epsilon_{\bf k+q})$), the constraints
reduce to $0<v_F q_x + q_y^2/2m^* < \omega-v_{rot} |q_y|$.  This
is approximately solved for small $\omega$ by $0< q_y <
(\omega-v_{rot}|q_x|)/v_F$ and $ |q_x|<\omega/v_F$. Thus both
$q_x,q_y$ are again bounded and $O(\omega)$, so the above Fermi
liquid-like scaling applies.

For the second term (involving
$\tilde{R}''(q_x,\omega-\epsilon_{\bf k+q})$), the constraints reduce
to $0<v_F q_x  + q_y^2/2m^* < \omega - v_{rot} |q_x|$.  This is solved
by taking $-\omega/v_{rot}<q_x<\omega/(v_{rot}+v_F)$ and ${\rm
  max}(0,-v_F q_x) < q_y^2/2m^* < \omega -v_F q_x - v_{rot}|q_x|$.
Hence for this term, $q_x$ is $O(\omega)$ while $q_y$ is
$O(\sqrt{\omega})$.  Since with this scaling, $|q_y| \gg \omega \sim
|q_x|$, one has the significant simplification
\begin{equation}
  \label{eq:upsilonkybig}
  \upsilon_y({\bf q},\omega) \sim \frac{\pi v_0^2 v_1^2}{2v_+^2}
  \frac{q_x \lambda_0}{v_{rot}^2q_x^2-\omega^2} \qquad {\rm for}\;|q_y| \gg
  \omega\sim |q_x| .
\end{equation}
Since only $\lambda_0$ appears, in this limit,
$U^{(2)\prime\prime}_{00} \gg U^{(2)\prime\prime}_{ii},
U^{(2)\prime\prime}_{0i}$, i.e. the fluctuations of $\beta_0$ are much
stronger than those of $\beta_i$.  One may therefore approximate
\begin{eqnarray}
  \label{eq:thetaf0}
\left.U^{(2)\prime\prime}_{\mu\nu}({\bf q},\omega)\right|_{\theta_F=0}
& \!\!\! \sim & \!\!\!
  \frac{A(\Delta_v)\pi^2}{2} \frac{t_v^2 v_0^4 v_1^4}{v_+^4}
  q_x^2(\omega^2-v_{rot}^2q_x^2)^{\Delta_v-3} \nonumber \\
  & & \times \Theta(|\omega|-v_{rot}|q_x|) {\rm sgn}\omega \delta_{\mu
    0}\delta_{\nu 0} .
\end{eqnarray}
Inserting this into Eq.~(\ref{eq:Imself}), one may integrate over
$q_y$ (to yield a constant multiplying $\sqrt{|q_x|}$ and rescale $q_x
\rightarrow \omega q_x$, to obtain the result
\begin{equation}
  \label{eq:hotspots}
  \Sigma''_{cf}({\bf k}_F,\omega) \sim \omega^{2\Delta_v-\frac{5}{2}}\qquad
  {\rm for}\, \theta_F=0,\frac{\pi}{2},\cdots .
\end{equation}
At the end point of the RFL, for which $\Delta_v=2$,
this gives the anomalous lifetime $\Sigma''({\bf k}_F,\omega) \sim
\omega^{3/2}$ at these special ``hot spots'' for which the Fermi
velocity is parallel to one of the principle axes of the square
lattice.  For the conventional Fermi surface believed to apply to the
cuprates, this corresponds to the points on the Fermi surface closest
to $(\pi,0)$, $(0,\pi)$.

Comparing the scattering rate at these hot spots to elsewhere on the
Fermi surface, one finds
\begin{equation}
  { \Sigma^{\prime \prime}_{cf;hot}(\omega) \over \Sigma^{\prime
      \prime}_{cf;typ}(\omega) } \sim \sqrt{  m^* v_F^2   \over \hbar
    \omega }  .
\end{equation}
Since the Fermi surface in the cuprates is particularly ``flat''
near $(\pi,0)$, the effective mass would be large, $m^* \gg m_e$,
and a significantly enhanced scattering rate at such ``hot spots''
would be expected in the RFL phase.

In parameter regimes where the RFL phase is unstable (via ``charge
hopping") at low temperatures to a conventional superconductor,
the enhanced scattering at the hot spots will be significantly
suppressed upon cooling below the transition temperature.  Indeed,
the low energy roton excitations are gapped out in the
superconducting phase, and at temperatures well below $T_c$ will
not be appreciably thermally excited. Unimpeded by the rotons, the
electron lifetime will be greatly enhanced relative to that in the
normal state, particularly at the hot spots on the Fermi surface
with tangents along the $\hat{x}$ or $\hat{y}$ axes, for example
at wavevectors near $(\pi,0)$ in the cuprates.

\subsubsection{Scattering mediated by ``superconducting fluctuations'',
  i.e. boson-fermion pair exchange} 
\label{sec:scatt-medi-boson}

A second mechanism of fermion decay is by the ``pairing term'' ${\cal
  H}_\Delta$ in Eq.~(\ref{eq:explicit}).  We supposed $|\Delta|$ is small,
and so consider the first perturbative contribution to the quasiparticle
lifetime, at $O(\Delta^2)$.  In general, for $d$-wave pairing, this
takes the form
\begin{eqnarray}
  \label{eq:sigmadwave}
  \Sigma_{sf}({\bf k},\omega_n) & = & \sum_{i,j=1}^2
  \Delta_i\Delta_j\int_{{\bf q}\Omega_n}\!\!
  \frac{e^{ik_i}\!+\!e^{i(q_i-k_i)}}{2}
  \frac{e^{-ik_j}\!+\!e^{-i(q_j-k_j)}}{2}\nonumber \\
  & & \times
  \frac{G^{cp}_{ij}({\bf
      q},\Omega_n)}{-i(\Omega_n-\omega_n)+\epsilon_{\bf q-k}},
\end{eqnarray}
where 
\begin{eqnarray}
  \label{eq:Gcpijft}
  G^{cp}_{ij}({\bf q},\Omega_n) & = & \sum_{\bf r}\int_\tau
  G^{cp}_{ij}({\bf r},\tau)e^{i {\bf q}\cdot{\bf r}-i\Omega_n\tau},
\end{eqnarray}
with $G^{cp}_{ij}({\bf r},\tau)$ from Eq.~(\ref{eq:Gcpij}).

Our uncertainties in the details of the Cooper pair propagator
(originating from ambiguities in the string geometry) do not allow a
thorough calculation of the resulting self energy.  However, as we will
now show, we can obtain a rough understanding of its scaling properties
by some simple approximations, which we believe do not significantly
effect the results.  In particular, we will assume that the lifetime is
controlled by the small wavevector $|q|\ll \pi$ portion of the Cooper
pair propagator.  In this regime, we expect $G^{cp}_{ij}({\bf
  q},\Omega_n) \approx G^{cp}({\bf q},\Omega_n)$, the Fourier transform
of the {\sl local} Cooper pair propagator studied in depth in
Sec.~\ref{sec:quasi-diagonal-long}.  Making this approximation, one has
$|q|\ll |k|$ for ${\bf k}$ near the quasiparticle Fermi surface, and
one may write
\begin{eqnarray}
  \label{eq:sigmadwave2}
  \Sigma_{sf}({\bf k},\omega_n) & \approx & |\Delta_{\bf k}|^2
  \int_{{\bf q}\Omega_n}
    \frac{G^{cp}({\bf
      q},\Omega_n)}{-i(\Omega_n-\omega_n)+\epsilon_{\bf q-k}},
\end{eqnarray}
with $\Delta_{\bf k}=|\Delta|(\cos k_x-\cos k_y)$.  Note that this form
immediately implies that scattering due to this mechanism is strongly
suppressed upon approaching the nodal regions of the Fermi surface.  

A detailed analysis is now possible based on a spectral representation
of $G^{cp}$, as in Sec.~\ref{sec:scatt-medi-vort}.\cite{BFunpub}\ Unlike
the above case, however (since the power law decay of $G^{cp}({\bf
  r},\tau)$ is approximately isotropic) a much simpler scaling analysis
suffices to obtain the qualitative behavior of the lifetime.  In particular, the scaling form
of Eq.~(\ref{eq:scaling}) implies that
\begin{eqnarray}
  \label{eq:Gcpscalingqw}
  G^{cp}({\bf q},\Omega_n) \sim |\Omega_n|^{2\Delta_c-3} \tilde{\cal
    G}({\bf q}/\Omega_n),
\end{eqnarray}
up to non-singular additive corrections, for small $|q|$ and $\Omega_n$.
Furthermore, for small ${\bf q}$ and ${\bf k}$ on the Fermi surface, one
may write $\epsilon_{\bf q-k} \approx v_{\scriptscriptstyle F}
q_\parallel + q_\perp^2/2m$, where $q_\parallel,q_\perp$ are the
components of ${\bf q}$ parallel and perpendicular to the local Fermi
velocity at ${\bf k}$, and $v_{\scriptscriptstyle F}$ and $m$ are the
magnitude of the local Fermi velocity and effective mass.  The
singularity in the integrand in Eq.~(\ref{eq:sigmadwave2}) is then cut
off by the external frequency $\omega_n$, and rescaling ${\bf
  q}\rightarrow {\bf q}\Omega_n$, one sees that the effective mass term
is negligible, and obtains by power counting,
\begin{eqnarray}
  \label{eq:Sigmapc}
  \Sigma_{sf}({\bf k},\omega_n) \sim |\Delta_{\bf k}|^2
|\omega_n|^{2\Delta_c-1},
\end{eqnarray}
again with possible analytic and sub-dominant corrections.  Upon analytic
continuation to obtain the lifetime, only non-analytic terms contribute,
and we expect 
\begin{eqnarray}
  \label{eq:sigmapppc}
  \Sigma''_{sf}({\bf k},\omega) \sim |\Delta_{\bf k}|^2
  |\omega|^{2\Delta_c-1} {\rm sign}(\omega),
\end{eqnarray}
for ${\bf k}$ on the Fermi surface.  This result can be verified by
more detailed calculations using the spectral representation of
$G^{cp}$.\cite{BFunpub}  Moreover, we expect that for
$k_{\scriptscriptstyle B}T \gg \omega$, $\omega$ can be replaced by
$k_{\scriptscriptstyle B}T$ in this formula.  

As expected, for sufficiently large $\Delta_c$, this lifetime vanishes
rapidly at low energies, and in particular for $\Delta_c>3/2$, this
contribution is sub-dominant to the usual Fermi liquid one.  However, in
the regime with $\Delta_c<\Delta_v$, this is never the case ($\Delta_c <
1/\sqrt{2}$), at least within our simple model without dramatic
corrections to the RL exponents.  Indeed, for $\Delta_c<1$, as supposed
herein, this ``scattering rate'' is much {\sl larger} than the
quasiparticle energy at low frequency.  Taken literally, this implies
increasingly incoherent behavior away from the nodes as temperature is
lowered.  Near the nodes, the amplitude of this strong scattering
contribution vanishes rapidly, similar to the idea of ``cold spots''
proposed by Ioffe and Millis.\cite{IoffeMillis} \

\section{Discussion}

\label{sec:Discussion}

We close with a discussion of some theoretical issues concerning
the vortex-quasiparticle formulation of interacting electrons that
we have been employing throughout. In particular, we contrast our
approach with the earlier $Z_2$ formulation\cite{Z2} and mention
the connection with more standard and more microscopic
formulations of correlated electrons. We then address the possible
relevance of the Roton Fermi Liquid phase to the cuprate phase
diagram and the associated experimental phenomenology.

\subsection{Theoretical Issues}

Since the discovery of the cuprate superconductors, the many
attempts to reformulate theories of two-dimensional strongly
correlated electron in terms of new collective or composite
degrees of freedom, have been fueled on the one hand by related
theoretical successes and on the other by cuprate phenomenology.
The remarkable and successful development of {\it
bosonization}\cite{Bosonization} both as a  reformulation of
interacting one-dimensional electron models in terms of bosonic
fields and as a means to extract qualitatively new physics outside
of the Fermi liquid paradigm, played an important role in a number
of early theoretical approaches to the cuprates\cite{PWAbook}. The
equally impressive successes of the {\it composite boson} and {\it
composite fermion} approaches in the fractional quantum Hall
effect\cite{FQHEbook} were also influential in early high $T_c$
theories, most notably the {\it anyon} theories\cite{Anyon}. Gauge
theories of the Heisenberg and $t-J$ models were notable early
attempts to reformulate 2d interacting electrons in terms of
``spin-charge'' separated variables\cite{U1A,U1B} -- electrically
neutral fermionic spin one-half ``spinons'' and charge $e$ bosonic
holons  -- and were motivated both by analogies with 1d
bosonization and by resonating valence bond ideas\cite{PWAbook}.
More recent approaches to 2d spin-charge separation have
highlighted the connections with superconductivity\cite{NL,Z2}, by
developing a formulation in terms of vortices, Bogoliubov
quasiparticles and plasmons: the three basic collective
excitations of a 2d superconductor. These latter theories were
primarily attempting to access the pseudo-gap regime by approaching
from the superconducting phase\cite{NL} - focusing on low energy
physics where appreciable pairing correlations were manifest.

\subsubsection{Vortex-fermion formulation}

In this paper, although we are employing a formulation with the same
field content -- $hc/2e$ vortices, fermionic quasiparticles and
collective plasmons -- we are advocating a rather different philosophy.
In particular, we wish to use these same fields to describe higher
energy physics in regimes where the physics is decidedly non-BCS-like
even at short distances, most importantly the cuprate normal state near
optimal doping.  The philosophy of this approach is very similar to that
of the $Z_2$ gauge theory proposal of a fractionalized under-doped normal
state.  Indeed, as demonstrated in Sec.~\ref{sec:u1-vortex-spinon} and
Apps.~\ref{ap:enslaveZ2}-\ref{ap:enslaveU1}, our $U(1)$ formulation is
completely (unitarily) equivalent to a $Z_2$ gauge theory.  However,
because the unitary transformation relating the two formulations is
non-trivial, the $U(1)$ formulation is (much) more convenient for the
types of manipulations and approximations we employ here, largely
because the $U(1)$ gauge fields are continuous variables.  The price
paid for the use of the $U(1)$ formulation is that the ``spinon pairing
term'' of the $Z_2$ gauge theory appears non-local in the $U(1)$
Hamiltonian.  

Fortunately, this non-locality and its consequences are readily
understood.  In particular, although the $U(1)$ vortex-quasiparticle
Hamiltonian is microscopically equivalent to a $Z_2$ gauge theory, it is
{\sl also} equivalent (as described in App.~\ref{ap:elecform}) to a
theory of electronic (i.e. charge $e$) quasiparticles coupled to charge
$2e$ bosons (with the latter described in dual vortex variables).  The
``two-fluid'' point of view of this vortex-electron formulation is
convenient for understanding the qualitative properties of the RFL phase
and its descendents, although it should be emphasized that the
current-current couplings (embodied by the gauge fields of the $U(1)$
formulation) between the two fluids are not expected to be weak.  In the
vortex-electron language, the non-local term simply represents coherent
``Josephson coupling'' of electron pairs and the bosons.  The
non-locality of the pairing term arises simply from the non-local
representation of boson operators in the usual $U(1)$ boson-vortex
duality.  The crucial feature which allows us to handle the pairing term
despite its non-locality is the strength of vortex fluctuations in the
bosonic Roton Liquid, which persists even with these strong
current-current couplings.  The resulting power-law decay of bosonic
pair-field correlations (ODQLRO) opens up a regime, corresponding to the
RFL phase, in which the pairing term is irrelevant and can be treated
perturbatively.  We note in passing that, although it is not relevant
for the cuprates, the above discussion makes clear that a RL phase (for
which fermions need never be introduced) should be possible in a purely
bosonic model, which would be interesting in and of itself.

A second important feature of the present formulation is the retention
of the lattice scale structure.  Appropriate to the cuprates, we have
carefully defined the theory on a 2d square lattice which we wish to
identify with the microscopic copper lattice. Since our theory obviously
does not similarly retain physics on atomic {\it energy} scales, our
starting ``bare'' Hamiltonian should be viewed as a {\it low energy} but
not spatially coarse grained effective theory (or at most an effective
theory coarse-grained spatially only insofar as to remove e.g. the $O$
$p$-orbitals).  It is important to emphasize that the very existence of
the Roton Fermi liquid phase {\it requires} the presence of a square
lattice.  In the RFL ground state there is an infinite set of
dynamically generated symmetries, corresponding to a conserved number of
vortices on every row and column of the lattice.  If we study the same
model on a different lattice, say the triangular lattice, the quantum
ground state analogous to the RFL phase (obtained, again, by taking the
roton hopping amplitude $\kappa_r \rightarrow \infty$) is highly
unstable, being destroyed by the presence of an arbitrarily small single
vortex hopping amplitude.

More generally, the importance of employing the dual
vortex-quasiparticle field theory reformulation cannot be
overemphasized.  The novel and unusual RFL phase emerges quite simply
when one takes a large amplitude for the roton hopping term, and the
fixed point theory is also quite simple consisting of a Fermi sea of
quasiparticles minimally coupled to an ``electric'' field with Gaussian
dynamics. It is very difficult to see how one could adequately describe
such a phase working with bare electron operators.  Indeed, we do not
yet understand the simpler task of constructing a microscopic
(undualized) boson model which enters the RL phase, even without the
complications of fermions.  However, previous experience with dual
vortex formulations for bosonic and electronic systems strongly argues
that the RL and RFL theories properly describe physically accessible
(albeit microscopically unknown) models.

An unfortunate drawback of the present formulation, however, is
the apparent lack of {\it direct} connection between the starting
lattice Hamiltonian and any microscopic electron model.  In
particular, it is presently unclear what microscopic electron
physics could be responsible for generating such a large roton
hopping term.  In the absence of a microscopic foundation, one
must resort to developing phenomenological implications of the
Roton Fermi Liquid phase and comparing with the experimentally
observed cuprate phenomenology.  We take a preliminary look in
this direction in Sec.~\ref{sec:cuprates-rfl-phase} below.

\subsubsection{Related approaches}

It is instructive to contrast the RFL phase with earlier notions of
spin-charge separation.  Anderson's original picture of the cuprate
normal state at optimal doping consisted of gas of spinons with a
Fermi surface co-existing and somehow weakly interacting with a gas of
holons\cite{PWA1,PWAbook} In the early gauge theory implementations of
this picture, both the spinons and holons carried a $U(1)$ (or
$SU(2)$) gauge charge\cite{U1A,U1B}.  Mean-field phase diagrams were
obtained by pairing the spinons and/or condensing the holons, and a
pseudo-gap regime with d-wave paired spinons yet uncondensed holons was
predicted\cite{U1A,U1B} -- several years before experiment (of course
the d-wave nature of the superconducting state was also predicted by
other approaches\cite{Scalapino}). Within this approach, the normal
state at optimal doping was viewed as an incoherent gas of uncondensed
holons strongly interacting with unpaired spinons.  Some efforts were
made to use the neglected gauge field fluctuations to stop the holons
condensing at inappropriately high energy scales, with some
success\cite{Bosongauge}.
A serious worry is that these gauge fluctuations, ignored at the
mean-field level, will almost certainly drive
confinement\cite{Polyakov} (gluing the spinons and the holons
together) and thus at low temperatures invalidate the assumed
stability of the initial mean-field saddle point.

A few early papers emphasized that spinon pairing would break the
continuous $U(1)$ (or $SU(2)$) gauge symmetry down to a discrete $Z_2$
subgroup\cite{U1toZ2,earlyvison2}, and this might be a way to avoid
confinement.  These ideas were later considerably amplified\cite{Z2},
and the stability of genuinely spin-charge separated ground states
established\cite{Frac1,Frac2}.  Such ground states are exotic
electrical insulators which support fractionalized excitations
carrying separately the spin and charge of the electron.  Each
fragment carries a $Z_2$ gauge charge, but in contrast to the $U(1)$
gauge theory saddle points, the $Z_2$ spinons are electrically neutral
and do not contribute additively to the resistance.  Such
fractionalized insulators necessarily support an additional $Z_2$
vortex-like excitation: the vison\cite{vison1}.

Within the present formulation, these $Z_2$ fractionalized insulators
can be readily accessed by condensing {\it pairs of
  vortices}\cite{NL}, rather than condensing rotons.  Since
vortex-{\it pairs} only have statistical interactions with charged
particles and not the spinons, the resulting pair-vortex condensate is
an electrical {\it insulator} with deconfined spinon, ``chargon'' and
vison excitations\cite{NL,Z2} -- dramatically different from the
Roton Fermi Liquid.  Recent experiments on very under-doped cuprate
samples have failed to find evidence of a gapped
vison\cite{visexp1,visexp2}, suggesting that the pseudo-gap phase is
not fractionalized.  But a negative result in these vison detection
experiments does not preclude the RFL phase, which supports mobile
single vortices even at very low temperatures.  A different approach supposing
unbound and mobile single vortices is the QED$_3$ theory of
Ref.~\onlinecite{tesanovic}. 

At the most basic level, a roton is a small pattern of swirling
electrical currents.  Much recent attention has focused on the
possibility of a phase in the under-doped cuprates with non-vanishing
orbital currents\cite{ddw}, counter rotating about elementary plaquettes
on the two sub-lattices of the square lattice.  Such a phase was
initially encountered as a mean-field state within a slave-Fermion gauge
theory approach\cite{stagflux} - the so-called ``staggered-flux phase'',
but has been resurrected as the ``d-density wave'' ordered state of a
Fermi liquid\cite{ddw}.  In either case, such a phase within the present
formulation would be described as a ``vortex-antivortex lattice'' - a
checkerboard configuration on the plaquettes of the 2d square lattice.
Ground states with {\it short-ranged} ``orbital antiferromagnetic''
order have also been suggested recently\cite{stagshort}.  Surprisingly,
the Roton liquid phase also has appreciable short-ranged orbital order.
A ``snapshot view'' of the orbital current correlations in the RL phase
can be obtained by examining the vortex-density structure function.  For
simplicity, consider the charge sector of the RFL theory (the RL) in the
limit of large plasmon velocity, $v_0 \rightarrow \infty$, which
suppresses the longitudinal charge density fluctuations.  The transverse
electrical currents are then described by the Hamiltonian $H_{rot}$ in
Eq.~(\ref{eq:Hrot}) with $\tilde{a}_j \rightarrow 0$.  The
vortex-density structure function, $S_{NN}$, can be readily obtained
from this Gaussian theory:
\begin{equation}
S_{NN}({\bf k}) \equiv  \langle |\hat{N}({\bf k})|^2 \rangle
= \frac{1}{2} \sqrt{\frac{\kappa_r}{u_v}} |{\cal K}_x({\bf k})
{\cal K}_y({\bf k})| \sqrt{{\cal K}^2({\bf k})}   ,
\end{equation}
with $|{\cal K}_j({\bf k})| = 2 |\sin(k_j/2)|$ and ${\cal
K}^2({\bf k})  = \sum_j |{\cal K}_j({\bf k})|^2$ as before. The
vortex structure function is analytic except on the $k_x=0$ and
$k_y=0$ axes and is {\it maximum} at ${\bf k} = (\pi,\pi)$,
indicative of short-ranged orbital-antiferromagnetism. As such,
the RFL/RL phase can perhaps be viewed as a quantum-melted
staggered-flux (or vortex-antivortex) state, with only residual
short-ranged orbital current correlations - the rapid motion of
the rotons being responsible for the melting.  If the fermions are
paired, the amplitude for the basic roton hopping process which
generates the structure function above vanishes upon approaching
half filling, as is apparent from Eq.~(\ref{rotamp}).  A
preliminary analysis\cite{EBL} suggests that the further
neighbor roton hopping processes which do survive at half-filling,
lead to a vortex structure function which vanishes as $|{\bf k}-(\pi,\pi)|$
for wavevectors near $(\pi,\pi)$.

\subsection{Cuprates and the RFL phase?}
\label{sec:cuprates-rfl-phase}

In applying BCS theory to low temperature superconductors, one
implicitly assumes that the normal state above $T_c$ is adequately
described by Fermi liquid theory.  Within a modern renormalization
group viewpoint\cite{Shankar}, this is tantamount to presuming that
the effective Hamiltonian valid below atomic energy scales (of say
10eV) is sufficiently ``close'' (in an abstract space of Hamiltonians)
to the Fermi liquid ``fixed point'' Hamiltonian (actually an invariant
or ``fixed'' manifold of Hamiltonians characterized by the marginal
Fermi liquid parameters).  In practical terms, ``close'' means that
under a renormalization group transformation which scales down in
energies, the renormalized Hamiltonian arrives at the Fermi liquid
fixed point on energy scales which are still well above $T_c$.  BCS
theory then describes the universal crossover flow between the Fermi
liquid fixed point (which is marginally unstable to an attractive
interaction in the Cooper channel) and the superconducting fixed point
which characterizes the universal low energy properties of the
superconducting state well below $T_c$.  Four orders of magnitude
between 10eV and $T_c$ gives the RG flows plenty of ``time'' to
accomplish the first step to the Fermi liquid fixed point, and is the
ultimate reason behind the amazing quantitative success of BCS theory.

\subsubsection{Assumptions underlying the RFL approach}
\label{sec:assumpt-underly-rfl}

In what follows, our working hypothesis is that the effective
electron Hamiltonian on the 10eV scale appropriate for the 2d
copper-oxygen planes at doping levels within and nearby the
superconducting ``dome'', is ``close'' to the Roton Fermi liquid
``fixed manifold'' of Hamiltonians (characterized by Fermi/Bose
liquid parameters for the quasiparticles/rotons). More specifically, we
presume that when renormalized down to the scale of say one-half
of an eV, the effective Hamiltonian can be well approximated by a
Hamiltonian {\it on the RFL fixed manifold} - up to {\it small}
perturbations.  The important small perturbations are those that
are {\it relevant} over appreciable portions of the fixed
manifold.  As established in Sections IV and V, there are three
such perturbations: (i) Single vortex hopping, (ii) ``charge''
hopping and (iii) attractive quasiparticle interactions in the Cooper
channel. When relevant, these three processes destabilize the RFL
phase and cause the RG flows to cross over to different fixed
points which determine the asymptotic low temperature behavior.  From
our calculations, we find that at least one, and often more of these
three perturbations is always relevant, regardless of the roton liquid
parameters.  Hence the RFL should be regarded as a {\sl critical}, and
usually {\sl multi-critical} phase, rather than a stable one.  
For these three processes, the resulting quantum ground states
are, respectively: (i) A (``confined'' and) conventional Fermi
liquid phase, (ii) a conventional superconducting phase with
singlet $d_{x^2-y^2}$ pairing and gapless nodal Bogoliubov
quasiparticles and (iii) a ``rotonic superconductor'' with the
properties of a conventional superconductor but coexisting gapless roton
excitations.  The rotonic superconductor, as described in
Sec.~\ref{sec:selectr-bcs-inst}, has further potential instabilities
driven by either single vortex hopping and ``charge'' hopping/explicit
pairing.  Both perturbations, if relevant, will generate an energy gap
for the rotons.  It seems likely that the domains of relevance of these
two perturbations overlap, so that there no regime of true stability of
the rotonic superconductor.  The true ground state of the system in this
regime is then a conventional superconductor, and its ``rotonic'' nature
is evidenced only as an intermediate energy crossover.

\subsubsection{Effective parameters}
\label{sec:effective-parameters}

In order to construct a phase diagram within this scenario, it is
necessary to specify the various parameters (eg. Bose and Fermi
liquid parameters) of the RFL model as a function of the doping
level, $x$.  Because the RFL is multi-critical, we cannot rely upon
``universality'' to validate ad-hoc requirements of smallness of
perturbations, as might be the case e.g. for renormalized perturbations
around a stable fixed point.  Instead, we will make some assumptions 
based (partly) on physics.  First, our main assumption is the basic
validity at high energies of the roton dynamics, and of Fermi liquid-like
quasiparticles.  Second, we assume that superconductivity is never
strong, i.e. always occurs below the rotonic/Fermi liquid energy scales.
Mathematically, these two assumptions are encompassed in the
inequalities
\begin{eqnarray}
  \label{eq:inequalities}
  \kappa_r, u_v, t_s \gg t_v \gg t_c, \Delta_j.
\end{eqnarray}
Reading from left to right, this corresponds to the first and second
assumptions above.  In practice, we can at best hope for a factor of a
few between e.g. $t_v$ and $\kappa_r$, so the ``$\gg$'' symbols above
should not be taken too strongly.  Although it is not important to our
discussion, it is natural to assume that $v_0 \sim \kappa_r, u_v, t_s$
(since Coulombic energies at the lattice scale are comparable to the
electronic bandwidth).  A third assumption, which is not needed for
consistency of the approach, but seems desirable empirically, is that
the fermion dynamics is primarily by ``spinon'' rather than electron
hopping, $t_s \gg t_e$.  Many of the parameters of the RFL phase can be
fixed empirically from the observed behavior of the cuprates on or above
the eV scale.  For example, the $k-$space location of the quasiparticle
Fermi surface can be chosen to coincide with the electron Fermi surface
as measured via ARPES\cite{ARPES}. The value of other parameters, such
as the bare velocity $v_0$ which appears in ${\cal L}_a$ and sets the
scale of the plasmon velocity, can be roughly estimated from the basic
electronic energy scales, and in any case does not greatly effect the
relevance/irrelevance of the three important perturbations.  For our
basic lattice Hamiltonian introduced in Sec.~\ref{sec:model}, the two
most important parameters characterizing the RFL phase are the vortex
core energy, $u_v$ and the roton hopping strength, $\kappa_r$. Indeed,
as shown in Sec.~\ref{sec:inst-roton-liqu}, the scaling dimensions
which determine the relevance of the vortex and ``charge'' hopping
perturbations at the RFL fixed point, denoted $\Delta_v$ and $\Delta_c$
respectively, depended sensitively on the ratio $u_v/\kappa_r$.  For
example, ignoring renormalizations from the gapless fermions we found,
\begin{equation}
\Delta_v = {1 \over 2 \Delta_c} = \frac{1}{4\pi}
\sqrt{\frac{u_v}{\kappa_r}} \frac{v_+}{\tilde{v}_0}  , 
\end{equation}
with $v^2_+ = \tilde{v}_0^2 + {1 \over 4} u_v \kappa_r = v_0^2 + {1
  \over 2} u_v \kappa_r$.

\begin{figure}
\begin{center}
\vskip-2mm
\hspace*{0mm}
\centerline{\fig{2.8in}{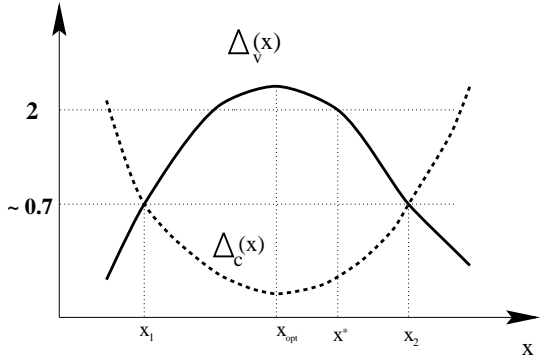}}
\vskip-2mm
\caption{Proposed values for the vortex and ``charge'' hopping scaling
  dimensions, $\Delta_v(x)$ and $\Delta_c(x)$, as a function of the
  doping $x$.  These hopping perturbations are relevant at the RFL
  fixed point when their $\Delta <2$.  Within the doping range, $x_1 <
  x < x_2$, the ``charge'' hopping is {\it more} relevant, and the RFL
  phase is unstable to superconductivity.  }
\label{fig:scaledim}
\end{center}
\end{figure}

Rather than specifying the doping dependence of $u_v, \kappa_r, v_0$
and the Bose/Fermi liquid parameters, though, it is simpler and
suffices for our purposes to specify the final $x-$dependence of
$\Delta_v$ and $\Delta_c$.  Shown in Figure 2 is a proposed form for
$\Delta_{v/c}(x)$, primarily chosen to fit the gross features of
the cuprate phase diagram.  For example, since the RFL phase is
strongly unstable to the Fermi liquid phase when $\Delta_v \ll 2$, we
have taken $\Delta_v$ decreasing to small values in the strongly
over-doped limit.  On the other hand, to account for the observed {\it
  non-Fermi liquid} behavior in the normal state near {\it optimal}
doping\cite{PWAbook}, requires that we take $\Delta_v \ge 2$ in
that regime.  On the under-doped side of the dome we can use the
observed linear $x$-dependence of the superfluid density to guess
the behavior of $\Delta_v$ for small $x$.  In this regime the
(renormalized) vortex core energy in the superconducting state is
presumably tracking the transition temperature, varying linearly
with $x$.  It seems plausible that the ``bare'' vortex core energy
$u_v$ in the lattice Hamiltonian, while perhaps significantly
larger, also tracks this $x-$ dependence. This implies $\Delta_v
\sim \sqrt{u_v} \sim \sqrt{x}$ as depicted in the Figure.
Moreover, to recover (conventional) insulating behavior when $x
\rightarrow 0$, requires that vortex condensation (rather than
charge hopping) be more relevant in this limit, i.e. $\Delta_v<
\Delta_c$ (see below).

\subsubsection{Phase diagram}
\label{sec:phase-diagram}

Under the above assumptions, we now discuss in some detail the
resulting phase diagram and predicted behaviors.  Consider first the
ground states upon varying $x$.  In the extreme over-doped limit with
$\Delta_v \ll 2$, vortex hopping will be a strongly relevant
perturbation at the RFL fixed point.  The vortices will condense at
$T=0$, leading to a conventional Fermi liquid ground state.  Upon
decreasing $x$ there comes a special doping value ($x_2$ in Fig. 2)
where $\Delta_c$ becomes {\it smaller} than $\Delta_v$.  At $x=x_2$
the RFL phase is unstable to {\it both} vortex and ``charge'' hopping
processes, since both $\Delta_v = \Delta_c = 1/\sqrt{2} < 2$.  It seems
reasonable to assume that in this situation, the more strongly
relevant process ultimately dominates at low energies.  This implies
that $x_2$ demarcates the boundary between a Fermi liquid and a
superconducting ground state, as illustrated in Figure 3.  Upon
further decreasing $x$, the ``charge'' hopping becomes even more
strongly relevant, tending to increase $T_c$ until it reaches a
maximum at ``optimal doping'', denoted $x_{opt}$ in Figures 2 and 3.
As one further decreases $x$, $T_c$ should start decreasing.  But as
shown in Sec. VA, with decreasing vortex core energy, $u_v \sim x$,
the enhanced vortex density fluctuations generate an increasing
antiferromagnetic exchange interaction $J \sim t_s^2 / u_v$.  This
{\it attractive} interaction between fermions will mediate
quasiparticle pairing, with a pairing energy scale growing rapidly as $x
\rightarrow 0$.  It is natural to associate this ``quasiparticle pairing''
temperature scale with the crossover temperature into the pseudo-gap
regime\cite{PWAbook}, shown as $T^*$ in Figure 3.  As discussed in
Sec.~\ref{sec:selectr-bcs-inst}, under the assumption that quasiparticle
kinetic energy arises primarily through spinon hopping, $t_s \gg t_e$,
the superfluid density associated with the quasiparticle pairing is
small, so that potential true superconductivity as a consequence of this
pairing is suppressed to a low or zero temperature.   

Since $\Delta_v < \Delta_c$ for doping levels with $x < x_1$, vortex
hopping should again dominate over ``charge'' hopping.  This is the same
condition which we argued leads to the Fermi liquid state for $x>x_2$
above.  However, the physics for small $x$ is more complex, owing to the
strong antiferromagnetic interactions and proximity to the commensurate
filling $x=0$ at which antiferromagnetic order is probable.  In the
$U(1)$ vortex-quasiparticle formalism of this paper, this difference
arises from the freedom to choose (as $\Delta_j\rightarrow 0$) the
fermion density to minimize the total (free) energy of the system.  In
the majority of this paper we have taken $n_f=\rho_0$, in order to
minimize the vortex kinetic energy.  However, if antiferromagnetic
quasiparticle interactions are large, and $x\ll 1$, another possibility
arises.  To optimally benefit from the antiferromagnetic interactions,
one may instead choose the fermion density commensurate, $n_f=1$, for
which the Fermi surface is optimally nested and the quasiparticles can
become fully gapped, gaining the maximal ``condensation energy'' from
antiferromagnetic ordering.  The cost of this choice is some loss of
vortex kinetic energy as the vortex motion becomes somewhat frustrated
by the resulting dual ``flux'' $\pi x$.  Although we assume vortex energy
scales are large, this flux itself is small for small $x$, so eventually
as $x\rightarrow 0$ this rise in kinetic energy becomes smaller than the
lowering of fermionic energy due to antiferromagnetism and such a choice
becomes favorable.  The ultimate physics of the remaining vortices in
this limit is  difficult to analyze reliably, but it
seems evident that for $\Delta_v<\Delta_c$, the more relevant vortex
hopping will lead to an insulating state.  Since the quasiparticles are
then gapped, the remaining uncompensated flux $\pi x$ is expected to
lead to incommensurate charge ordering when the vortices condense -- the
dual analog of the Abrikosov lattice.

\begin{figure}
\begin{center}
\vskip-2mm
\hspace*{0mm}
\centerline{\fig{2.8in}{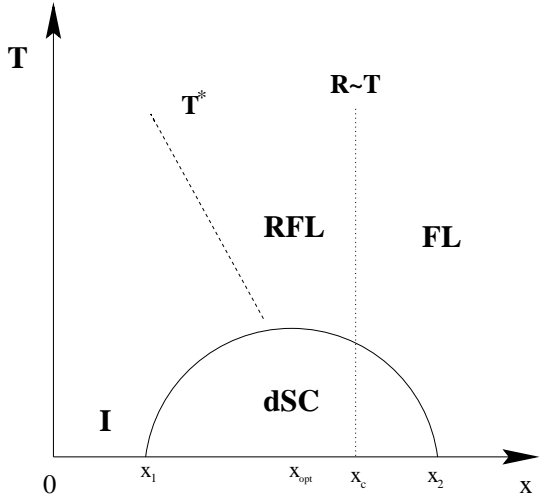}}
\vskip-2mm
\caption{Schematic phase diagram that follows from the doping dependence
  of $\Delta_{v,c}$ shown in Fig. 2.  With increasing doping $x$, the
  ground state evolves from a charge ordered insulator (I), into a
  superconductor (dSC) and then back into a Fermi liquid (FL).  The
  normal state behavior near optimal doping is controlled by the
  Roton Fermi liquid (RFL) ground state.  Below $T^*$, the
  quasiparticles pair.  A quantum phase transition between the RFL and
  FL ground states occurs at $x_c$ - but is preempted by
  superconductivity.  At $x=x_c$, a normal state resistance {\it linear}
  in temperature is predicted.}
\label{fig:phasediag}
\end{center}
\end{figure}

\subsubsection{The RFL Normal state}

Having established the phase diagram which follows from the assumed
doing dependence of $\Delta_v(x)$, we turn to a discussion of the
normal state properties above $T_c$.  Under our working assumption, in
the energy range above $T_c$ we can ignore the small (but ultimately
relevant) ``charge'' hopping perturbation, and use the RFL fixed point
Hamiltonian to describe the physics of the normal state.  First
consider the special doping value, $x = x_c$, where $\Delta_v=2$ on
the over-doped side of the superconducting dome (see Figures 2 and 3).
For $x < x_c$ the vortex hopping strength $t_v$ decreases upon scaling
down in energy, so that it can be treated perturbatively.  As shown in
Sec. VIA, at second order in the vortex hopping, the electrical
conductivity is additive in the roton and quasiparticle conductivities.  With
impurities present the quasiparticle conductivity, $\sigma^f$, will saturate
to low temperatures, whereas $\sigma_{rot} \sim T^{-\gamma}$ (with
$\gamma = 2 \Delta_v -3 $) diverges as $T \rightarrow 0$.  The low
temperature normal state electrical resistance is thus predicted to
behave as a power law,
\begin{equation}
R(T) \sim t_v^2 T^\gamma ,
\end{equation}
with a {\it vanishing residual resistivity}.  Moreover, right {\it at}
$x=x_c$, since $\gamma=1$ a {\it linear} temperature dependence is
predicted.  Notably, this transport behavior is a due to the presence
of the quasi-condensate in the RFL phase, and as such is completely
independent of the single particle scattering rate.

The Hall conductivity, however, will be largely determined by the
fermionic quasiparticle contribution, as detailed in Sec.
\ref{sec:RFLproperties}. Specifically, the cotangent of the Hall
angle, $\cot(\Theta_H) \equiv \rho_{xx}/\rho_{xy}$, was found to
vary as $\cot(\Theta_H) \sim 1/ (T^\gamma \tau_f^2)$ in the RFL
phase, with $\tau_f^{-1}$ the spinon momentum relaxation rate.
Moreover, at the ``hot spots'' on the Fermi surface, $\tau_f^{-1}
\sim T^{\gamma + \frac{1}{2}}$, due to scattering off the gapless
rotons. Under the assumption that these same processes dominate
the temperature dependence of the quasiparticle {\it transport}
scattering rate, we deduce that,
\begin{equation}
\cot(\Theta_H) \sim { 1 \over T^\gamma \tau_f^2 } \sim T^{1
+\gamma} ,
\end{equation}
and right at $x=x_c$, a quadratic dependence $\cot(\Theta_H) \sim
T^2$. In striking contrast to conventional Drude theory which
predicts $\cot(\Theta_H) \sim \tau^{-1} \sim R$ (with $\tau$ the
electron's momentum relaxation time), in the RFL phase the
cotangent of the Hall angle varies with a {\it different power} of
temperature than for the electrical resistance, $R \sim T^\gamma$.
This non-Drude behavior is consistent with the electrical
transport generally observed in the optimally doped
cuprates\cite{expRlin,expHall1,expHall2}, where $R \sim T$ and
$\cot(\Theta_H) \sim T^2$.

Consider next the thermal conductivity $\kappa$ near optimal doping
within the RFL normal state.  One of the most important defining
characteristics of a conventional Fermi liquid is the Wiedemann-Franz
law - the universal low temperature ratio of thermal and electrical
conductivities, $L \equiv \kappa/\sigma T$.  In a Fermi liquid,
electron-like Landau quasiparticles carry both the conserved charge
and the heat, and since the energy of the individual quasiparticles
becomes conserved as $T \rightarrow 0$, the Lorenz ratio is universal,
$L^{FL} = L_0 = \pi^2 k_B^2 / 3 e^2$.  In contrast, the electrical
conductivity is infinite in a superconductor, but the condensate is
ineffective at carrying heat so that the Lorenz ratio vanishes,
$L^{sc} = 0$.  Within the RFL phase, heat can also be transported by
the single fermion excitations, with a contribution to the
thermal conductivity linear in temperature: $\kappa_s = L_0 \sigma^f
T$.  At low enough temperatures this will dominate over the phonon
contribution, $\kappa_{phon} \sim T^3$.  But the roton excitations,
which have a quasi-one dimensional dispersion at low energies, will
presumably also contribute a linear temperature dependence,
$\kappa_{rot} \sim T$.  Thus, the {\it total} thermal conductivity in
the RFL phase is expected to vanish linearly in temperature, $\kappa
\sim T$.  But since the roton contribution to the {\it electrical}
conductivity diverges as $T \rightarrow 0$, the RFL phase is predicted
to have a vanishing Lorenz ratio:
\begin{equation}
L^{RFL} = \frac{\kappa}{\sigma T} \sim T^{\gamma} .
\end{equation}
The quasi-condensate in the RFL phase is much more effective at
transporting charge than heat, much as in a superconductor.  Electron
doped cuprates near optimal doping, when placed in a strong magnetic
field to quench the superconductivity, do exhibit a small Lorenz ratio
at low temperatures\cite{Taillefer}, $L \approx L_0/5$.  But
extracting the zero field Lorenz ratio is problematic, since above
$T_c$ the phonon contribution to $\kappa$ is non-negligible.

It is instructive to consider the electrical resistance also in the {\it
  under-doped regime}, particularly upon cooling below the fermion
pairing temperature.  Above this crossover line the predicted electrical
resistance varies with a power of temperature, $R \sim T^\gamma$.
Since, as we assume, the fermion's kinetic energy comes primarily in the
form of spinon hopping, $t\approx t_s \gg t_e$, the resulting superfluid
density is however very small (see Sec.~\ref{sec:selectr-bcs-inst}), and
phase coherent superconductivity does not result, at least not in this
temperature range.  Nevertheless, one would expect a dramatic increase
in the fermion conductivity, $\sigma^f$, upon cooling through $T^*$ --
much as seen in superconducting thin films upon cooling through the
materials bulk transition temperature.  Since the conductivity is
additive in the roton and spinon contributions, $R(T)^{-1} \sim
\sigma_{rot}(T)+ \sigma^f(T)$, a large and rapid increase in
$\sigma^f(T)$ should be detectable as a drop in the electrical
resistance relative to the ``critical'' power law form, i.e.
\begin{equation}
{ R(T) \over T^\gamma } \sim {1 \over 1 + cT^\gamma \sigma^f(T) }
.
\end{equation}
This behavior is generally consistent with that seen in the under-doped cuprates\cite{PWAbook,expHall2}, provided
we take $\gamma  \approx 1$.

\subsubsection{Entering the superconductor}

We finally discuss the predicted behavior upon cooling from the
RFL normal state into the superconducting phase.  The main change
occurs in the spectrum of roton excitations, which become gapped
inside the superconductor.  The roton gap, $\Delta_{rot}$, should
be manifest in optical measurements, since the optical
conductivity will drop rapidly for low frequencies, $ \omega <
\Delta_{rot}$.  As in BCS theory, the ratio of the (roton) gap to
the superconducting transition temperature $2 \Delta_{rot}/T_c$
should be of order one.  This ratio is determined by the RG
crossover flow between the RFL and superconducting fixed points.
It seems likely that this ratio will be {\it non-universal},
depending on the marginal Bose liquid parameters characterizing
the RFL fixed point (this is distinct from changes in this ratio
in strong coupling Eliashberg theories, since here variations in
the ratio are due to marginal parameters of the RFL fixed manifold
even at arbitrarily weak coupling).  Nevertheless, it would be
instructive to compute this ratio for the simple near-neighbor RFL
model we have been studying throughout, and to analyze the
behavior of the optical conductivity above the gap.

Another important consequence of gapped rotons below $T_c$, is that
the electron lifetime should rapidly increase upon cooling into the
superconducting phase.  With reduced scattering from the rotons, the
ARPES line width should narrow, most dramatically near the normal state
``hot spots'' (with tangents parallel to the $x$ or $y$ axes).  This
behavior is consistent with the ARPES data in the
cuprates\cite{ARPES}, which upon cooling into the superconductor does
show a significant narrowing of the quasiparticle peak, particularly
so at the Fermi surface crossing near momentum $(\pi,0)$.

Despite these preliminary encouraging similarities between various
properties of the RFL phase and the cuprate phenomenology, much more
work is certainly needed before one can establish whether this exotic
non-Fermi liquid ground state might actually underlie the physics of
the high temperature superconductors.  We have already emphasized the
strengths of this proposition, but there are, of course, some
experimental features which seem challenging to explain from this point
of view.  The ``quasiparticle charge'', i.e. temperature derivative of
the superfluid density $\left.\partial K_s/\partial T\right|_{T=0}$
appears, based upon a small number of experimental data points, to be
large and roughly independent of doping $x$, in apparent conflict with
the RFL prediction.  The linear temperature dependence of the
electron lifetime $1/\tau_f \sim k_B T/\hbar$ observed for nodal
quasiparticles near optimal doping in ARPES also does not seem natural
in the RFL.  However, it seems likely that it may be possible to explain
a small number of such deviations from the {\sl most naive} RFL
predictions by more detailed considerations.  Further investigations of
the RFL proposal should confront other experimental probes, such as
interlayer transport and tunneling.
Towards this end, it will be
necessary to generalize the present approach to three-dimensions.
More detailed predictions, such as for the optical conductivity upon
entering the superconductor and the thermal Hall effect in the RFL
normal state, might also be helpful in this regard.  It
would of course be most appealing to identify a {\it new ``smoking gun''}
experiment for the Roton Fermi liquid state, analogous to the
vison-trapping experiment\cite{vison1} for detecting 2d spin-charge
separation.  However, given the critical nature of the RFL, with copious
gapless excitations with varieties of quantum numbers, finding such an
incontrovertible experimental signature may be difficult.

\begin{acknowledgments}
  
  We are grateful to Arun Paramekanti and Ashvin Vishwanath for a number
  of helpful conversations, and to Patrick Lee for commenting on a draft
  of the manuscript.  We are especially indebted to T.  Senthil for his
  many clarifying remarks, particularly pointing to the importance of
  ``spinon pairing'' terms in the Hamiltonian.  This work was generously
  supported by the NSF; M.P.A.F. under Grants DMR-0210790 and
  PHY-9907949, and L.B. by grant DMR-9985255. L.B. also acknowledges
  support from the Sloan and Packard foundations.

\par

\end{acknowledgments}

\appendix

\section{Fermi Liquid Phase} \label{sec:fermiliquid}

We expect the Fermi Liquid phase to occur upon complete proliferation
and unbinding of vortices.  To obtain the Fermi liquid in our
formulation, we therefore in this appendix consider the limit of large
vortex hopping $t_v \rightarrow \infty$ and small vortex energy
$u_0\rightarrow 0$.  We have previously demonstrated the equivalence
of the U(1) gauge theory to a $Z_2$ gauge theory, and it is
this latter formulation which is most convenient in this limit.

To observe the Fermi liquid, we analyze the $Z_2$ gauge theory in its
Hamiltonian form.  Although this limit is very straightforward, based
on previous work on $Z_2$ gauge theory, it is instructive to go
through it in some detail here, in order to observe the effects of the
spinon pairing term.  The Hamiltonian density can be separated into
pure gauge, charge, and spin parts, ${\cal H}={\cal H}_h+{\cal
  H}^{Z_2}_c+{\cal H}^{Z_2}_s$, with
\begin{equation}
  {\cal H}_h = -h \sum_j \sigma_j^x({\bf r}),
\end{equation}
where $\sigma^x_j$ is the usual Pauli matrix in the space of states on
link in the $j$ direction coming from site ${\bf r}$ (and hence
anticommutes with $\sigma^z_j$).  In the charge sector,
\begin{equation}
  {\cal H}^{Z_2}_c = -2t_c \sum_j \sigma_j^z({\bf r}) \cos(\phi_{{\bf r} +
    {\bf \hat{x}}_j}-\phi_{\bf r} ) + u_c n_{\bf r}[n_{\bf r}-1],
\end{equation}
where $n_{\bf r}$ is the number operator conjugate to $\phi_{\bf r}$,
satisfying $[\phi_{\bf r},n_{\bf r'}]=i\delta_{\bf rr'}$.  For
simplicity, in the spin sector we consider a local $s$-wave pair field
instead of a $d$-wave one. This is not essential, but simplifies the
presentation and still addresses the essential issue of the relevance
of spinon pairing in the Fermi liquid.  Hence,
\begin{eqnarray}
  {\cal H}^{Z_2}_s & = & -  t_s \sigma_j^z({\bf r})
  [ f^\dagger_{{\bf r} +  {\bf \hat{x}}_j \sigma}
    f^{\vphantom\dagger}_{{\bf r} \sigma}+{\rm h.c.}] + \epsilon_0
    f^\dagger_{{\bf r}\sigma}f^{\vphantom\dagger}_{{\bf r}\sigma}
    \nonumber \\
    && +
  \Delta(f_{{\bf r}\uparrow}f_{{\bf r}\downarrow} + {\rm h.c.}) + u_f
  (f^\dagger_{{\bf r}\sigma}f^{\vphantom\dagger}_{{\bf r}\sigma})^2.
\end{eqnarray}
Note that we have added a local on-site energy $\epsilon_0$ and
interaction $u_f$, allowed in general by symmetry.  The Hamiltonian
commutes with the gauge generators,
\begin{equation}
  G_{\bf r}=(-1)^{n_{\bf r}+ f^\dagger_{{\bf
        r}\sigma}f^{\vphantom\dagger}_{{\bf r}\sigma}} \prod_{j}
  \sigma^x_j({\bf r}) \sigma^x_j({\bf r}-{\bf\hat{x}}_j).
\end{equation}
We require $G_{\bf r}=1$ to enforce gauge invariance.

In the analysis of the large vortex hopping limit above, we obtained
the action for a $Z_2$ gauge theory with zero kinetic term.  This
corresponds in the Hamiltonian to large $h$ (in particular we will
take $h \gg t_c, t_s$.  In this limit $\sigma_j^x \approx 1$, and the
gauge constraint becomes
\begin{equation}
(-1)^{n_{\bf r}+ f^\dagger_{{\bf
      r}\sigma}f^{\vphantom\dagger}_{{\bf r}\sigma}} =1,
\end{equation}
i.e. requiring an even number of bosons and fermions on each site.
Further, for large $h$, the chargon and spinon hopping terms are
strongly suppressed, and can be considered perturbatively.  At zeroth
order in $t_c,t_s$, then, the charge and spin sectors are decoupled at
each site and decoupled also from one another except by the gauge
constraint.  In the charge sector, for $u_c>0$, it is energetically
favorable to have only either zero or one chargon per site $n_{\bf
  r}=0,1$.  If $n_{\bf r}=0$, then we must have either zero or two
spinons per site.  Note than in this subspace, even for small
$\Delta$, these two local spinon singlet states are non-degenerate:
the energy of the two-spinon state differs from the zero-spinon state
by the energy $\epsilon_0+4u_f$.  With non-zero $\Delta$, one obtains
as eigenstates simply two different linear combination of these two
states on each site.  The lower energy of the two will be realized in
the ground state, and the upper energy state has no physical
significance.  Physically, the lower energy state, which is neutral
($n_{\bf r}=0$) and is a spin singlet, corresponds to the local
vacuum, i.e. a site with no electron on it.  If $n_{\bf r}=1$, then we
must have one spinon on this site, which may have either spin
orientation.  This state has thus the quantum numbers of a physical
electron.  Fixing the total charge of the system $Q=\sum_{\bf r}
n_{\bf r} \neq 0$ will require some number of electrons in the
system.  At zeroth order these are localized, but at O($t_c t_s/h$),
the electrons acquire a hopping between sites, and one obtains a
system clearly in a Fermi liquid phase (it is not non-interacting,
since one has a hard-core constraint in the limit considered).

The above considerations can be applied for zero or non-zero $\Delta$,
and there are no qualitative differences in the results (the detailed
nature of the vacuum state depends smoothly on $\Delta$, leading to a
weak dependence of the effective electron hopping on the ratio
$\Delta/u_f$) in either case.  This strongly suggests that $\Delta$ is
not a ``relevant'' (in the renormalization group sense) perturbation
in the Fermi liquid phase.  This notion can be confirmed more formally
by considering the limit of very weak $\Delta \ll u_f$, in which it
may be treated perturbatively.  At zeroth order in this perturbation
theory, the vacuum state (on a single site) is just the state with
zero spinons.  Formally, the perturbative relevance of $\Delta$ is
determined by the behavior of the two-point function of the pair-field
operator, e.g.
\begin{equation}
  C_\Delta(\tau) = \langle f^{\vphantom\dagger}_{\bf
    r\uparrow}(\tau)f^{\vphantom\dagger}_{\bf
    r\downarrow}(\tau) f^\dagger_{\bf r\downarrow}(0)f^\dagger_{\bf
    r\uparrow}(0) \rangle.
\end{equation}
Since the pair-field operator creates a site doubly-occupied by
spinons, the energy of the intermediate states encountered in the
imaginary time evolution from $0$ to $\tau$ is increased by $u_f$, so
that the spinon pair-field correlator decays exponentially,
$C_\Delta(\tau) \sim e^{-4u_f\tau}$.  This indicates that the spinon
pair field is strongly irrelevant (formally with infinite scaling
dimension).  The importance of this observation in the context of this
paper is that it provides an example in which the spinon pair field --
which naively has a special significance because it alone violates
spinon number conservation -- is irrelevant.  This irrelevance is a
consequence of strong vorticity fluctuations, which bind (confine)
charge to the spinons to form electrons.  Since charge is conserved,
electron number {\sl must} be conserved in the resulting effective
theory.  Similar (but not quite so large) vorticity fluctuations in
the RFL have the effect of rendering the spinon pair field irrelevant.

\section{Enslaving the $Z_2$ gauge fields} \label{ap:enslaveZ2}

 To enslave the $Z_2$ gauge fields, we employ
two sequential unitary transformations, $U_{12}=U_1 U_2$, with
\begin{eqnarray}
  \label{eq:Z2slavers}
  U_1 & \!\!= &\!\!\prod_{\bf r} \left(\prod_{x'=0}^\infty \sigma_1^z({\bf
      r}+ x' \hat{\bf x}) \right)^{n^f_{\bf r}}\!\! \!\!=\!\! \prod_{\bf r}
  (\sigma_1^z({\bf r}))^{\sum_{x'=0}^\infty n^f_{{\bf r}-x'\hat{\bf
        x}}}, \\
  U_2 & \!\!= & \!\!\prod_{\sf r} \left(\prod_{x'=0}^\infty \overline{\sigma}_1^x({\sf
      r}+ x' \hat{\bf x}) \right)^{N_{\sf r}} \!\! \!\!= \!\!\prod_{\sf r}
  (\overline{\sigma}_1^x({\sf r}))^{\sum_{x'=0}^\infty N_{{\sf r}-x'\hat{\bf
        x}}}.
\end{eqnarray}
The two operators $U_1$ and $U_2$ are mutually commuting. Roughly,
$U_1$ transforms to a gauge in which $\sigma_1^z=1$, and $U_2$
transforms to a gauge with $\overline{\sigma}_1^x=1$. More
precisely, applying the first unitary transformation, $H_{pl}$ and
$H_N$ are invariant, while the fermion Hamiltonian transforms to
\begin{eqnarray}
  \label{eq:unitary1}
  U_1^\dagger H^{Z_2}_{f} U_1^{\vphantom\dagger} & = &\left. H^{Z_2}_{f}
  \right|_{\sigma_j^z({\bf r}) \rightarrow \sigma_j^{z,{\rm
        slave}}({\bf r}; N)},
\end{eqnarray}
where
\begin{eqnarray}
  \label{eq:sigmazslave}
  \sigma_1^{z,{\rm slave}}({\bf r}) & = & 1,  \\
  \sigma_2^{z,{\rm slave}}({\bf r}) & = & \prod_{x'=0}^\infty \prod_{\Box({\bf
      r+w}+x'\hat{\bf x})} \sigma^z = \prod_{x'=0}^\infty (-1)^{N_{{\bf
      r+w}+x'\hat{\bf x}}}.   \nonumber
\end{eqnarray}
The vortex kinetic terms also transform
\begin{equation}
  \label{eq:unitary1a}
  U_1^\dagger H^{Z_2}_{kin} U_1^{\vphantom\dagger}  =  \left. H^{Z_2}_{kin}
  \right|_{\overline{\sigma}_j^x({\sf r}) \rightarrow
    \overline{\sigma}_j^x \overline{\sigma}_j^{x,{\rm
        slave}}({\sf r}; n)},
\end{equation}
with
\begin{eqnarray}
  \label{eq:sigmaxslave}
  \overline{\sigma}_1^{x,{\rm slave}}({\sf r}) & = & 1,  \\
  \overline{\sigma}_2^{x,{\rm slave}}({\sf r}) & = & = \prod_{x'=0}^\infty (-1)^{n^f_{{\sf
      r}-\overline{\bf w}-x'\hat{\bf x}}}.   \nonumber
\end{eqnarray}
Simultaneously, the first constraint is rendered trivial
\begin{equation}
  \label{eq:trivialz2constraint1}
 {\cal C}_{\bf r}^{1, {\rm slave}}= U_1^\dagger \tilde{\cal C}_{\bf r}^1
   U_1^{\vphantom\dagger}  =  \prod_{\Box({\bf r})} \overline{\sigma}^x = 1.
\end{equation}
Further transformation with $U_2$ removes $\overline{\sigma}^x_j$
from $H^{Z_2}_{kin}$ and trivializes the remaining constraint,
i.e.
\begin{equation}
  \label{eq:trivialz2constraint2}
 {\cal C}_{\sf r}^{2, {\rm slave}}= U_2^\dagger \tilde{\cal C}_{\sf r}^2
   U_2^{\vphantom\dagger}  =  \prod_{\Box({\sf
        r})} \sigma^z = 1.
\end{equation}
The final transformed Hamiltonian no longer involves any dynamical
gauge fields (whose state is uniquely specified by
Eqs.~(\ref{eq:trivialz2constraint1},~\ref{eq:trivialz2constraint2})),
and is simply given by
\begin{eqnarray}
  \label{eq:z2slavefinal}
  H_{Z_2}^{\rm slave} & = & U_{12}^\dagger
  \tilde{H}_{Z_2} U_{12}^{\vphantom\dagger}
 = \left. \tilde
 {H}_{Z_2}\right|_{\sigma_j^\mu \rightarrow \sigma_j^{\mu,{\rm
        slave}}}.
\end{eqnarray}
The electron destruction operator in the $Z_2$ vortex-spinon
theory, $\tilde{c}_{{\bf r}\sigma}= \tilde{b}_{\bf r} f_{{\bf
r}\sigma}$ with $\tilde{b}_{\bf r}$ in Eq.~(\ref{btilde}),
transforms upon enslaving in an identical fashion,
\begin{equation}
U^\dagger_{12} \tilde{c}_{{\bf r}\sigma} U_{12}^{\vphantom\dagger}
= \left. \tilde{c}_{{\bf r}\sigma}\right|_{\sigma_j^z \rightarrow
\sigma_j^{z,{\rm slave}}}. \label{slaveZ2electron}
\end{equation}

\section{Enslaving the $U(1)$ gauge theory} \label{ap:enslaveU1}

As for the $Z_2$ case, to enslave the gauge fields in the $U(1)$
formulation we apply two sequential unitary transformations,
$U_{ab} = U_a U_b$, with
\begin{equation}
U_a= e^{i\sum_{{\sf r},{\sf r}^\prime} N_{\sf r} V({\sf r} - {\sf
r}^\prime) (\vec{\nabla}\cdot \vec{\alpha})({\sf r}^\prime)}
e^{i\sum_{{\bf r},{\bf r}^\prime} n^f_{\bf r} V({\bf r} - {\bf
r}^\prime) (\vec{\nabla}\cdot \vec{\beta})({\bf r}^\prime)} ,
\end{equation}
where $\nabla^2 V({\bf r} - {\bf r}^\prime) = \delta_{{\bf r}{\bf
r}^\prime}$ and,
\begin{equation}
U_b = e^{{i \over 2} \sum_{{\bf r},{\sf r}^\prime} n^f_{\bf r}
\Theta({\bf r} - {\sf r}^\prime) N_{{\sf r}^\prime}} ,
\end{equation}
where the lattice Laplacian is
\begin{equation}
  \label{eq:laplacian}
  \nabla^2 f({\bf r}) = \sum_j \partial_j^2 f({\bf r}-{\bf\hat{x}}_j)
  = \sum_j f({\bf r}+{\bf\hat{x}}_j) +  f({\bf r}-{\bf\hat{x}}_j)
  -2f({\bf r}).
\end{equation}

Here we have introduced a ``angle'' function $\Theta({\bf r}-{\sf
r}')$, to be determined later.
The two unitary transformations commute with one another. Again,
both $H_{pl}$ and $H_N$ are invariant commuting with $U_a$ and
$U_b$, whereas under application of $U_a$ the vortex kinetic
energy transforms to,
\begin{equation}
  U_a^\dagger H_{kin} U_a^{\vphantom\dagger}  =  \left. H_{kin}
  \right|_{e^{i\alpha_j} \rightarrow
    e^{i\alpha^t_j} e^{i\alpha_j^{slave}(n^f)} }.
\end{equation}
Here, $\alpha_j^t$ is the transverse part of $\alpha_j$ and
$\alpha_j^{slave}$ satisfies,
\begin{equation}
\epsilon_{ij} \partial_i \alpha_j^{slave}({\bf r} - {\bf w}) = \pi
n^f_{\bf r} ; \hskip1cm \vec{\nabla}\cdot\vec{\alpha}^{slave} = 0
.
\end{equation}
Similarly, the spinon and electron hopping Hamiltonians transform
to,
\begin{eqnarray}
  U_a^\dagger H_{s/e} U_a^{\vphantom\dagger} & = &\left. H_{s/e}
  \right|_{e^{i\beta_j} \rightarrow
    e^{i\beta^t_j} e^{i\beta_j^{slave}(N)}},
\end{eqnarray}
with $\beta_j^t$ the transverse part of $\beta_j$ and
\begin{equation}
\epsilon_{ij} \partial_i \beta_j^{slave}({\sf r} - {\bf w}) = \pi
N_{\sf r} ; \hskip1cm \vec{\nabla}\cdot\vec{\beta}^{slave} = 0
.\label{eq:betaslave}
\end{equation}
Notice that the longitudinal parts of $\alpha$ and $\beta$ have
been eliminated, and the remaining transverse pieces commute with
one another and can be treated as c-numbers. Simultaneously, the
two $U(1)$ constraints are rendered trivial,
\begin{equation}
{\cal G}_f^{slave} = U_a^\dagger {\cal G}_f U_a^{\vphantom\dagger}
= e^{-{i \over \pi} \sum_{\bf r} \Lambda_{\bf r} \epsilon_{ij}
\partial_i \alpha^t_j({\bf r} - {\bf w})} = 1  ,
\end{equation}
\begin{equation}
{\cal G}_v^{slave} = U_a^\dagger {\cal G}_v U_a^{\vphantom\dagger}
= e^{{i \over \pi} \sum_{\sf r} \chi_{\sf r} \epsilon_{ij}
\partial_i \beta^t_j({\sf r} - {\bf w})} = 1  ,
\end{equation}
implying that $\alpha_j^t=\beta^t_j =0$, and fully eliminating
both gauge fields from the full transformed Hamiltonian,
\begin{equation}
U_a^\dagger H U_a^{\vphantom\dagger} = \left. H
  \right|_{e^{i\alpha_j},e^{i\beta_j} \rightarrow e^{i\alpha_j^{slave}},
  e^{i\beta_j^{slave}}}  .
  \end{equation}

Further transformation with $U_b$, which is essentially a
non-singular gauge transformation of both the vortices and
spinons, modifies the enslaved gauge fields (so that they vanish
on the horizontal bonds and are integer multiples of $\pi$ on the
vertical bonds) while leaving the gauge fluxes invariant.
Specifically, we require
\begin{eqnarray}
  \label{eq:thetarequire}
  \partial_x \Theta({\bf r}) & = &
  2\tilde\beta_x({\bf r}), \qquad \forall {\bf r},\\
  \partial_y \Theta({\bf r}) & = & 2\tilde\beta_y({\bf r}) \qquad \forall
  {\bf r} \neq x{\bf\hat{x}}+{\bf\overline{w}}, x\geq 0.
\end{eqnarray}
 Here $\tilde\beta_j({\bf r})$ is the enslaved gauge field
configuration for a vortex located at ${\sf r}=0$ (i.e. determined
from Eqs.~(\ref{eq:betaslave}) with $N_{\sf r}=\delta_{\sf r,0}$).
The transverseness of $\tilde\beta_j$ implies then
\begin{equation}
  \label{eq:ThetaLaplace}
  \nabla^2 \Theta({\bf r}+{\bf\overline{w}}) = \psi_x
  (\delta_{y,0}-\delta_{y,-1}),
\end{equation}
with an unknown function $\psi_x$ such that $\psi_x=0$ for $x<0$.
Taking a line sum around the origin requires then
\begin{equation}
  \label{eq:bcslave}
  \partial_y \Theta(x{\bf\hat{x}}+{\bf\overline{w}}) = -2\pi
  +2\tilde\beta_y(x{\bf\hat{x}}+{\bf\overline{w}}),
\end{equation}
for $x\geq 0$.  Eqs.\ref{eq:ThetaLaplace},\ref{eq:bcslave} are the
lattice analog of Laplace's equation and the condition that
$\Theta$ jumps by $2\pi$ across the positive $x$-axis.  These
conditions determine $\psi_x$ and hence $\Theta$.  After some
algebra, the solution is expressible as a Fourier integral
\begin{equation}
  \label{eq:Thetafour}
  \Theta({\bf r}) = \int_{\bf k} \Theta({\bf k}) e^{i{\bf k\cdot r}},
\end{equation}
where
\begin{equation}
  \label{eq:Thetasol}
  \Theta({\bf k}) = -\frac{{2\pi} e^{i {\bf
        k\cdot\overline{w}}}}{{\cal K}^2 F(k_x)} \left[ \frac{{\cal
        K}_y^*}{{\cal K}_x^*}- {{\cal
        K}_x^*}{{\cal K}_y^*} I(k_x)\right],
\end{equation}
with
\begin{eqnarray}
  \label{eq:integrals}
  F(k_x) & = & 1 - \frac{|\sin(k_x/2)|}{\sqrt{\sin^2(k_x/2)+1}}, \\
  I(k_x) & = & \frac{1}{4\sin(k_x/2) \sqrt{\sin^2(k_x/2)+1}}.
\end{eqnarray}
For large arguments, the asymptotic behavior can be obtained,
\begin{equation}
\Theta({\bf r}) \sim {\rm arctan}[y/x],
\end{equation}
for $\sqrt{x^2+y^2}\gg 1$, with the ${\rm arctan}$ defined on the
interval $[0,2\pi]$.  Hence $\Theta({\bf r})$ gives a proper
lattice version of the continuum angle function.

With this definition, one finds
\begin{equation}
H^{slave} = U_{ab}^\dagger H U_{ab}^{\vphantom\dagger} = \left. H
  \right|_{e^{i\alpha_j},e^{i\beta_j} \rightarrow \overline{\sigma}_j^{x,slave},
  \sigma_j^{z,slave}}  ,
  \end{equation}
with $\overline{\sigma}_j^{x,slave}({\sf r};n^f)$ and $
\sigma_j^{z,slave}({\bf r};N)$ as defined in
Eq.~(\ref{eq:sigmaxslave}) and Eq.~(\ref{eq:sigmazslave}),
respectively.  Remarkably, the enslaved $U(1)$ Hamiltonian is {\it
identical} to the enslaved $Z_2$ Hamiltonian in
Eq.~(\ref{eq:z2slavefinal}); $H^{slave} \equiv H_{Z_2}^{slave}$.

We have thereby established the formal equivalence between the
$Z_2$ and $U(1)$ formulations of the vortex-spinon field theory -
the unitarily transformed enslaved versions of the original
Hamiltonians $\tilde{H}_{Z_2}$ and $H$ in
Eqs.~(\ref{HamZ2vs},\ref{HamU1vs}) are identical to one another.
Finally, we can verify that the enslaved versions of the electron
operators in the $U(1)$ and $Z_2$ formulations also coincide. From
the definition of the electron operator in the $U(1)$ formulation,
$c_{{\bf r}\sigma}$ in Eq.~(\ref{electronU1}), one can readily
show that,
\begin{equation}
 U_{ab}^\dagger c_{{\bf r}\sigma}
U_{ab}^{\vphantom\dagger} = \left.  c_{{\bf r}\sigma}
  \right|_{e^{i\beta_j} \rightarrow
  \sigma_j^{z,slave}}  .
  \end{equation}
With the analogous expression for the enslaved $Z_2$ electron
operator in Eq.~(\ref{slaveZ2electron}), and upon comparing the
defining expressions of the electron operators in the $Z_2$ and
$U(1)$ formulations in Eqs.~(\ref{electronZ2}) and
(\ref{electronU1}), respectively, one thereby establishes the
desired formal equivalence: $U^\dagger_{12} \tilde{c}_{{\bf
r}\sigma} U_{12} \equiv U_{ab}^\dagger c_{{\bf r}\sigma}U_{ab}$.

\section{The Vortex-electron formulation} \label{ap:elecform}

In this Appendix we briefly discuss a third Hamiltonian
formulation of the vortex-fermion field theory.  The
vortex-electron Hamiltonian will be expressed in terms of
``electron operators'', or more correctly operators which create
excitations having a non-vanishing overlap with the bare electron.
In order to transform to this formulation, we start with the
enslaved version of the $U(1)$ vortex-spinon Hamiltonian as
obtained in Appendix \ref{ap:enslaveU1}:
\begin{equation}
U_a^\dagger H U_a^{\vphantom\dagger} = \left. H
  \right|_{e^{i\alpha_j},e^{i\beta_j} \rightarrow e^{i\alpha_j^{slave}},
  e^{i\beta_j^{slave}}}  ,
  \end{equation}
where $\alpha_j^{slave}$ and $\beta_j^{slave}$ satisfy,
\begin{equation}
\epsilon_{ij} \partial_i \alpha_j^{slave}({\bf r} - {\bf w}) = \pi
n^f_{\bf r} ; \hskip1cm \vec{\nabla}\cdot\vec{\alpha}^{slave} = 0
,
\end{equation}
\begin{equation}
\epsilon_{ij} \partial_i \beta_j^{slave}({\sf r} - {\bf w}) = \pi
N_{\sf r} ; \hskip1cm \vec{\nabla}\cdot\vec{\beta}^{slave} = 0 .
\end{equation}
Now consider the unitary transformation,
\begin{equation} U_{el}= e^{{i \over 2} \sum_{{\bf r},{\bf
r}^\prime} n^f_{\bf r} V({\bf r} - {\bf r}^\prime) \epsilon_{ij}
\partial_i e_j({\bf r}^\prime - {\bf w})} ,
\end{equation}
where again $\nabla^2 V({\bf r} - {\bf r}^\prime) = \delta_{{\bf
r}{\bf r}^\prime}$.  As is apparent from
Eqs.~(\ref{SvortSphi},\ref{elecspinon}) this transformation takes
one from the spinon operator to the electron operator,
\begin{equation}
U^\dagger_{el} f_{{\bf r}\sigma} U_{el}  = {\cal S}_\phi({\bf r})
f_{{\bf r}\sigma} = c_{{\bf r} \sigma} .
\end{equation}
Here ${\cal S}_{\phi}({\bf r}) = \prod_{\bf r}^\infty
 e^{-i \pi
e_j^t}$ is defined in Eq.~(\ref{SvortSphi}) and the last equality
follows from Eq.~(\ref{elecspinon}) in the enslaved gauge with
purely transverse gauge field, $\beta^{\ell}=0$. The electrical
charge density in the vortex-spinon formulation transforms to
include the {\it electron} density:
\begin{equation}
U^\dagger_{el} \epsilon_{ij} \partial_i a_j({\bf r} - {\bf w})
U_{el} = \epsilon_{ij} \partial_i a_j({\bf r} - {\bf w})  + \pi
c^\dagger_{{\bf r}\sigma} c_{{\bf r}\sigma} .
\end{equation}

The full Hamiltonian density within the vortex-electron
formulation is readily obtained from the enslaved vortex-spinon
Hamiltonian:  ${\cal H}_{ve} = U^\dagger_{el} U^\dagger_a {\cal H}
U_a U_{el}$.  It can be compactly expressed as,
\begin{eqnarray}
{\cal H}_{ve} &=& {u_v \over 2} \sum_j e^2_j + {v_0^2  \over 2 u_v
} [ \epsilon_{ij} \partial_i a_j + c^\dagger_{{\bf r}\sigma}
c_{{\bf
r}\sigma} - \pi \rho_0 ]^2 \nonumber \\
 &+& {\cal H}_v(a_j) + {\cal H}_r(a_j) +
{\cal H}_f  ,
\end{eqnarray}
where ${\cal H}_v$ and ${\cal H}_r$ denote the vortex and roton
hopping terms, respectively, and are given explicitly as,
\begin{equation}
{\cal H}_v(a_j) = -t_v \sum_j \cos(a_j)  ,
\end{equation}
\begin{equation}
{\cal H}_{r}(a_j) = -{\kappa_r \over 2} \sum_i \cos(\epsilon_{ij}
\partial_i a_j)  .
\end{equation}
The fermionic Hamiltonian is
\begin{eqnarray}
  \label{eq:Hfelect}
{\cal H}_f  & = &  - \sum_j t_e c^\dagger_{{\bf r} + {\bf
\hat{x}}_j \sigma}
    c^{\vphantom\dagger}_{{\bf r} \sigma} \\
     & & \hspace{-0.5in} -  \sum_j e^{i\pi \overline{e}_j ({\bf r}) } [t_s
c^\dagger_{{\bf r} + {\bf \hat{x}}_j \sigma}
    c^{\vphantom\dagger}_{{\bf r} \sigma} + \Delta_j B_{\bf r}^\dagger c_{{\bf r}+\hat{\bf
        x}_j \sigma} \epsilon_{\sigma\sigma'}c_{{\bf r}\sigma'} + {\rm
      h.c.}] . \nonumber
\end{eqnarray}
The electric field appearing in the spinon hopping term yields the
same physical effects as the gauge field in the $U(1)$
formulation. Specifically, when the electron hops from one site to
a neighboring site,  the factor $e^{i\pi \overline{\epsilon}}$
which shifts the gauge field $a_j$ by $\pi$, effectively hops a
compensating chargon in the opposite direction. In addition, this
minimal coupling form encodes the requisite minus sign when a
spinon is hopped around a vortex and vice versa. The full
Hamiltonian must be supplemented with the constraint that
$\vec{\nabla} \cdot \vec{e} = N$, with integer $N$. We emphasize
that the total spin,
\begin{equation}
\vec{S} = {1 \over 2} \sum_{\bf r} c^\dagger_{{\bf r}\sigma}
\vec{\tau}_{\sigma \sigma^\prime} c^{\vphantom\dagger}_{{\bf
r}\sigma^\prime} ,
\end{equation}
and electric charge,
\begin{equation}
Q = \sum_{\bf r} [ {1 \over \pi} \epsilon_{ij} \partial_i a_j({\bf
r} - {\bf w}) + c^\dagger_{{\bf r}\sigma}
c^{\vphantom\dagger}_{{\bf r}\sigma} ] ,
\end{equation}
are conserved, commuting with ${\cal H}_{ve}$.

It is of course also possible to pass to a Euclidean path integral
representation of the partition function associated with the above
vortex-electron Hamiltonian. Specifically, the corresponding
Euclidean Lagrangian can be readily expressed as,
\begin{equation}
{\cal L}_{ve} = i e_j \partial_0 a_j + c^\dagger_{\bf r}
\partial_0 c^{\vphantom\dagger}_{\bf r} + ia_0(\vec{\nabla} \cdot
\vec{e} - N_{\sf r}) + {\cal H}_{ve} ,
\end{equation}
where the time component of the gauge field $a_0({\sf r})$ lives
on the sites of the dual lattice.  In the partition function, the
vortex number $N_{\sf r}$ is a continuous field running over the
real numbers, but the integration only contributes when $N$ is
integer.  To see why, it is instructive to let $a_\mu \rightarrow
a_\mu - \partial_\mu \theta$, and to integrate the vortex phase
variable $\theta$ over the reals.  Since the Hamiltonian is $2\pi$
periodic in $\theta$, upon splitting the integration as
$\partial_0 \theta = 2\pi \ell + \partial_0 \tilde{\theta}$ with
$\tilde{\theta}=[0,2\pi]$, the summation of $exp(i2\pi \ell N)$
over integer $\ell$ vanishes unless the vortex number $N$ is an
integer.

To obtain a more tractable representation of the Lagrangian, we
introduce a Hubbard-Stratonovich field, $e_0({\bf r})$, to decouple the
Coulomb interaction term above.  Here $i e_0$ has the physical meaning of
a dynamical electrostatic potential.  In this way the full
Euclidean Lagrangian can be conveniently decomposed as the sum of a
bosonic charge sector and a fermionic spin and charge carrying sector,
${\cal L}_{ve} = {\cal L}_c + {\cal L}_f$.  The full bosonic sector is
given by,
\begin{eqnarray}
{\cal L}_c &=& {u_v \over 2} [e_j^2 + {1 \over {\pi^2 v_0^2}} e_0^2 ] +
ie_j(\partial_0 a_j - \partial_j a_0) - \frac{i}{\pi} e_0 \epsilon_{ij}
\partial_i a_j \nonumber \\
&+& i (\partial_0 \theta - a_0) N + {\cal L}_v + {\cal L}_r ,
\end{eqnarray}
with vortex and roton hopping terms,
\begin{equation}
{\cal L}_v = -t_v \sum_j \cos(\partial_j \theta - a_j)  ,
\end{equation}
\begin{equation}
{\cal L}_{r} = -{\kappa_r \over 2} [ \cos(\Delta_{xy} \theta -
\partial_x a_y) + (x \leftrightarrow y)] .
\end{equation}
The Lagrangian density in the fermionic sector is,
\begin{equation}
{\cal L}_f = c^\dagger_{\bf r} (\partial_0 - i e_0) c_{\bf r} +
{\cal H}_f + i e_0 \rho_0  .
\end{equation}

In addition to the global symmetries corresponding to spin and
charge conservation, the full Lagrangian has a local gauge
symmetry, being invariant under,
\begin{eqnarray}
    \label{eq:gauge1}
    \theta_{\sf r} & \rightarrow & \theta_{\sf r} + \Theta_{\sf r},
    \nonumber \\ 
    a_\mu({\sf r}) & \rightarrow & a_\mu({\sf r}) + \partial_\mu \Theta_{\sf
    r},
  \end{eqnarray}
with $\Theta_{\sf r}$ an arbitrary function of space and imaginary
time.  Because of this gauge invariance we are free to choose an
appropriate gauge.

\section{Polarization tensor}
\label{sec:appendix}

In this appendix, we calculate the corrections to the polarization
tensor $\Pi_{ij}$ at $O(t_v^2)$ including fluctuations of both
$\tilde{a}$ and $\theta$, for the general case $v_0<\infty$.
Integrating out all dynamical fields to $O(t_v^2)$, one finds that the
effective action as a functional of $A_j$ takes the
form $S_A^{\rm eff} = S_A^0 + S_A^{(2)}$, where
\begin{equation}
  \label{eq:sa2}
  S_{A}^{(2)}  =  -\frac{t_v^2}{2}e^{S_A^0}    \sum_{\sf r,r'}
  \int_{\tau\tau'}
 \bigg\langle  C_i({\sf r},\tau) C_j({\sf r}^\prime,\tau^\prime)   e^{-S_A} \bigg\rangle_{A=0},
\end{equation}
with the shorthand notation, $C_i({\sf r},\tau) = \cos(\partial_i \theta -
    \tilde{a}_i)_{{\sf r}\tau}$ and with
the $\langle \cdot\rangle_{A=0}$ indicating a Gaussian average
with respect to the RL action.  This can be written as
\begin{equation}
  \label{eq:sa2a}
  S_A^{(2)} = -\frac{t_v^2}{4} e^{S_A^0} \left( e^{\langle \Gamma_x^2
      \rangle_{A=0}} + e^{\langle \Gamma_y^2 \rangle_{A=0}}\right),
\end{equation}
with
\begin{eqnarray}
  \label{eq:gammadef}
  \Gamma_x & = & \int_{{\bf k},\omega_n}
[ (\psi_{{\bf k}\omega_n}  \frac{{\cal
      K}_x^*}{{\cal K}} + \frac{i\omega_n {\cal K}_j^* A_j}{\pi{\cal
      K}}) a({\bf k},\omega_n)  \nonumber \\
& & - \psi_{{\bf k}\omega_n} {\cal K}_y
  \theta({\bf k},\omega_n) ] ,
\end{eqnarray}
$\psi_{{\bf k}\omega_n} = e^{i({\bf k}\cdot{\sf r}-\omega_n\tau)}
-  e^{i({\bf k}\cdot{\sf r'}-\omega_n\tau')}$, and $\Gamma_y$ obtained
from $\Gamma_x$ by $x\leftrightarrow y$.  Evaluating the
expectation value gives
\begin{eqnarray}
  \label{eq:sa2b}
  S_A^{(2)} & \sim & -t_v^2 \sum_{xx'y}\int_{\tau\tau'}
    \frac{1}{[x^2+v_{rot}^2(\tau-\tau')^2]^{\Delta_v}} \\
    & \times & e^{\int_{{\bf k}\omega_n} \frac{i\omega_n
        u_v}{\pi{\cal K}^2} \psi_{-{\bf k},-\omega_n} (\tilde{\cal
        K}_y G_{12} + \tilde{\cal K}_x G_{11}) \tilde{K}_j A_j({\bf
        -k},-\omega_n)} \nonumber \\ & & + x\leftrightarrow y. \nonumber
\end{eqnarray}
Since we are interested in the polarization tensor, we may expand the
exponential in Eq.~(\ref{eq:sa2b}) to $O(A^2)$ to obtain $\Pi_{ij} =
\Pi_{ij}^0 + \Pi_{ij}^{(2)}$, with
\begin{eqnarray}
  \label{eq:polar2}
  \Pi_{ij}^{(2)} & \sim & -\frac{t_v^2\omega_n^2 u_v^2}{2{\cal K}^4}
  (\omega_n^2+v_{rot}^2 k_x^2)^{\Delta_v-1} \nonumber \\ &  \times &
   |\tilde{\cal K}_y G_{12} + \tilde{\cal K}_x G_{11}|^2 \tilde{K}_i
   \tilde{K}_j + (x \leftrightarrow y).
\end{eqnarray}
In the limit of interest for the conductivity, $|{\bf k}|\rightarrow
0$ at fixed frequency, $G_{11} \gg G_{12}$, and we obtain
Eq.~(\ref{eq:polar3}) of the main text.

\end{document}